\documentclass{iopart}

\usepackage{amsfonts,amssymb}

\newtheorem{remark}{Remark}

\newtheorem{example}{Example}

\def\bbbe{\Bbb E}
\def\bbbr{{\Bbb R}}
\def\bbbt{{\Bbb T}}
\def\bbbo{{\Bbb O}}
\def\bbbc{{\Bbb C}}
\def\bbbe{{\Bbb E}}
\def\bbbz{{\Bbb Z}}
\def\bbbd{{\Bbb D}}
\def\wedgecomma{\mathop{\wedge}\limits_{'}}
\def\ad{\mbox{ad}\,}

\def\re{{\rm Re}\,}
\def\im{{\rm Im}\,}
\def\htt{{\rm ht}\,}
\def\Ad{\mbox{Ad}\,}
\def\Aut{\mbox{Aut}\,}

\def\fr#1{{\mathfrak{#1}}}
\def\openone{\leavevmode\hbox{\small1\kern-3.3pt\normalsize1}}

\eqnobysec

\begin{document}

\title[Reductions of $N $-wave interactions...I]
{Reductions of $N $-wave interactions related to low--rank
simple Lie algebras.\; I:
$\bbbz_2$--reductions }
\author[V. S. Gerdjikov et all.]{V. S. Gerdjikov$^\dag$, G. G.
Grahovski$^\dag$ and N. A. Kostov$^ \ddag$}

\address{\dag\ Institute for Nuclear Research and Nuclear Energy, Bulgarian
Academy of Sciences, 72 Tzarigradsko chaussee, 1784 Sofia, Bulgaria}
\address{\ddag\ Institute of Electronics, Bulgarian
Academy of Sciences, 72 Tzarigradsko chaussee, 1784 Sofia, Bulgaria}

\begin{abstract}
The analysis and the classification of all reductions for the nonlinear
evolution equations solvable by the inverse scattering method is
an interesting and still open problem. We show how the second order
reductions of the $N $--wave interactions related to low--rank simple Lie
algebras $\fr{g} $ can be embedded also in the Weyl group of $\fr{g} $.
This allows us to display along with the well known ones a number of new
types of integrable $N $-wave systems. Some of the reduced systems
find applications to nonlinear optics.  \end{abstract}

\jl{1}


\maketitle

\section{Introduction}

It is well known that the $N $--wave equations \cite{ZM,ZM1,K,1,FaTa,KRB}
\begin{eqnarray}\label{eq:1.4}
i[J,Q_t] - i[I,Q_x] +[[I,Q],[J,Q]] = 0,
\end{eqnarray}
are solvable by the inverse scattering method (ISM) \cite{1,FaTa}
applied to the generalized system of Zakharov--Shabat type
\cite{1,Sh,G*86}:
\begin{eqnarray}\label{eq:1.1}
\fl &&L(\lambda )\Psi(x,t,\lambda ) = \left(i{d \over dx} + [J,Q(x,t)]
- \lambda J \right)\Psi(x,t,\lambda ) = 0, \quad J\in \fr{h},\\
\label{eq:1.3.1}
\fl && Q(x,t)= \sum_{\alpha \in\Delta } Q_\alpha (x,t) E_\alpha \equiv
\sum_{\alpha \in \Delta _+} (q_{\alpha }(x,t)
E_{\alpha } + p_{\alpha}(x,t) E_{-\alpha })
\in \fr{g \backslash \fr{h}},
\end{eqnarray}
where $Q(x,t) $ and $J $ take values in the simple Lie algebra $\fr{g} $.
Here $ \fr{h}$ is the Cartan subalgebra of $\fr{g} $, $\Delta  $
(resp. $\Delta _+ $) is the system of roots (resp., the system of positive
roots) of $\fr{g} $;  $E_{\alpha } $ are the root vectors of the simple
Lie algebra $\fr{g} $.  Indeed, equation (\ref{eq:1.4}) is the
compatibility condition
\begin{equation}\label{eq:LM}
[L(\lambda ),M(\lambda )] = 0,
\end{equation}
where
\begin{eqnarray}\label{eq:1.3}
\fl M(\lambda )\Psi(x,t,\lambda ) = \left( i{d \over dt} + [I,Q(x,t)] -
\lambda I\right) \Psi(x,t,\lambda ) = 0, \quad I \in \fr{h} .
\end{eqnarray}

Here and below $r = \mbox{rank}\,\fr{g} $, and $\vec{a}, \vec{b} \in
\bbbe^r$ are the vectors corresponding to the Cartan elements $J, I \in
\fr{h}$.

The inverse scattering problem for (\ref{eq:1.1}) with real valued $J $
\cite{ZM} was reduced to a Riemann-Hilbert problem for the (matrix-valued)
fundamental analytic solution of (\ref{eq:1.1}) \cite{1,Sh,G*86}; the
action-angle variables for the $N $-wave equations with $\fr{g}\simeq
sl(n)$ were obtained in the preprint \cite{ZM}, see also \cite{BS}.
Most of these results were derived first for the simplest non-trivial case
when $J $ has pair-wise distinct real eigenvalues.

However often the reduction conditions require that $J$ be
complex-valued, see \cite{G*86, BS, 2}. Then the
construction of the fundamental analytic solutions and the solution of the
corresponding inverse scattering problem for (\ref{eq:1.1}) becomes more
difficult \cite{VYa,BC}.

The interpretation of the ISM as a generalized Fourier transform and the
expansions over the ``squared solutions'' of (\ref{eq:1.1}) were derived
in \cite{G*86} for real $J $ and in \cite{VYa} for complex $J $. This
interpretation allows one also to prove that all $N $-wave type equations
are Hamiltonian and possess a hierarchy of pair-wize compatible
Hamiltonian structures \cite{G*86,VYa} $\{H^{(k)}, \Omega ^{(k)}\} $,
$k=0,\pm 1,\pm 2, \dots$.

Indeed, as a phase space ${\cal M}_{\rm ph} $ of these equations one can
choose the space spanned by the complex-valued functions $\{ q_\alpha
,p_\alpha , \alpha \in \Delta _+\}$, $\dim_\bbbc {\cal M}_{\rm ph}=
|\Delta | $.  The the corresponding nonlinear evolution equation (NLEE)
as, e.g.  (\ref{eq:1.4}) and its higher analogs can be formally written
down as Hamiltonian equations of motion:
\begin{equation}\label{eq:cH}
\Omega^{(k)} (Q_t,\cdot ) = dH^{(k)} (\cdot), \qquad k=0, \pm1, \pm 2,
\dots,
\end{equation}
where both $\Omega ^{(k)} $ and $H^{(k)} $ are complex-valued. The
simplest Hamiltonian formulation of (\ref{eq:1.4}) is given by
$\{H^{(0)} $, $\Omega^{(0)}\}$ where $H^{(0)}=H_0 + H_{\rm I} $ and
\begin{eqnarray}\label{eq:1.5}
H_0 ={c_0 \over 2i}\int_{-\infty }^{\infty } \, dx\, \left\langle
Q,[I,Q_x] \right\rangle = \sum_{\alpha \in \Delta _+}\,H_0(\alpha ), \\
\label{eq:1.5'}
H_0(\alpha )=ic_0 \int_{-\infty }^{\infty }dx\, {(\vec{b},\alpha )\over
(\alpha ,\alpha )} (q_{\alpha }p_{\alpha ,x} - q_{\alpha ,x}p_{\alpha }),
\\
\label{eq:H-om}
\fl H_{\rm I} = {c_0\over 3} \int_{-\infty }^{\infty }\, dx \, \left\langle
[J,Q],[Q,[I,Q]] \right\rangle = \sum_{[\alpha ,\beta ,\gamma ]\in {\cal
M}}\omega _{\beta ,\gamma } H(\alpha ,\beta ,\gamma ) ;\\
\fl H(\alpha ,\beta ,\gamma ) = c_0 \int_{-\infty}^{\infty } dx\,
(q_\alpha p_\beta p_\gamma  - p_\alpha q_\beta q_\gamma ), \quad \omega
_{\beta ,\gamma } ={4N_{\beta ,\gamma }\over (\alpha ,\alpha )}\,
\mbox{det}\left(\begin{array}{cc} (\vec{a},\beta ) & (\vec{a},\gamma )
\\ (\vec{b},\beta ) & (\vec{b},\gamma ) \end{array} \right),\nonumber
\end{eqnarray}
and the symplectic form $\Omega ^{(0)}$ is equivalent to a canonical one
\begin{eqnarray}\label{eq:Ome}
\Omega^{(0)} = {ic_0 \over 2} \int_{-\infty }^{\infty } dx\, \left\langle
[J, \delta Q(x,t)] \wedgecomma \delta Q(x,t) \right\rangle =
\sum_{\alpha \in \Delta _+} \Omega ^{(0)}(\alpha ),\\
\label{eq:Ome'}
\Omega ^{(0)}(\alpha )=ic_0 {2(\vec{a},\alpha ) \over (\alpha ,\alpha )}
\int_{-\infty }^{\infty }dx\, \delta q_{\alpha }(x,t) \wedge
\delta p_{\alpha }(x,t).
\end{eqnarray}
Here $c_0 $ is a constant to be explained below, $\langle \, \cdot\, ,
\cdot \, \rangle $ is the Killing form of $\fr{g}$ and the triple
$[\alpha ,\beta ,\gamma ] $ belongs to ${\cal M} $ if $\alpha,\beta,
\gamma \in \Delta _+ $ and $\alpha =\beta  +\gamma $; $N_{\beta , \gamma
} $ is defined in (\ref{eq:31.2}), (\ref{eq:32.1}) below.

The Hamiltonian of  the $N $-wave equations and their higher analogs
$H^{(k)} $ depend analytically on $q_\alpha ,p_\alpha $. That allows one
to rewrite the equation (\ref{eq:cH}) as a standard Hamiltonian equation
with real-valued $\Omega ^{(k)} $ and $H^{(k)} $. The phase space then is
viewed as the manifold of real-valued functions $\{ \re q_\alpha ,\re
p_\alpha ,\im q_\alpha ,\im p_\alpha \}$, $\alpha \in \Delta _+ $, so
$\dim _\bbbr {\cal M}_{\rm ph}=2|\Delta | $. Such treatment is rather
formal and we will not explain it in more details here.

Another better known way to make $\Omega ^{(k)} $ and $H^{(k)} $ real is
to impose reduction on them involving complex or hermitian conjugation;
below we list several types of such reductions. Obviously we can multiply
both sides of (\ref{eq:cH}) by the same constant $c_k $. We will use this
freedom below and whenever possible will adjust the constant $c_0 $ (or
$c_k $) in such a way that both $\Omega ^{(0)} $, $H^{(0)} $ (or
$\Omega ^{(k)} $, $H^{(k)} $) are real.

Physically to each term $H(\alpha ,\beta ,\gamma ) $ we relate part of a
wave-decay diagram which shows how the $\alpha  $-wave decays into
$\beta$-  and $\gamma  $-waves.  In other words we assign to each root
$\alpha $ a wave with wave number $k_{\alpha } $ and frequency $\omega
_{\alpha } $.  Each of the elementary decays preserves them, i.e.
\[
k_{\alpha } = k_{\beta } + k_{\gamma }, \quad \omega (k_{\alpha }) =
\omega (k_{\beta }) + \omega (k_{\gamma }).
\]

Our aim is to investigate all inequivalent $\bbbz_2 $ reductions of the $N
$-wave type equations related to the low-rank simple Lie algebras. Thus we
exhibit new examples of integrable $N $-wave type interactions some of
which have applications to physics.

{}From the definition of the reduction group $G_R $ introduced by A.
V. Mikhailov in \cite{2} and further developed in
\cite{ForGib*80b,Za*Mi,ForKu*83} it is natural that we have to have
realizations of $G_R $ as: i)~finite subgroup of the group
$\Aut(\fr{g}) $ of automorphisms of the algebra $\fr{g} $; and ii)~finite
subgroup of the conformal mappings on the complex $\lambda $-plane. We
impose also the natural restriction that the reduction preserves the form
of the Lax operator (\ref{eq:1.1}). In particular that means that we have
to limit ourselves only to those elements of $\Aut(\fr{g}) $ that
preserve the Cartan subalgebra $\fr{h} $ of $\fr{g} $. This condition
narrows our choice to $V_0\otimes \Ad(\fr{h})\otimes W(\fr{g}) $ where
$W(\fr{g} )$ is the Weyl group of $\fr{g} $ and $V_0 $ is the group of
external automorphisms of $\fr{g} $. As a result we find that the
reduction is sensitive to the way $G_R$ is embedded into
$\Aut(\fr{g}) $.

We start with the $\bbbz_2 $-reductions which provide the richest class of
interesting examples and display a number of non-trivial and inequivalent
reductions for the $N $-wave type equations.

In the next Section~2 we briefly outline the main idea of the reduction
group \cite{2}. In Section~3 we introduce convenient notations and
describe how generic $\bbbz_2 $-reductions act on $H^{(0)} $ and
$\Omega^{(0)}$. We also list the properties of the Weyl groups for
the algebras ${\bf A}_k $, ${\bf B}_k $, ${\bf C}_k $, $k=2,3 $ and ${\bf
G}_2 $. The main attention here is paid to their equivalence classes.
Obviously if two reductions generated by two different automorphisms $A $
and $A' $ are inequivalent then $A $ and $A' $ must belong to different
equivalence classes of $W(\fr{g}) $.

The list of the inequivalent reductions is displayed in Section~4. For
each of the cases we list the restrictions imposed on the potential $Q $
and the Cartan elements $J $ and $I $. Whenever the reduction preserves
the simplest Hamiltonian structure (\ref{eq:1.5}), (\ref{eq:H-om}),
(\ref{eq:Ome}) we write down only the Hamiltonian. In the
cases when the reduction makes the simplest Hamiltonian structure
degenerate we write down the system of equations.

Whenever $G_R $ acts on $iU(x,\lambda ) $ as Cartan involution
the result of the reduction is to get a real form of the algebra
$\fr{g} $. These cases are discussed in Section~5.

In Section~6 we formulate the effect of $G_R $ on the scattering data
of the Lax operator. In the next Section~7 these facts are used to
explain how in certain cases the reduction makes `half' of the symplectic
forms and `half' of the Hamiltonians in the hierarchy degenerate.
The paper finishes with several conclusions and two appendices which
contain some subsidiary facts about the root systems of the Lie algebras
(Appendix A) and the typical interaction terms in $H_I $ (Appendix B).

The present paper is an extended exposition of a part of our reports
\cite{GGK} with some misprints corrected.

\section{Preliminaries and general approach}\label{2}

The main idea underlying Mikhailov's reduction group \cite{2} is to impose
algebraic restrictions on the Lax operators $L $ and $M $ which will be
automatically compatible with the corresponding equations of motion
(\ref{eq:LM}). Due to the purely Lie-algebraic nature of the Lax
representation (\ref{eq:LM}) this is most naturally done by imbedding the
reduction group as a subgroup of $\Aut \fr{g} $ -- the group of
automorphisms of $\fr{g} $. Obviously to each reduction imposed on $L $
and $M $ there will correspond a reduction of the space of fundamental
solutions $\fr{S}_\Psi \equiv \{\Psi (x,t,\lambda )\} $ of (\ref{eq:1.1})
and (\ref{eq:1.3}).

Some of the simplest $\bbbz_2 $-reductions of $N $-wave systems have been
known for a long time (see \cite{2}) and are related to external
automorphisms of $\fr{g} $ and $\fr{G} $, namely:
\begin{equation}\label{eq:C-1}
\fl C_1\left( \Psi (x,t,\lambda ) \right) = A_1 \Psi^\dag (x,t,\kappa_1
(\lambda )) A_1^{-1} = \tilde{\Psi}^{-1}(x,t,\lambda ),
\qquad \kappa _1(\lambda )=\pm\lambda ^*,
\end{equation}
where $A_1 $ belongs to the Cartan subgroup of the group $\fr{G} $:
\begin{equation}\label{eq:A-1}
A_1 = \exp \left( \pi i H_1 \right),
\end{equation}
and $H_1 \in \fr{h}$ is such that $\alpha (H_1)\in \bbbz $ for all roots
$\alpha \in \Delta $ in the root system $\Delta $ of ${\frak g} $.
Note that the reduction condition relates the fundamental solution
$\Psi(x,t,\lambda )\in \fr{G} $ to a fundamental solution
$\tilde{\Psi}(x,t,\lambda ) $ of (\ref{eq:1.1}) and (\ref{eq:1.3}) which
in general differs from $\Psi(x,t,\lambda ) $.

Another class of $\bbbz_2 $ reductions are related to external automorphisms
of the type:
\begin{equation}\label{eq:C_2}
\fl C_2\left( \Psi (x,t,\lambda ) \right) = A_2 \Psi^T (x,t,\kappa_2
(\lambda )) A_2^{-1} = \tilde{\Psi}^{-1}(x,t,\lambda ),
\qquad \kappa _2(\lambda )=\pm\lambda ,
\end{equation}
where $A_2 $ is again of the form (\ref{eq:A-1}). The best known examples
of NLEE obtained with the reduction (\ref{eq:C_2}) are the sine-Gordon and
the MKdV equations which are related to $\fr{g}\simeq sl(2) $. For higher
rank algebras such reductions to our knowledge have not been studied.
Generically reductions of type (\ref{eq:C_2}) lead to degeneration of the
canonical Hamiltonian structure, i.e. $\Omega ^{(0)}\equiv 0 $; then
we need to use some of the higher Hamiltonian structures (see
\cite{2,VYa}) for proving their complete integrability.

In fact the reductions (\ref{eq:C-1}) and (\ref{eq:C_2}) provide us
examples when the reduction is obtained with the combined use of external
and inner automorphisms.

Along with (\ref{eq:A-1}), (\ref{eq:C-1}) one may use also reductions
with inner automorphisms:
\begin{eqnarray}\label{eq:C_3}
C_3\left( \Psi (x,t,\lambda ) \right) = A_3 \Psi^* (x,t,\kappa_1
(\lambda )) A_3^{-1} = \tilde{\Psi}(x,t,\lambda ),
\end{eqnarray}
and
\begin{eqnarray}\label{eq:C_4}
C_4\left( \Psi (x,t,\lambda ) \right) = A_4 \Psi (x,t,\kappa_2
(\lambda )) A_4^{-1} = \tilde{\Psi}(x,t,\lambda ).
\end{eqnarray}

Since our aim is to preserve the form of the Lax pair we limit ourselves
by automorphisms preserving the Cartan subalgebra ${\frak h} $. This
condition is obviously fulfilled if $ A_k $, $k=1,\dots,4 $ is in the form
(\ref{eq:A-1}). Another possibility is to choose $A_1 $, \dots, $A_4 $ so
that they correspond to  Weyl group automorphisms.

In fact (\ref{eq:C-1}) and (\ref{eq:C_2}) are related to external
automorphisms only if $\fr{g} $ is from the ${\bf A}_r $ series. For the
${\bf B}_r $, ${\bf C}_r $ and ${\bf D}_r $ series (\ref{eq:C-1}) is
equivalent to an inner automorphism (\ref{eq:C_3}) with the special choice
for the Weyl group element $w_0 $ which maps all highest weight vectors
into the corresponding lowest weight vectors (see Remark (\ref{rem:w0})).
{}Finally $\bbbz _2 $ reductions of the form (\ref{eq:C-1}) in fact
restrict us to the corresponding real form of the algebra ${\frak g} $.

\subsection{The reduction group}

The reduction group $G_R $ is a finite group which preserves the
Lax representation (\ref{eq:LM}), i.e. it ensures that the reduction
constraints are automatically compatible with the evolution. $G_R $ must
have two realizations: i) $G_R \subset {\rm Aut}\fr{g} $ and ii) $G_R
\subset {\rm Conf}\, \Bbb C $, i.e. as conformal mappings of the complex
$\lambda $-plane. To each $g_k\in G_R $ we relate a reduction
condition for the Lax pair as follows \cite{2}:
\begin{equation}\label{eq:2.1}
C_k(L(\Gamma _k(\lambda ))) = \eta _k L(\lambda ), \quad
C_k(M(\Gamma _k(\lambda ))) = \eta _k M(\lambda ),
\end{equation}
where $C_k\in \mbox{Aut}\; \fr{g} $ and $\Gamma _k(\lambda )\in
\mbox{Conf\,} \bbbc $ are the images of $g_k $ and $\eta _k =1 $ or $-1 $
depending on the choice of $C_k $. Since $G_R $ is a finite group then for
each $g_k $ there exist an integer $N_k $ such that $g_k^{N_k} =\openone
$. In all the cases below $ N_k=2 $ and the reduction group is isomorphic
to $\bbbz_2 $.

More specifically the automorphisms $C_k $, $k=1,\dots,4 $ listed above
lead to the following reductions for the matrix-valued functions
\begin{equation}\label{eq:U-V}
U(x,t,\lambda ) = [J,Q(x,t)] - \lambda J, \qquad
V(x,t,\lambda ) = [I,Q(x,t)] - \lambda I,
\end{equation}
of the Lax representation:
\numparts
\begin{eqnarray}\label{eq:U-V.a}
\fl \mbox{1)}& \qquad C_1(U^{\dagger}(\kappa _1(\lambda )))= U(\lambda ),
\qquad &C_1(V^{\dagger}(\kappa _1(\lambda )))= V(\lambda ), \\
\label{eq:U-V.b}
\fl \mbox{2)} & \qquad C_2(U^{T}(\kappa _2(\lambda )))= -U(\lambda ), \qquad
&C_2(V^{T}(\kappa _2(\lambda )))= -V(\lambda ), \\
\label{eq:U-V.c}
\fl \mbox{3)}& \qquad C_3(U^{*}(\kappa _1(\lambda )))= -U(\lambda ), \qquad
&C_3(V^{*}(\kappa _1(\lambda )))= -V(\lambda ), \\
\label{eq:U-V.d}
\fl \mbox{4)}& \qquad C_4(U(\kappa _2(\lambda )))= U(\lambda ), \qquad
&C_4(V(\kappa _2(\lambda )))= V(\lambda ),
\end{eqnarray}
\endnumparts\label{eq:UV}

\subsection{Finite groups}\label{ssec:2.2}

The condition (\ref{eq:2.1}) is obviously compatible with the group action.
Therefore it is enough to ensure that (\ref{eq:2.1}) is fulfilled for the
generating elements of $G_R $.

In fact (see \cite{FG}) every finite group $G $ is determined uniquely by
its generating elements $g_k $ and genetic code, e.g.:
\begin{equation}\label{eq:30.1}
g_k^{N_k}=\openone , \quad (g_jg_k)^{N_{jk}}=\openone , \quad
N_k, N_{jk} \in \bbbz.
\end{equation}
{}For example the cyclic $\bbbz_N $ and the dihedral $\bbbd_N $ groups
have as genetic codes
\begin{equation}\label{eq:30.2}
g^N=\openone , \quad N\geq 2 \quad \mbox{for}\; \bbbz_N,
\end{equation}
and
\begin{equation}\label{eq:30.3}
g_1^2=g_2^2=(g_1g_2)^N= \openone , \quad N\geq 2 \quad \mbox{for}\;
\bbbd_N.
\end{equation}

\subsection{Cartan-Weyl basis and Weyl group}\label{ssec:2.3}

Here we fix the notations and the normalization conditions for the
Cartan-Weyl generators of $\fr{g} $. We introduce $h_k\in \fr{h} $,
$k=1,\dots,r $ and $E_\alpha $, $\alpha \in \Delta $ where $\{h_k\} $
are the Cartan elements dual to the orthonormal basis $\{e_k\}$ in the
root space $\bbbe^r $. Along with $h_k $ we introduce also
\begin{equation}\label{eq:31.1}
H_\alpha = {2 \over (\alpha ,\alpha ) } \sum_{k=1}^{r} (\alpha ,e_k) h_k,
\quad \alpha \in \Delta ,
\end{equation}
where $(\alpha ,e_k) $ is the scalar product in the root space $\bbbe^r $
between the root $\alpha $ and $e_k $. The commutation relations are
given by \cite{Helg,LA}:
\begin{eqnarray}\label{eq:31.2}
&& [h_k,E_\alpha ] = (\alpha ,e_k) E_\alpha , \quad [E_\alpha ,E_{-\alpha
}]=H_\alpha , \nonumber\\
&& [E_\alpha ,E_\beta ] = \left\{ \begin{array}{ll}
N_{\alpha ,\beta } E_{\alpha +\beta } \quad & \mbox{for}\; \alpha +\beta
\in \Delta \\ 0 & \mbox{for}\; \alpha +\beta \not\in \Delta
\cup\{0\}. \end{array} \right.
\end{eqnarray}

We will denote by $\vec{a}=\sum_{k=1}^{r}a_k e_k $ the $r $-dimensional
vector dual to $J\in \fr{h} $; obviously $J=\sum_{k=1}^{r}a_k h_k $. If $
J $ is a regular real element in $\fr{h} $ then without restrictions we
may use it to introduce an ordering in $\Delta $. Namely we will
say that the root $\alpha \in\Delta _+ $ is positive (negative) if
$(\alpha ,\vec{a})>0 $ ($(\alpha ,\vec{a})<0 $ respectively).
The normalization of the basis is determined by:
\begin{eqnarray}\label{eq:32.1}
&& E_{-\alpha } =E_\alpha ^T, \quad \langle E_{-\alpha },E_\alpha \rangle
={2 \over (\alpha ,\alpha ) }, \nonumber\\
&& N_{-\alpha ,-\beta } = -N_{\alpha ,\beta }, \quad N_{\alpha ,\beta } =
\pm (p+1),
\end{eqnarray}
where the integer $p\geq 0 $ is such that $\alpha +s\beta \in\Delta $ for
all $s=1,\dots,p $ and $ \alpha +(p+1)\beta \not\in\Delta $.
The root system $\Delta $ of $\fr{g} $ is invariant with respect to the
Weyl reflections $S_\alpha $; on the vectors $\vec{y}\in \bbbe^r $ they
act as
\begin{equation}\label{eq:32.2}
S_\alpha \vec{y} = \vec{y} - {2(\alpha ,\vec{y}) \over (\alpha ,\alpha )}
\alpha , \quad \alpha \in \Delta .
\end{equation}
All Weyl reflections $S_\alpha $ form a finite group $W_{\fr{g}} $ known
as the Weyl group. One may introduce in a natural way an action of the
Weyl group on the Cartan-Weyl basis, namely:
\begin{eqnarray}\label{eq:32.3}
&& S_\alpha (H_\beta ) \equiv A_\alpha H_\beta A^{-1}_{\alpha } =
H_{S_\alpha \beta }, \nonumber\\
&& S_\alpha (E_\beta ) \equiv A_\alpha E_\beta A^{-1}_{\alpha } =
n_{\alpha ,\beta } E_{S_\alpha \beta }, \quad n_{\alpha ,\beta }=\pm 1.
\end{eqnarray}

It is also well known (see \cite{RS}) that the matrices
$A_\alpha $ are given (up to a
factor from the Cartan subgroup) by
\begin{equation}\label{eq:32.4}
A_\alpha =e^{E_\alpha } e^{-E_{-\alpha }} e^{E_\alpha } H_A,
\end{equation}
where $H_A $ is a conveniently chosen element from the Cartan subgroup
such that $H_A^2=\openone $. The formula (\ref{eq:32.4}) and the explicit
form of the Cartan-Weyl basis in the typical representation will be used
in calculating the reduction condition following from (\ref{eq:2.1}).

\subsection{Graded Lie algebras}\label{ssec:2.4}

One of the important notions in constructing integrable equations and
their reductions is the one of graded Lie algebra and Kac-Moody algebras
\cite{Helg}. The standard construction is based on a finite order
automorphism $C\in \Aut \fr{g} $, $C^N=\openone $. Obviously the
eigenvalues of $C $ are $\omega ^k $, $k=0,1,\dots , N-1 $, where $\omega
=\exp(2\pi i/N) $. To each eigenvalue there corresponds a linear subspace
$\fr{g}^{(k)} \subset \fr{g}$ determined by
\begin{equation}\label{eq:33.1}
\fr{g}^{(k)} \equiv \left\{ X\colon X\in \fr{g}, \quad C(X)=\omega ^k X
\right\} .
\end{equation}
Obviously $\fr{g}=\mathop{\oplus}\limits_{k=0}^{N-1} \fr{g}^{(k)} $ and
the grading condition holds
\begin{equation}\label{eq:34.1}
\left[\fr{g}^{(k)} , \fr{g}^{(n)} \right] \subset \fr{g}^{(k+n)},
\end{equation}
where $k+n $ is taken modulo $N $. Thus to each pair $\{\fr{g} , C\} $ one
can relate an infinite-dimensional algebra of Kac-Moody type
$\widehat{\fr{g}}_C $ whose elements are
\begin{equation}\label{eq:34.2}
X(\lambda ) = \sum_{k}^{} X_k \lambda ^k, \quad X_k \in \fr{g}^{(k)} .
\end{equation}
The series in (\ref{eq:34.2}) must contain only finite number of negative
(positive) powers of $\lambda $ and $\fr{g}^{(k+N)} \equiv \fr{g}^{(k)}$.
This construction is a most natural one for Lax pairs; we see that due to
the grading condition (\ref{eq:34.1}) we can always impose a reduction on
$L(\lambda ) $ and $M(\lambda ) $ such that both $U(x,t,\lambda ) $ and
$V(x,t,\lambda )\in \widehat{\fr{g}}_C $. So one of the generating
elements of $G_R$ will be used for introducing a grading in $\fr{g} $; then
the reduction condition (\ref{eq:2.1}) gives
\begin{equation}\label{eq:34.3}
U_0, V_0 \in \fr{g}^{(0)} , \quad I, J \in \fr{g}^{(1)}\cap \fr{h}.
\end{equation}
If in particular $N=2 $, the automorphism $C $ has the form (\ref{eq:C-1})
and $\kappa (\lambda )=\lambda ^* $ then all $X_k $ in (\ref{eq:34.2})
must be elements of the real form of $\fr{g} $ defined by $C $. We will
pay special attention to this situation in Section 5 below.

A possible second reduction condition will enforce additional constraints
on $U_0 $, $V_0 $ and $J $, $I $.

\subsection{Realizations of $G_R\subset \Aut \fr{g} $.}\label{ssec:2.5}

It is well known that $\Aut \fr{g} \equiv V\otimes \Aut_0 \fr{g} $ where
$V $ is the group of external automorphisms (the symmetry group of the
Dynkin diagram) and $\Aut_0 \fr{g} $ is the group of inner automorphisms.
Since we start with $I,J\in \fr{h} $ it is natural to consider only those
inner automorphisms that preserve the Cartan subalgebra $\fr{h} $. Then
$\Aut_0 \fr{g} \simeq \Ad_H \otimes W $ where $\Ad_H $ is the group of
similarity transformations with elements from the Cartan subgroup:
\begin{equation}\label{eq:35.1}
\Ad_C X = CXC^{-1}, \qquad C=\exp \left( {2\pi i H_{\vec{c}}
\over N }\right),
\qquad X \in \fr{g},
\end{equation}
and $W $ is the Weyl group of $\fr{g} $. Its action on the Cartan-Weyl
basis was described in (\ref{eq:32.3}) above. From (\ref{eq:31.2}) one
easily finds
\begin{equation}\label{eq:35.2}
CH_\alpha C^{-1} = H_\alpha , \quad
CE_\alpha C^{-1} = e^{2\pi i (\alpha ,\vec{c})/N} E_\alpha ,
\end{equation}
where $\vec{c}\in \bbbe^r $ is the vector corresponding to $H_{\vec{c}}
\in
\fr{h}$ in (\ref{eq:35.1}). Then the condition $C^N=\openone $ means that
$(\alpha ,\vec{c})\in \bbbz $ for all $\alpha \in \Delta $. Obviously
$H_{\vec{c}} $ must be chosen so that $\vec{c}=\sum_{k=1}^{r} 2c_k\omega
_k/(\alpha _k,\alpha _k) $ where $\omega _k $ are the fundamental weights
of $\fr{g} $ and $c_k $ are integer. In the examples below we will use
several possibilities by choosing $C_k $ as appropriate compositions of
elements from $V $, $\Ad_{\fr{H}} $ and $W $. In fact if $\fr{g} \simeq
{\bf G}_2 $ or belongs to ${\bf B}_r $ or ${\bf C}_r $ series then
$V\equiv \openone$.

\subsection{Realizations of $G_R\subset\mbox{Conf}\,
\bbbc$.}\label{ssec:2.6}

Generically each element $g_k\in G_R $ maps $\lambda $ into a
fraction-linear function of $\lambda $. Such action however is
appropriate for a more general class of Lax operators which are fraction
linear functions of $\lambda $. Since our Lax operators are linear in
$\lambda $ then we have the following possibilities for $\bbbz _2$:
\begin{eqnarray}\label{eq:37.1}
&& \Gamma _1 (\lambda )= a_0+\eta \lambda , \quad \eta = \pm 1
, \nonumber\\
&& \Gamma _2 (\lambda )= b_0 + \epsilon \lambda^* , \quad \epsilon =
\pm 1 ,\quad b_0 + \epsilon b_0^*=0.
\end{eqnarray}
In the examples below $a_0=b_0=0 $.

\section{Inequivalent reductions}\label{sec:3}

We will consider two substantially different types of reductions
(\ref{eq:UV}). The first and best known type of $\bbbz_2 $-reductions
corresponds to inner automorphisms $C_j $ from the Cartan subgroup which
have the form (\ref{eq:35.1}) with $N=2 $.

{}For each of these reductions we will describe the structure of $\Omega
^{(0)} $, $H_0 $ and $H_{\rm I} $. To make the notations more convenient
we will introduce $\left\{ \begin{array}{c} \alpha  \\ \beta , \gamma
\end{array} \right\} $ and ${\cal  H}_{\rm I}(\alpha ) $ as follows
\begin{eqnarray}\label{eq:22.1}
\fl \left\{ \begin{array}{c} \alpha  \\ \beta , \gamma \end{array} \right\}
= \omega _{\beta ,\gamma } H(\alpha ,\beta ,\gamma )
= c_0 \omega _{\beta ,\gamma } \int_{-\infty }^{\infty } dx \,
\left( Q _{\alpha } Q_{-\beta } Q_{-\gamma } -
Q _{-\alpha } Q_{\beta } Q_{\gamma } \right), \\
\label{eq:23.6}
{\cal  H}_{\rm I} (\alpha ) = \sum_{[\beta,\gamma] \in {\cal  M}_\alpha }
\left\{ \begin{array}{c} \alpha \\ \beta ,\gamma \end{array} \right\},
\end{eqnarray}
where $Q_\alpha  $ are introduced in (\ref{eq:1.3.1}) and ${\cal
M}_\alpha  $ is the set of pairs of roots $[\beta,\gamma ] $ such that
$\beta +\gamma =\alpha  $, see Appendix~A.  In the last summation we do
not require $\beta $ and $\gamma $ to be positive.  The explicit
expression for $\left\{ \begin{array}{c} \alpha \\ \beta ,\gamma
\end{array} \right\} $ allows us to check that
\begin{equation}\label{eq:23.5}
\left\{ \begin{array}{c} \alpha \\ \beta ,\gamma \end{array} \right\} =
\left\{ \begin{array}{c} \gamma \\ -\beta ,\alpha \end{array} \right\} =
\left\{ \begin{array}{c} -\alpha \\ -\beta ,-\gamma \end{array} \right\}.
\end{equation}

In proving (\ref{eq:23.5}) we used (\ref{eq:32.1}), (\ref{eq:32.3}) and
the properties of the structure constants $N_{\beta ,\gamma } $ of the
Chevalley basis; namely, if $\alpha -\beta -\gamma =0 $ then
\begin{equation}\label{eq:?}
{N_{\alpha ,-\beta }  \over (\gamma ,\gamma ) }=
{N_{-\beta ,-\gamma  }  \over (\alpha ,\alpha ) }=
{N_{-\gamma ,\alpha }  \over (\beta ,\beta ) } =
-{N_{\beta ,\gamma  }  \over (\alpha ,\alpha ) }.
\end{equation}
and as a consequence
\begin{equation}\label{eq:om_bg}
\omega _{\alpha ,-\beta } = \omega _{-\beta ,-\gamma } =
\omega _{-\gamma ,\alpha } = -\omega _{\beta ,\gamma }.
\end{equation}
{}From (\ref{eq:H-om}) it also follows that
\begin{equation}\label{eq:23.7}
H_{\rm I}  = \sum_{[\alpha ,\beta ,\gamma ]\in {\cal  M}}
\left\{ \begin{array}{c} \alpha \\ \beta ,\gamma \end{array} \right\}
= {1  \over 3 } \sum_{\alpha \in \Delta _+} {\cal  H}_{\rm I}(\alpha ).
\end{equation}
Indeed, using (\ref{eq:23.5}) we find that each triple $\left\{
\begin{array}{c} \alpha \\ \beta ,\gamma \end{array} \right\} $ is equal
to $\left\{ \begin{array}{c} \tilde{\alpha} \\ \tilde{\beta} ,
\tilde{\gamma} \end{array} \right\} $ where the triple of roots
$[\tilde{\alpha }, \tilde{\beta }, \tilde{\gamma }] \in {\cal  M} $.

\subsection{Reductions with Cartan subgroup elements}\label{ssec:3.1}

{}For the first type the action of the reduction group on the Cartan-Weyl
basis is given by (\ref{eq:35.2}) and the corresponding constraints on
$q_\alpha $, $p_\alpha $ have the form:
\numparts
\begin{eqnarray}\label{eq:rc-1}
\mbox{1)} \qquad & p_\alpha =-\eta s_\alpha  q_\alpha ^*, \qquad
s_\alpha =e^{-\pi i (\vec{c},\alpha )}   \\
\label{eq:rc-2}
\mbox{2)} \qquad & p_\alpha = s_\alpha q_\alpha , \qquad \eta =-1,
\\
\label{eq:rc-3}\mbox{3)} \qquad  & q_\alpha =\eta s_\alpha q_\alpha^* ,
\qquad p_\alpha =\eta s_\alpha  p_\alpha^* ,  \\
\label{eq:rc-4}
\mbox{4)} \qquad & q_\alpha = s_\alpha  q_\alpha , \qquad p_\alpha
= s_\alpha p_\alpha , \qquad \eta =1.
\end{eqnarray}
\endnumparts

Let us describe how each of these constraints simplify the Hamiltonian
$H^{(0)} = H_0 + H_{\rm I} $ and the symplectic form $\Omega ^{(0)} $.
We can write them down in the form:
\begin{equation}\label{eq:hom-0}
\fl H_0 = \sum_{\alpha \in \Delta _+}^{}s_\alpha H_{0*}(\alpha ), \qquad
H_{\rm I} = \sum_{\alpha \in \Delta _+}^{}s_\alpha H_{\rm I}(\alpha ),
\qquad \Omega^{(0)} = \sum_{\alpha \in \Delta _+}^{} s_\alpha
\Omega^{(0)} _{*}(\alpha ).
\end{equation}

{}For the case 1) we easily find
\begin{eqnarray}\label{eq:h0-al}
H_{0*}(\alpha ) = -i\eta c_0 {(b,\alpha ) \over (\alpha ,\alpha ) }
\int_{-\infty }^{\infty } dx \left( q_{\alpha } q^*_{\alpha ,x} -
q_{\alpha ,x} q^*_{\alpha} \right) , \\
\label{eq:hI-al}
H_{\rm I}(\alpha ) = \sum_{\beta +\gamma =\alpha } c_0
\omega _{\beta ,\gamma}
s_{\alpha } \int_{-\infty }^{\infty } dx \left( q_{\alpha } q^*_{\beta }
q^*_{\gamma } + \eta q^*_{\alpha } q_{\beta } q_{\gamma } \right) , \\
\label{eq:om0-al}
\Omega^{(0)} _{*}(\alpha ) = -i\eta c_0 {2(a,\alpha) \over (\alpha
,\alpha ) } \int_{-\infty}^{\infty } dx\delta q_{\alpha}\wedge \delta
q^*_{\alpha } ,
\end{eqnarray}
We have $|\Delta _+| $ complex valued fields $q_\alpha  $ and the above
expressions for $\Omega ^{(0)} $ and $H_0 $ show that $q_\alpha  $ is
dynamically conjugated to $q_\alpha ^* $.

In the case 2) we consider only $\eta =-1 $; the choice $\eta=1 $ means
that $\vec{a}=-\vec{a} $ and as a consequence we have $J=0 $, i.e. no
$N $-wave equations are possible for $\eta=1 $. For $\eta=-1 $
we have $|\Delta _+| $ complex valued fields $q_\alpha  $
which up to a sign coincide with $p_\alpha  $. Then
$H_0(\alpha ) $, $\Omega^{(0)} (\alpha ) $ and $H_{\rm I} $ become
identically zero.  The corresponding set of equations is
nontrivial but is does not allow a canonical Hamiltonian formulation.
However it allows Hamiltonian description using other members in the
hierarchy of Hamitonian structures.

In the case 3) we have $2|\Delta _+| $ `real'-valued\footnote{Here and
below we count as `real' also the fields that are in fact purely
imaginary.} fields $q_\alpha  $ and $p_\alpha  $. The formulae for
$H_0(\alpha ) $, $H_{\rm I}(\alpha ) $ and $\Omega^{(0)}(\alpha ) $
(\ref{eq:1.5})--(\ref{eq:Ome'})
do not change, only each of the summands in them becomes real due to the
fact that all fields $q_\alpha $ and $p_\alpha $ are now simultaneously
either real or purely imaginary.

In the last case 4) we consider only $\eta =1 $; the choice $\eta=-1 $
here means that $\vec{a}=-\vec{a} $ and as a consequence we have $J=0 $,
i.e. like in case 2) no $N $-wave equations are possible for $\eta=-1 $.
One easily finds that now the positive roots
$\Delta _+ $ split into two subsets $\Delta _+ = \Delta _+^0 \cup \Delta
_+^1 $ such that $(\vec{c},\alpha ) $ is even for all $\alpha \in \Delta
_+^0 $ and odd for all $\alpha \in \Delta _+^1$. Obviously if $\alpha \in
\Delta _+^1 $ then $s_\alpha =-1 $ and the fields $q_\alpha$, $p_\alpha $
must vanish due to (\ref{eq:rc-4}). As a result the effect of the
reduction is to restrict us to an $N $-wave system related to the
subalgebra $\fr{g}_0 \subset \fr{g} $ with root system $\Delta ^0 =
\Delta_+^0 \cup (-\Delta _+^0) $. Such reductions are out of the scope of
the present paper.

\subsection{Reductions with Weyl group elements}\label{ssec:3.2}

{}For the second type of $\bbbz_2 $-reductions $C_j$'s are related to Weyl
group elements $w $ such that $w^2=\openone  $. Generically $w $ is a
composition of several Weyl reflections $S_{\beta _1} S_{\beta _2}\dots $
where the roots $\beta _1,\beta _2,\dots $ are pair-wise orthogonal. Fixing
up $w $ we can split $\Delta _+ $ into a union of four subsets:
\begin{equation}\label{eq:4-split}
\Delta _+ \equiv \Delta^\perp _+ \cup \Delta^{||} _+ \cup \Delta^{+} _+
\cup \Delta^{-} _+,
\end{equation}
where
\numparts
\begin{eqnarray}\label{eq:4-perp}
&w(\alpha ) \equiv \alpha' =\alpha  &\qquad \mbox{for all \ } \alpha \in
\Delta^{\perp} _+ ,\\
\label{eq:4-||}
&w(\alpha ) \equiv \alpha' = -\alpha &\qquad \mbox{for all \ } \alpha \in
\Delta^{||}_+ ,\\
\label{eq:4-p}
&w(\alpha ) \equiv \alpha' >0,  &\qquad \mbox{for all \ } \alpha \in
\Delta^{+}_+ \; \mbox{and} \; \alpha \neq \alpha ' ,\\
\label{eq:4-m}
&w(\alpha) \equiv \alpha' <0, &\qquad  \mbox{for all \ } \alpha \in
\Delta^{-}_+, \; \mbox{and} \; \alpha \neq -\alpha '
\end{eqnarray}
\endnumparts
Depending on the choice of $w $ one or more of these subsets may be empty.
{}From now on we will denote by $\prime $ the action of $w $ on the
corresponding root: $w(\alpha )=\alpha ' $ and $w(\alpha')= \alpha $.

The subsets $\Delta _+^+ $ and $\Delta _+^- $ always
contain even number of roots. Indeed if $\alpha \in \Delta _+^+ $
then $\alpha ' $ also belongs to $\Delta _+^+ $. Analogously  if $\alpha
\in \Delta _+^- $ then $-\alpha ' $ also belongs to $\Delta _+^- $.  As we
shall see below the reduction relates the coefficients $Q_\alpha  $ with
$Q_{\alpha '} $ or $Q_{-\alpha '} $. Therefore we will introduce the
subsets $\tilde{\Delta }_+^\pm \subset \Delta _+^\pm$ satisfying:
\begin{eqnarray}\label{eq:12a-a}
\tilde{\Delta }_+^+ \cup w(\tilde{\Delta }_+^+)=\Delta _+^+ \\
\tilde{\Delta }_+^- \cup \left(-w(\tilde{\Delta }_+^-)\right)
=\Delta _+^-
\end{eqnarray}
In other words out of each pair $\{\alpha ,\alpha '\}\in\Delta _+^+ $
(resp. $\{\alpha ,-\alpha '\}\in\Delta _+^- $) only one element belongs to
$\tilde{\Delta }_+^+ $ (resp. $\tilde{\Delta }_+^- $). For definiteness
below we will choose the element whose height is lower, i.e. $\alpha \in
\tilde{\Delta }_+^\pm  $ if $\htt(\alpha )< \htt(\pm\alpha ') $.

We will also make use of the sets
\begin{equation}\label{eq:12a-b}
\Delta _+^{0} = \Delta _+^{\perp}\cup \Delta _+^{+}, \qquad
\Delta _+^{1} = \Delta _+^{||}\cup \Delta _+^{-},
\end{equation}
which obviously satisfy $w(\Delta _+^0 )=\Delta _+^0 $ and
$w(\Delta _+^1 )=-\Delta _+^1 $.

The reduction conditions corresponding to each of the four types are most
easily written down in terms of $Q_\alpha $, see (\ref{eq:1.3.1}).
Indeed we have:
\numparts
\begin{eqnarray}\label{eq:21.1c}
\mbox{1)} \qquad
Q_{\alpha '} =  - \eta n_{\alpha ,\alpha '} Q_{-\alpha }^*, \qquad
&\vec{a} = \eta w_1(\vec{a}^*),  \\
\label{eq:21.2c}
\mbox{2)} \qquad  Q_{\alpha '} =  - \eta n_{\alpha ,\alpha '} Q_{-\alpha
}, \qquad &\vec{a} = -\eta w_2(\vec{a}),  \\
\label{eq:21.3c}
\mbox{3)} \qquad  Q_{\alpha '} =  \eta n_{\alpha ,\alpha '} Q_{\alpha }^*,
\qquad &\vec{a} = -\eta w_3(\vec{a}^*),   \\
\label{eq:21.4c}
\mbox{4)} \qquad  Q_{\alpha '} =  \eta n_{\alpha ,\alpha '} Q_{\alpha },
\qquad &\vec{a} = \eta w_4(\vec{a}).
\end{eqnarray}
\endnumparts

We will also describe the effect of the reduction on $H $ and $\Omega
^{(0)} $. Using the notations defined in Eqs.
(\ref{eq:1.5})-(\ref{eq:H-om}) we can write
\begin{eqnarray}\label{eq:s3.14H}
\fl H = \sum_{\alpha \in \tilde{\Delta }_+^+} \left (H(\alpha ) + H(\alpha
') \right) +\sum_{\alpha \in \tilde{\Delta }_+^-} \left (H(\alpha ) +
H(-\alpha ') \right) +\sum_{\alpha \in \Delta _+^\perp \cup \Delta
_+^{||}} H(\alpha ) \\
\label{eq:s3.14Om}
\fl \Omega ^{(0)} =\sum_{\alpha \in \tilde{\Delta }_+^+} \left (
\Omega (\alpha ) + \Omega (\alpha ') \right) +\sum_{\alpha \in
\tilde{\Delta }_+^-} \left (\Omega(\alpha ) + \Omega (-\alpha ') \right)
+\sum_{\alpha \in \Delta _+^\perp \cup \Delta _+^{||}} \Omega (\alpha )
\\
\label{eq:s3.14Ha}
H(\alpha )  =H_0(\alpha ) + H_{\rm I}(\alpha ).
\end{eqnarray}

{}For the reductions (\ref{eq:35.2}) the restrictions on the potential
matrix $Q(x,t) $ read as follows:
\numparts
\begin{eqnarray}
\label{eq:rc-II.1c}
\fl \mbox{1)} \qquad
& q_\alpha ^* = - \eta n_{\alpha ,\alpha '} p_{\alpha '}, \qquad
p_\alpha ^* = - \eta n_{\alpha ,\alpha '} q_{\alpha '}, \quad \mbox{for
} \alpha, \alpha '\in \Delta _+^{0}, \nonumber\\
\label{eq:rc-II.1b}
\fl & q_{-\alpha '}^* = - \eta n_{\alpha ,\alpha '} q_\alpha , \qquad
p_{-\alpha '}^* = - \eta n_{\alpha ,\alpha '}  p_\alpha , \quad
\mbox{for  } \alpha, -\alpha '\in \Delta _+^{1} \\
\label{eq:rc-II.2a}
\fl \mbox{2)}        \qquad
& q_\alpha  =  n_{\alpha ,\alpha '} p_{\alpha'} , \qquad
p_\alpha =  n_{\alpha ,\alpha '}  q_{\alpha'} , \quad \mbox{for \
} \alpha ,\alpha '\in \Delta _+^0, \quad \eta =-1,\nonumber\\
\label{eq:rc-II.2b}
\fl &q_\alpha  = n_{\alpha ,\alpha '} q_{-\alpha'} , \qquad p_\alpha
= n_{\alpha ,\alpha '}  p_{-\alpha '} , \quad \mbox{for \ }
\alpha , -\alpha '\in \Delta _+^{1},  \\
\label{eq:rc-II.3b}
\fl  \mbox{3)}  \qquad
& q_{\alpha '}^* =  \eta n_{\alpha ,\alpha '} q_\alpha , \qquad
p_{\alpha '}^* =  \eta n_{\alpha ,\alpha '}  p_\alpha , \quad  \mbox{for \
} \alpha , \alpha ' \in \Delta _+^{0}, \nonumber\\
\label{eq:rc-II.3c}
\fl & q_{\alpha }^* =  \eta n_{\alpha ,\alpha '} p_{-\alpha '},
\qquad p_\alpha ^* =  \eta n_{\alpha ,\alpha '} q_{-\alpha '}, \quad
\mbox{for \ } \alpha, - \alpha '\in \Delta _+^{1}, \\
\label{eq:rc-II.2d}
\fl \mbox{4)} \qquad
& q_\alpha  =  n_{\alpha ,\alpha '} q_{\alpha'} , \qquad p_\alpha
=  n_{\alpha ,\alpha '}  p_{\alpha'} , \quad \mbox{for \ } \alpha
, \alpha ' \in \Delta _+^0,\quad \eta =1, \nonumber\\
\label{eq:rc-II.2dd}
\fl & p_\alpha  =  n_{\alpha ,\alpha '} q_{-\alpha'} , \qquad q_\alpha
= n_{\alpha , -\alpha '}  p_{-\alpha '} , \quad \mbox{for \ } \alpha
\in \Delta _+^{1}.
\end{eqnarray}
\endnumparts

The set of independent fields for each of these reductions are collected
in the tables \ref{tab:indep.c} and \ref{tab:indep.w}.

\begin{table}{}
\caption{The set of independent fields for the reductions with Cartan
subgroup elements.\label{tab:indep.c}}

\vspace*{0.1in}

\begin{center}

\begin{tabular}{|l|l|l|l|}
\hline \hline
Type & Complex & Real & Redundant \\
\hline \hline
1) & $q_{\alpha }, \qquad \alpha \in \Delta _+ $ & -- & -- \\  \hline
2) & $q_{\alpha }, \qquad \alpha \in \Delta _+ $ & -- & -- \\  \hline
3) & -- & $q_{\alpha },p_{\alpha } \qquad \alpha \in \Delta _+ $  & -- \\
\hline
4) & $q_{\alpha },p_{\alpha } \qquad \alpha \in \Delta _+ $ &
-- & $q_{\alpha },p_{\alpha } \qquad \alpha \in \Delta _+ $ \\
\qquad & with $s_\alpha =1 $ & \qquad & with $s_\alpha =-1 $\\
\hline
\end{tabular}
\end{center}
\end{table}

\begin{table}{}
\caption{The set of independent fields for the reductions with Weyl
subgroup elements. By $\Delta _{+,\varepsilon }^{\perp} $  and
$\Delta _{+,\varepsilon }^{||}  $, $\varepsilon =\pm 1 $ we denote the
subsets of roots $\alpha \in \Delta _{+}^{\perp }  $ (resp., $\alpha \in
\Delta _{+}^{||}  $) for which $n_{\alpha \alpha '}=\varepsilon  $.
\label{tab:indep.w}}

\vspace*{0.1in}

\begin{center}

\begin{tabular}{|l|l|l|l|}
\hline \hline
Type & Complex & Real & Redundant \\
\hline \hline
1) & $q_{\alpha }, \qquad \alpha \in \Delta_+ ^{\perp} \cup
\tilde{\Delta }_+^+ $
& $q_{\alpha }, p_{\alpha }, \qquad \alpha \in \Delta_+^{||} $
& -- \\
\qquad  & $q_{\alpha }, p_{\alpha }, \qquad \alpha \in
\tilde{\Delta }_+^{-} $& \qquad  & \qquad  \\
\hline
2) & $q_{\alpha }, \qquad \alpha \in \Delta_+ ^{\perp} \cup \Delta_+^+ $
& -- & $q_{\alpha }, p_{\alpha }, \qquad \alpha \in
\Delta_{+,-1} ^{||}$ \quad\\
\qquad  & $q_{\alpha }, p_{\alpha }, \qquad \alpha \in
\tilde{\Delta }_+^{-}\cup \Delta_{+,1} ^{||} $ & & \\ \hline
3) & $q_{\alpha }, \qquad \alpha \in \Delta _+^{||} \cup \Delta_+^- $
& $q_{\alpha }, p_{\alpha }, \qquad \alpha \in \Delta _+^{\perp} $
& -- \\
\qquad  & $q_{\alpha }, p_{\alpha }, \qquad \alpha \in
\tilde{\Delta }_+^{+} $& \qquad  & \qquad  \\
\hline
4) & $q_{\alpha }, \qquad \alpha \in \Delta_+ ^{||} \cup \Delta_+^- $
& -- & $q_{\alpha }, p_{\alpha }, \qquad \alpha \in
\Delta_{+,-1} ^{\perp }$ \quad  \\
\qquad  & $q_{\alpha }, p_{\alpha }, \qquad \alpha \in
\tilde{\Delta }_+^{+}\cup \Delta_{+,1} ^{\perp }  $& \qquad  & \\ \hline
\end{tabular}
\end{center}
\end{table}

{}For the reductions of types 1) and 3) while the first two sets of
variables are complex-valued, the fields related to the roots $\alpha
\in \Delta _+^{||} $ for 1) and the fields related to roots $\alpha
\in \Delta _+^{\perp} $ for 3) should be either real or purely imaginary
due to (\ref{eq:rc-II.1c}) and (\ref{eq:rc-II.3b}) respectively.
Below for the sake of brevity we will call them `real'.  In other words
after the reduction we get an $N $-wave system with $2|\Delta _+^{||}| $
`real' fields and $|\Delta _+^{\perp}| + |\Delta _+^{+}| + |\Delta_+^{-}|
$ complex fields for the first reduction in (\ref{eq:35.2}) and for the
third one we have $2|\Delta _+^{\perp}|$ real and
$|\Delta _+^{||}| + |\Delta _+^{+}| + |\Delta_+^{-}|$ complex functions.

The reduction conditions on $H(\alpha ) $ and $\Omega ^{(0)}(\alpha ) $
read:
\begin{eqnarray}\label{eq:10.2}
\fl H_{0,R}(\alpha )=-i\eta c_0n_{\alpha ,\alpha '} {(\vec{b},\alpha )\over
(\alpha ,\alpha ) } \int_{-\infty }^{\infty } dx\, \left( Q_\alpha
Q_{\alpha ',x}^* - Q_{\alpha ,x} Q_{\alpha '}^* \right), \\
\label{eq:10.4}
\fl \Omega _{R}^{(0)}(\alpha ) = -i\eta c_0n_{\alpha ,\alpha '}
{(\vec{a},\alpha )\over (\alpha ,\alpha ) } \int_{-\infty }^{\infty } dx\,
\delta  Q_\alpha \wedge \delta  Q_{\alpha '}^* , \\
\label{eq:10.3}
\fl \left\{ \begin{array}{c} \alpha \\ \beta ,\gamma \end{array}\right\}_R
=\left\{ \begin{array}{c}
\alpha '\\ \beta ',\gamma '\end{array}\right\}^*_R
= c_0 n_{\alpha ,\alpha '} \omega _{\beta ,\gamma } \int_{-\infty
}^{\infty } dx\, \left( Q_{\alpha } Q_{\beta '}^{*} Q_{\gamma '}^{*} +
\eta Q_{\alpha '}^{*} Q_\beta Q_\gamma \right), \end{eqnarray}
for the first reduction in (\ref{eq:35.2});
\begin{eqnarray}\label{eq:10.2b}
\fl H_{0,R}(\alpha )=-i\eta c_0n_{\alpha ,\alpha '} {(\vec{b},\alpha )\over
(\alpha ,\alpha ) } \int_{-\infty }^{\infty } dx\, \left( Q_\alpha
Q_{\alpha ',x} - Q_{\alpha ,x} Q_{\alpha '} \right), \\
\label{eq:10.4b}
\fl \Omega _{R}^{(0)}(\alpha ) = -i\eta c_0n_{\alpha ,\alpha '}
{(\vec{a},\alpha )\over (\alpha ,\alpha ) } \int_{-\infty }^{\infty } dx\,
\delta  Q_\alpha \wedge \delta  Q_{\alpha '} , \\
\label{eq:10.3b}
\fl \left\{ \begin{array}{c} \alpha \\ \beta ,\gamma \end{array}\right\}_R =
 \eta
\left\{ \begin{array}{c} \alpha '\\ \beta ',\gamma '\end{array}\right\}_R
=c_0 n_{\alpha ,\alpha '} \omega _{\beta ,\gamma } \int_{-\infty }^{\infty
} dx\, \left( Q_{\alpha } Q_{\beta '} Q_{\gamma '} + \eta
Q_{\alpha '} Q_\beta Q_\gamma \right).
\end{eqnarray}
for the second reduction in (\ref{eq:35.2});
\begin{eqnarray}\label{eq:10.2c}
\fl H_{0,R}(\alpha )=i\eta c_0n_{\alpha ,\alpha '} {(\vec{b},\alpha )\over
(\alpha ,\alpha ) } \int_{-\infty }^{\infty } dx\, \left( Q_\alpha
Q_{-\alpha ',x}^* - Q_{\alpha ,x} Q_{-\alpha '}^* \right), \\
\label{eq:10.4c}
\fl \Omega _{R}^{(0)}(\alpha ) = i\eta c_0n_{\alpha ,\alpha '}
{(\vec{a},\alpha )\over (\alpha ,\alpha ) } \int_{-\infty }^{\infty } dx\,
\delta  Q_\alpha \wedge \delta  Q_{-\alpha '}^* , \\
\label{eq:10.3c}
\fl \left\{ \begin{array}{c} \alpha \\ \beta ,\gamma \end{array}\right\}_R =
\left\{ \begin{array}{c} -\alpha '\\ -\beta ',-\gamma '\end{array}
\right\}^*_R
=c_0 n_{\alpha ,\alpha '} \omega _{\beta ,\gamma } \int_{-\infty }^{\infty
} dx\, \left( Q_{\alpha } Q_{-\beta '}^{*} Q_{-\gamma '}^{*} - \eta
Q_{-\alpha '}^{*} Q_\beta Q_\gamma \right),
\end{eqnarray}
for the third reduction in
(\ref{eq:35.2}); and
\begin{eqnarray}\label{eq:10.2d}
\fl H_{0,R}(\alpha
)=i\eta c_0n_{\alpha ,\alpha '} {(\vec{b},\alpha )\over (\alpha ,\alpha )
} \int_{-\infty }^{\infty } dx\, \left( Q_\alpha Q_{-\alpha ',x} -
Q_{\alpha ,x} Q_{-\alpha '} \right),
\\ \label{eq:10.4d}
\fl \Omega
_{R}^{(0)}(\alpha ) = i\eta c_0n_{\alpha ,\alpha '} {(\vec{a},\alpha
)\over (\alpha ,\alpha ) } \int_{-\infty }^{\infty } dx\, \delta  Q_\alpha
\wedge \delta  Q_{-\alpha '} , \\
\label{eq:10.3d}
\fl \left\{\begin{array}{c} \alpha \\ \beta ,\gamma \end{array}\right\}_R
= \eta \left\{ \begin{array}{c} -\alpha '\\ -\beta ',-\gamma '
\end{array}\right\}_R \nonumber\\
=c_0 n_{\alpha ,\alpha '} \omega _{\beta ,\gamma } \int_{-\infty }^{\infty
} dx\, \left( Q_{\alpha } Q_{-\beta '} Q_{-\gamma '} - \eta Q_{-\alpha '}
Q_\beta Q_\gamma \right)
\end{eqnarray}
for the last reduction.

The general properties of $H(\alpha ) $ and $\Omega ^{(0)}(\alpha ) $ are
as follows:
\begin{eqnarray}\label{eq:10.1}
\fl \mbox{1)} \qquad  H_0(\alpha )=H_0^*(\alpha '), \qquad \Omega
^{(0)}(\alpha ) = (\Omega^{(0)} (\alpha '))^*, \qquad {\cal  H}_{\rm
I}(\alpha ) ={\cal H}^*_{\rm I}(\alpha' ) ,\\
\fl  \mbox{2)} \qquad  H_0(\alpha )= \eta H_0(\alpha '), \qquad \Omega
^{(0)}(\alpha ) = \eta(\Omega^{(0)} (\alpha ')), \qquad {\cal  H}_{\rm
I}(\alpha ) = \eta{\cal H}_{\rm I}(\alpha' ) ,\\
\fl \mbox{3)} \qquad  H_0(\alpha)=H_0^*(\alpha '), \qquad \Omega
^{(0)}(\alpha ) = (\Omega^{(0)} (\alpha '))^*, \qquad {\cal  H}_{\rm
I}(\alpha ) ={\cal H}^*_{\rm I}(\alpha' ) , \\
\fl \mbox{4)} \qquad  H_0(\alpha )= \eta H_0(\alpha '), \qquad \Omega
^{(0)}(\alpha ) = \eta(\Omega^{(0)} (\alpha ')), \qquad {\cal  H}_{\rm
I}(\alpha ) = \eta{\cal H}_{\rm I}(\alpha' ) ,
\end{eqnarray}

Obviously if $\alpha \in \Delta _+^\perp $, i.e. $\alpha '=\alpha  $ then
the expressions in the right hand side of (\ref{eq:10.2}) and
(\ref{eq:10.4}) coincide with the ones in (\ref{eq:h0-al}) and
(\ref{eq:om0-al}). However if $\alpha '\neq \alpha >0 $ (i.e., if $\alpha
\in \Delta _+^+ $) the field variable $q_{\alpha } $ will be dynamically
conjugated not to $q_{\alpha }^{*} $ but to $q_{\alpha '}^{*} $.
The same holds true for the second reduction in (\ref{eq:35.2}): If $
\alpha \in \Delta ^{\perp} $  the expressions in left hand side of
(\ref{eq:10.2b}) and (\ref{eq:10.4b}) coincide with the general ones.
This fact makes these reduced $N $-wave systems substantially
different from the ones described in Subsection 3.1.

In Appendix A we list the sets ${\cal  M}_\alpha  $ of all pairs of roots
$\beta $, $\gamma  $ such that $\beta +\gamma =\alpha  $ and the
coefficients $\omega _{jk} $ (\ref{eq:H-om}). In Appendix B we give
explicit expressions for some of the specific reduced interaction
Hamiltonian terms.

\subsection{Inequivalent embeddings of $\bbbz_2 $ in $W_{\fr{g}}$}
\label{ssec:3.3}

The reduction group $G_R $ may be imbedded in the Weyl group $W({\frak g})
$ of the simple Lie algebra in a number of ways. Therefore it will be
important to have a criterium to distinguish the nonequivalent reductions.
As any other finite group, $W({\frak g}) $ can be split
into equivalence classes.
So one may expect that reductions with elements from the same equivalence
class would lead to equivalent reductions; namely the two systems of $N
$-wave equations will be related by a change of variables.

In what follows we will describe the equivalence classes of the Weyl
groups $W({\bf B}_2) $, $W({\bf G}_2) $ and $W({\bf B}_3) $; note that
$W({\bf B}_l)\simeq W({\bf C}_l) $.
This is due to two facts: 1) the system of positive roots for ${\bf B}_r $
is $\Delta _{{\bf B}_r}^+ \equiv \{ e_i\pm e_j, e_i\} $, $i<j $
while the one for ${\bf C}_r $ series is
$\Delta _{{\bf C}_r}^+ \equiv \{ e_i \pm e_j, 2e_i\} $, $i<j $;
and 2) the reflection $S_{e_j} $ with respect to the root $e_j $ coincide
with $S_{2e_j} $-- the one with respect to the root $2e_j $.
In the tables below we provide for each equivalence class: i)~the cyclic
group generated by each of the automorphisms in the class; ii)~the number
of elements in each class and iii)~a representative element in it.

\begin{remark}\label{rem:w0}
{}For ${\bf B}_r $ and ${\bf C}_r $ series and for ${\bf G}_2 $ the inner
automorphism $w_0 $ which maps the highest weight vectors into the lowest
weight vectors of the algebra acts on the Cartan-Weyl basis as follows:
\begin{eqnarray}\label{eq:w0.1}
w_0(E_{\alpha }) = n_{\alpha }E_{-\alpha }, \quad w_0(H_k)=-H_k,
\quad \alpha \in \Delta _+,
\quad n_{\alpha }= \pm 1.
\end{eqnarray}
\end{remark}

Let us list the genetic codes of the Weyl groups for these Lie algebras:
\begin{eqnarray}\label{eq:A2.2}
\fl W({\bf A}_2)\simeq \bbbd_3, \qquad
&S_{e_1-e_2}^2 = S_{e_2-e_3}^2 =\openone , \quad
&(S_{e_1-e_2}S_{e_2-e_3})^3=\openone ,\\
\label{eq:40.2}
\fl W({\bf B}_2) \simeq \bbbd_4, \qquad
&S_{e_1-e_2}^2 = S_{e_2}^2 =\openone , \quad
&(S_{e_1-e_2}S_{e_2})^4=\openone ,\\
\label{eq:G2.2}
\fl W({\bf G}_2)\simeq \bbbd_6, \qquad
&S_{e_1-e_2}^2 = S_{e_2}^2 =\openone , \quad
&(S_{e_1-e_2}S_{e_2})^6=\openone ,\\
\label{eq:A3.2}
\fl W({\bf A}_3) \simeq {\cal S}_4, \qquad
&S_{e_1-e_2}^2 = S_{e_2-e_3}^2 = S_{e_3-e_4}^2 =\openone ,\quad
&(S_{e_1-e_2}S_{e_2-e_3})^3= \openone \nonumber\\
&(S_{e_1-e_2}S_{e_2-e_3}S_{e_3-e_4})^4=\openone ,\\
\label{eq:42.2}
\fl W({\bf B}_3) \qquad
&S_{e_1-e_2}^2 = S_{e_2-e_3}^2 =S_{e_3}^2 =\openone , \qquad
&(S_{e_1-e_2}S_{e_2-e_3})^3= \openone \nonumber\\
\fl &(S_{e_2-e_3}S_{e_3})^4 = \openone , \quad
&(S_{e_1-e_2}S_{e_2-e_3}S_{e_3})^6=\openone,
\end{eqnarray}
where ${\cal S}_4 $ is the group of permutation of $4 $ elements.

Their equivalence classes are listed in the tables below, where in the
first line we denote the order of each of elements in the class, in the
second line we list the number of elements in each class and on the third
line give a representative element.

\[
\fl {\bf A}_2 \quad
\begin{array}{|c|c|c|} \hline
\openone & \bbbz_2 & \bbbz_3 \\
1 & 3 & 2 \\
\openone & S_{e_1-e_2} & S_{e_1-e_2} S_{e_2-e_3}\\ \hline
\end{array} \qquad
{\bf B}_2 \quad
\begin{array}{|c|c|c|c|c|} \hline
\openone & -\openone & \bbbz_2^{(1)} & \bbbz_2^{(2)} & \bbbz_4 \\
1 & 1 & 2 & 2 & 2 \\
\openone & w_0 & S_{e_1-e_2} & S_{e_1} & S_{e_1-e_2} S_{e_2}\\ \hline
\end{array}
\]

\[
\fl {\bf G}_2 \quad
\begin{array}{|c|c|c|c|c|c|} \hline
\openone & -\openone & \bbbz_2^{(1)} & \bbbz_2^{(2)} & \bbbz_3 &
\bbbz_6 \\
1 & 1 & 3 & 3 & 2 & 2 \\
\openone & w_0 & S_{\alpha _1} & S_{\alpha _2} &
(S_{\alpha _1}S_{\alpha _2})^2 &S_{\alpha _1}S_{\alpha _2} \\ \hline
\end{array}
\]

\[
\fl {\bf A}_3 \quad
\begin{array}{|c|c|c|c|c|} \hline
\openone & \bbbz_2^{(1)} & \bbbz_2^{(2)} & \bbbz_3 &\bbbz_4 \\
1 & 6 & 3 & 8 & 6 \\
\openone & S_{e_1-e_2} & S_{e_1-e_2} S_{e_3-e_4}&
S_{e_1-e_2}S_{e_2-e_3} &S_{e_1-e_2}S_{e_2-e_3}S_{e_3-e_4} \\
\hline \end{array} \]

\[
\fl {\bf B}_3 \quad
\begin{array}{|c|c|c|c|c|} \hline
\openone & -\openone & \bbbz_2^{(1)} & \bbbz_2^{(2)} & \bbbz_2^{(3)}
\\
1 & 1 & 6 & 3 & 6 \\
\openone & w_0 & S_{e_1-e_2} & S_{e_3} & S_{e_1-e_2}S_{e_3}\\ \hline
\bbbz_2^{(4)} & \bbbz_3 & \bbbz_4^{(1)} & \bbbz_4^{(2)} & \bbbz_6 \\
3 & 8 & 6 & 6 & 8 \\
S_{e_1} S_{e_2} & S_{e_1-e_2} S_{e_2-e_3} &
S_{e_1} S_{e_1-e_2} & S_{e_1} S_{e_3} S_{e_1-e_2} & S_{e_1-e_2}
S_{e_2-e_3} S_{e_3} \\ \hline\end{array}.
\]

We leave more detailed explanations of the general theory of finite groups
to other papers and turn now to the examples.

\begin{remark}\label{rem:2}
In all examples below we apply the reductions to $L $-operators of generic
form. This means that the unreduced $J $ is a generic element of $\fr{h} $
and therefore $(\vec{a},\alpha )\neq 0 $. In fact we have used above the
vector $\vec{a} $ for fixing up the order in the root system of $\fr{g} $.
The potential $Q $ is also generic, i.e. depends on $|\Delta | $
complex-valued functions where $|\Delta | $ is the number of roots of
$\fr{g} $.
However the reduction imposed on $J $ may lead to a qualitatively different
situation in which the reduced $J_{\rm r} $ is not generic, i.e. there
may exist a subset of roots $\Delta _0 $ such that $(\vec{a}_{\rm r},\alpha
)=0 $ for $\alpha \in \Delta _0 $. Then obviously the potential $[J,Q] $
in $L $ will depend only on $|\Delta |-|\Delta _0| $ complex-valued
fields, the other fields are redundant, see Tables 1 and 2.

In what follows whenever such situations arise we will provide the subset
$\Delta _0$ or, equivalently the list of redundant functions in $Q $.
Obviously both the corresponding $N $-wave equation and its Hamiltonian
structures will depend only on the fields labelled by the roots $\alpha $
such that $(\vec{a}_{\rm r},\alpha )\neq 0 $, see \cite{VSG*94}.
\end{remark}

\begin{remark}\label{rem:3}
Several of the $\bbbz_2 $-reductions below contain automorphisms which map
$J $ to $-J $. Then it is only natural that both the canonical
symplectic form $\Omega ^{(0)} $ and the Hamiltonian $H^{(0)} $ vanish
identically. In these cases we will write down the corresponding $N $-wave
systems of equations; their Hamiltonian formulation is discussed in
Section~5 below.
\end{remark}

\begin{remark}\label{rem:5}
Under some of the reductions the
corresponding Equation (\ref{eq:1.4}) becomes linear and trivial. This
happens when the Cartan subalgebra elements invariant under the reduction
form a one-dimensional subspace in $\fr{h} $ and therefore $J_{\rm r}
\propto I_{\rm r} $. For obvious reasons we have omitted these examples.
\end{remark}


\section{Description of the $\bbbz_2 $ reductions} \label{sec:descr}

\begin{remark}\label{rem:skip}
In what follows we will skip the leading zeroes in the notations of the
roots, e.g. by $\{1\}$ and $\{11\}$ we mean $\{001\}$ and $\{011\}$
respectively for the ${\bf A}_3 $, ${\bf B}_3 $ and ${\bf C}_3 $
algebras. For ${\bf A}_2 $, ${\bf C}_2 $ and ${\bf G}_2 $ algebra by
$\{1\} $ we mean $\{01\}$. We will also drop all indices R in the triples
$\left\{ \begin{array}{c} \alpha \\ \beta ,\gamma \end{array}\right\} $.

\end{remark}

\subsection{${\frak g} \simeq {\bf A}_2 = {\it sl(3)}$}\label{ssec:a.2}

This algebra has three positive roots $\Delta ^+=\{ 10, 01, 11\} $ where
$\alpha _1 = e_1 - e_2$, $\alpha _2=e_2-e_3 $ and $jk=j\alpha _1+k\alpha _2
$. Then $Q(x,t) $ contains six functions and the set ${\cal  M} $ contains
only one triple ${\cal  M}\equiv \{[11,01,10]\} $.

\begin{example}\label{exa:a2.1}
$C_{\ref{exa:a2.1}}=\openone $. $U^T(- \lambda )+U(\lambda )=0 $.
This reduction does not restrict the Cartan elements. We have
\begin{eqnarray}\label{eq:a.1.2}
p_{\alpha }=q_{\alpha }, \qquad \alpha \in \Delta _+ ;
\end{eqnarray}
and we obtain the next $3 $-wave system:
\begin{eqnarray}\label{eq:a2.1.1}
&&i(a_1-a_2)q_{10,t}-i(b_1-b_2)q_{10,x}-\kappa q_1q_{11}=0;
\nonumber\\
&&i(a_2-a_3)q_{1,t}-i(b_2-b_3)q_{1,x}-\kappa q_{10}q_{11}=0;
\\
&&i(a_1-a_3)q_{11,t}-i(b_1-b_3)q_{11,x}+\kappa q_{10}q_{1}=0;
\nonumber
\end{eqnarray}
with $\kappa =a_1b_2+a_2b_3+a_3b_1-a_2b_1-a_3b_2-a_1b_3 $. Due to the
reduction conditions for the elements of the potential matrix the
Hamiltonian vanishes, see Remark 3.

\end{example}

\begin{example}\label{exa:a2.4}
$C_{\ref{exa:a2.4}}=\openone $. $U^*(\eta \lambda^* ) +U(\lambda )=0 $.
Therefore:
\begin{eqnarray}\label{eq:a2.4.2}
a_i^*=-\eta a_i, \quad b_i^*=-\eta b_i, \quad
p_{\alpha }^*=\eta p_{\alpha }, \quad q_{\alpha }^*=\eta q_{\alpha } ;
\end{eqnarray}
and we obtain 6 'real' fields and the $6 $-wave system with the following
Hamiltonian:  \begin{eqnarray}\label{eq:a2.1.1a} H^{(0)} = H_0(10) +
H_0(01) + H_0(11) + \kappa H(11,1,10).  \end{eqnarray} Here again $\kappa
=a_1b_2+a_2b_3+a_3b_1-a_2b_1-a_3b_2-a_1b_3 $.  The case $\eta=1 $ leads
to the non-compact real form $sl(3, \bbbr) $ for the ${\bf A}_2 $- algebra.

\end{example}

\begin{example}\label{exa:a2.5}
$C_{\ref{exa:a2.5}}=S_{e_1-e_3} $. $C_{\ref{exa:a2.5}}(U^*(\eta
\lambda^* ))
+U(\lambda )=0 $. We have:
\begin{eqnarray}\label{eq:a2.5.2}
a_3=\eta a_1^*; \quad a_2^*=\eta a_2, \quad b_3=\eta b_1^*, \quad
b_2^*=\eta b_2 \nonumber\\
p_{10}^*=-\eta q_{1}^*, \quad p_{1}=-\eta q_{10}^* \quad
p_{11}=-\eta q_{11}^*;
\end{eqnarray}
and we obtain the $3 $-wave system with the Hamiltonian:
\begin{eqnarray}\label{eq:a2.5.1}
H^{(0)}= H_{0*}(10) + H_{0*}(01) + H_{0*}(11)-\kappa H_*(11,1,10).
\end{eqnarray}
where
\begin{eqnarray}\label{eq:H*}
H_*(\alpha ,\beta ,\gamma )=
{1  \over \sqrt{\eta } } \int_{-\infty }^{\infty }dx\,
(q_\alpha q_\beta ^*q_\gamma ^*+\eta q_\alpha ^* q_\beta q_\gamma ),
\end{eqnarray}
and again $\kappa =a_1b_2+a_2b_3+a_3b_1-a_2b_1-a_3b_2-a_1b_3 $.

\end{example}

\begin{example}\label{exa:a2.00}
$C_{\ref{exa:a2.00}}=\openone $. $U^{\dag}(\eta \lambda ^*)
-U(\lambda )=0 $. Therefore:
\begin{eqnarray}\label{eq:a2.0.2}
a_i^*=\eta a_i, \quad b_i^*=\eta b_i, \quad p_{\alpha }
=-\eta q_{\alpha }^*,
\end{eqnarray}
and we obtain the $3 $-wave system with the following Hamiltonian:
\begin{eqnarray}\label{eq:a2.1.1b}
H^{(0)} = H_{0*}(11) +  H_{0*}(10) +H_{0*}(01) + \kappa H_*(11,1,10).
\end{eqnarray}
Here again $\kappa =a_1b_2+a_2b_3+a_3b_1-a_2b_1-a_3b_2-a_1b_3 $ and
$H_*(11,01,10) $ is defined by (\ref{eq:H*}). The case $\eta=1 $ extracts
the compact real form $su(3) $ for the ${\bf A}_2 $- algebra.

\end{example}

\begin{example}\label{exa:a2.3}
$C_{\ref{exa:a2.3}}=S_{e_1-e_3} $.
$C_{\ref{exa:a2.3}}(U^{\dag}(\eta \lambda^* )) -U(\lambda )=0 $. We obtain:
\begin{eqnarray}\label{eq:a2.0.2a}
a_3=\eta a_1^*, \quad a_2^* = \eta a_2, \quad b_3=\eta b_1^*, \quad
b_2^*=\eta b_2 \nonumber\\
q_1=\eta q_{10}^*, \quad q_{11}^*=-\eta q_{11}, \quad
p_1=\eta p_{10}^*, \quad p_{11}^*=-\eta p_{11};
\end{eqnarray}
and we obtain the $4 $-wave (2 real and 2 complex) system with the
Hamiltonian:
\begin{eqnarray}\label{eq:a2.1.1c}
&& H^{(0)} = H_{0}(11) + 2\re (H_{0}(01) + H_{0}(10)) \nonumber\\
&&\qquad  - {1 \over \sqrt{-\eta} }
\int_{-\infty }^{\infty }dx\, (p_{11}|q_{10}|^2-q_{11}|p_{10}|^2).
\end{eqnarray}
Here again $\kappa =a_1b_2+a_2b_3+a_3b_1-a_2b_1-a_3b_2-a_1b_3 $ is real.

\end{example}

\begin{example}\label{exa:a2.hh}
$C_{\ref{exa:a2.hh}}=\Sigma =\mbox{{\rm diag}}\,(s_1, s_2, s_3) $.
$\Sigma U^{\dag}(\eta \lambda^* )\Sigma ^{-1}
-U(\lambda )=0 $ and $s_i=\pm 1 $.
This reduction restricts the Cartan elements to be real (purely imaginary)
for $\eta=1 $ ($\eta=-1 $) and:
\begin{eqnarray}\label{eq:a2.h.2}
p_{10}=-\eta {s_2 \over s_1}q_{10}^*, \quad
p_{1}=-\eta {s_3 \over s_2}q_{1}^*, \quad
p_{11}=-\eta {s_3 \over s_1}q_{11}^*.
\end{eqnarray}
Thus we get the $3 $-wave system with the Hamiltonian:
\begin{eqnarray}\label{eq:a2.1.1d}
H^{(0)} = {s_2 \over s_1 } H_{0*}(10) + {s_3 \over s_2 } H_{0*}(01) +
{s_3 \over s_1 }H_{0*}(11) + \kappa H_*(11,1,10).
\end{eqnarray}
Here again $\kappa =a_1b_2+a_2b_3+a_3b_1-a_2b_1-a_3b_2-a_1b_3 $ and
$H_*(i,j,k) $ is defined by (\ref{eq:H*}). The choice $\eta=1 $,
$s_1=s_2=s_3 $ reproduces the result of Example \ref{exa:a2.00}
while the choice $\eta=1 $, $s_1=-s_2=-s_3 $ extract
the non-compact real form $su(2,1) $ of the ${\bf A}_2 $- algebra.

\end{example}

\subsection{${\frak g} \simeq {\bf C}_2 = {\it sp(4)}$}\label{ssec:c.2}

This algebra has four positive roots $\Delta ^+=\{ 10, 01, 11,21\} $ where
$\alpha _1 = e_1 - e_2$, $\alpha _2=2e_2 $ and $jk=j\alpha _1+k\alpha _2
$. Then $Q(x,t) $ contains eight functions. The set ${\cal  M} $ consists
of two elements: ${\cal  M}=\{[21,11,10],[11,01,10]\} $.

\begin{example}\label{exa:c2.r}
$C_{\ref{exa:c2.r}}=\openone $. $U^*(\eta \lambda ^*)
+U(\lambda )=0 $, $\eta =\pm1 $.
Then all functions $q_{\alpha }, p_{\alpha } $ become real and the Cartan
elements become purely imaginary for $\eta =1 $ and vice versa for
$\eta =-1$; i.e.,
\begin{equation}\label{eq:c2.r.1*}
\fl a_i = -\eta a_i^*, \quad b_i = -\eta b_i^*, \quad i=1,2; \quad
q_\alpha =\eta q_\alpha ^*, \quad
p_\alpha =\eta p_\alpha ^*, \quad \alpha \in \Delta _+.
\end{equation}
Thus we obtain 8 'real' fields and the $8 $-wave system with the
Hamiltonian:
\begin{eqnarray}\label{eq:c2.r.1}
\fl H^{(0)}= H_0(10)+H_0(1) + H_0(11) + H_0(21)
+2\kappa (H(21,11,10) - H(11,1,10)),
\end{eqnarray}
with $\kappa =(a_1b_2-a_2b_1) $
which is related to the non-compact real form $sp(4, \bbbr) $ of the
${\bf C}_2$- algebra.
\end{example}

\begin{example}\label{exa:c2.4}
$C_{\ref{exa:c2.4}}=w_0 $. $C_{\ref{exa:c2.4}}(U^*(\eta \lambda ^*))
+U(\lambda )=0 $ and $\eta = \pm 1$. Then:
\begin{eqnarray}\label{eq:c2.4.1}
a_1^* = \eta a_1, \quad a_2^* = \eta a_2 ;\quad
b_1^* = \eta b_1, \quad b_2^* = \eta b_2 ;
\quad p_{\alpha } = - \eta q_{\alpha }^*
\end{eqnarray}
which leads to the general $4$--wave system on the compact real form
$sp(4,0)$ ($\eta=1 $) of ${\bf C}_2 $ algebra with the Hamiltonian:
\begin{eqnarray}\label{eq:c2.4.3}
H^{(0)} = H_{0*}(10)+H_{0*}(1) + H_{0*}(11) + H_{0*}(21)
\nonumber\\
+{2 \kappa \over \sqrt{\eta}} \int_{-\infty}^{\infty }dx\, [
q_{11}q_{10}^*q_{1}^* + q_{21}q_{11}^*q_{10}^* + \eta (q_{11}^*q_{10}q_1 +
q_{21}^*q_{11}q_{10})],
\end{eqnarray}
and $\kappa = a_1b_2 - a_2b_1 $.
\end{example}

\begin{example}\label{exa:c2.hh}
After a reduction of hermitian type $K^{-1}U^{\dagger}
(\eta \lambda ^*)K - U(\lambda)=0$, where $K = $ $\mathrm{{diag}}\,(s_1,
s_2, 1 /s_2, 1/s_1)$ and $\eta =\pm 1 $ we obtain
\begin{eqnarray}\label{eq:c2.hh*}
\fl && p_{10} = -\eta  s_1 /s_2 q_{10}^*, \quad p_1 = -\eta  s_2^2q_1^*,
\quad p_{11} =-\eta s_1s_2q_{11}^*, \quad p_{21} =
-\eta s_1^2q_{21}^*, \nonumber\\
\fl && a_i = \eta  a_i^*, \quad b_i = \eta  b_i^*,
\end{eqnarray}
and the next $4 $-wave system
\begin{eqnarray}\label{eq:3.1.3}
\fl && i(a_1 - a_2) q_{10;t} - i (b_1 - b_2) q_{10;x} +
2\eta \kappa (s_2^2 q_{11}
q_1^* - s_1s_2 q_{21} q_{11}^*) =0, \nonumber\\
\fl && ia_2q_{1;t} - ib_2 q_{1;x} +
2\eta \kappa (s_1 /s_2 ) q_{11} q_{10}^* =0,\\
\fl &&ia_1q_{21;t} -ib_4 q_{21;x} + 2\kappa q_{11} q_{10} =0, \nonumber\\
\fl && i(a_1 + a_2) q_{11;t} - i(b_1+b_2) q_{11;x} -2
\kappa \left( q_{10} q_1 +
\eta (s_1/s_2) q_{21} q_{10}^*\right) =0,\nonumber
\end{eqnarray}
where $\kappa =a_1b_2-a_2b_1 $. It is described by the following
Hamiltonian:
\begin{eqnarray}\label{eq:3.1.5}
H^{(0)} = {s_1 \over s_2}H_{0*}(10)+s_2^2H_{0*}(1) +
s_1s_2H_{0*}(11) + s_1^2H_{0*}(21) \nonumber\\
\fl +{2\kappa\over \sqrt{\eta }} \int_{-\infty }^{\infty } dx \,\left(
s_1s_2 \left( q_{11}q_1^*q_{10}^* + \eta  q_{11}^*q_1q_{10} \right)
- s_1^2 \left( q_{21}q_{11}^*q_{10}^* + \eta  q_{21}^*q_{11}q_{10}\right)
\right).
\end{eqnarray}
In the case $\eta =-1 $ if we identify $q_{10} = Q$, $q_{11} =
E_p$, $q_{21} = E_a$ and $q_{1} =E_s$, where $Q $ is the normalized
effective polarization of the medium and $E_p $, $E_s $ and $E_a $ are the
normalized pump, Stokes and anti-Stokes wave amplitudes respectively,
then we obtain the system of equations generalizing the one studied
in \cite{3} which describes Stokes--anti-Stokes wave generation.  This
approach allowed us to derive a new Lax pair for (\ref{eq:3.1.3}). A
particular case of (\ref{eq:3.1.3}) with $s_1=s_2=\pm 1 $ and $\eta =\pm 1
$ is equivalent to the $4 $-wave interaction, see \cite{1} and is related
to the compact real form $sp(4,0) $ of ${\bf C}_2 $.  For $s_1=-s_2= \pm 1
$ and $\eta  =1 $ the reduced system is related to the noncompact real
form $sp(2,2) $ of ${\bf C}_2 $- algebra.

\end{example}

\begin{example}\label{exa:c2.2}
$C_{\ref{exa:c2.2}}=S_{e_1-e_2} $. $C_{\ref{exa:c2.2}}(U^*(\eta \lambda
^*)) +U(\lambda )=0 $ and $\eta = \pm 1$.
This reduction gives the following restrictions:
\begin{eqnarray}\label{eq:c2.2.1}
&& a_2 = -\eta a_1^*, \quad b_2 = -\eta b_1^* ; \\
&& p_{10} = \eta q_{10}^*,\quad q_{11}^* = \eta q_{11}, \quad q_{21} =
-\eta q_{1}^*, \quad p_{11}^* = \eta p_{11},\quad p_{21} = -\eta p_1^*.
\nonumber
\end{eqnarray}
Then we obtain the $5$--wave (2 real and 3 complex) system which is
described by the Hamiltonian:
\begin{eqnarray}\label{eq:c2.2.3}
&&H^{(0)} = H_{0*}(10) + H_0(11) + 2 \re H_0(1) \nonumber\\
&&+ {2 \kappa \over \sqrt{\eta}} \int_{-\infty}^{\infty } dx \,[q_{11}
(q_{10}^*p_1 - q_{10}p_1^*) + \eta p_{11}(q_{10}^*q_1^* - q_{10}q_1)],
\end{eqnarray}
with $\kappa = a_1b_1^* - a_1^*b_1 $.
\end{example}

\begin{example}\label{exa:c2.3}
$C_{\ref{exa:c2.3}}=S_{2e_2} $. $C_{\ref{exa:c2.3}}(U^*(\eta \lambda ^*))
+U(\lambda )=0 $ and $\eta = \pm 1$.
Then we have:
\begin{eqnarray}\label{eq:c2.3.1}
&&a_1^* = -\eta a_1, \quad a_2^* = \eta a_2 ;\quad
b_1^* = -\eta b_1, \quad b_2^* = \eta b_2 ;\nonumber\\
&&q_{11} = -i\eta q_{10}^*,\quad p_{11} = i\eta p_{10}^*, \quad q_{21}^*
= -\eta q_{21},\nonumber\\
&&p_{21}^* = -\eta p_{21},\quad p_{1} = -\eta q_1^*.
\end{eqnarray}
which leads again to the $5 $--wave ($2$ real and $3$ complex) system with
the Hamiltonian:
\begin{eqnarray}\label{eq:c2.3.3}
\fl &&H^{(0)} = 2\re H_0(10) +H_{0*}(1)+ H_0(21) \nonumber\\
&&+{2 i \kappa \over \sqrt{\eta}}\int_{-\infty}^{\infty } dx \,[
p_{10}q_1^*q_{10}^*-\eta (p_{10}^*q_1q_{10} +
p_{21}|q_{10}|^2 + q_{21}|p_{10}|^2) ],
\end{eqnarray}
and $\kappa = a_1b_2 - a_2b_1 $.
\end{example}

\begin{example}\label{exa:c2.5}
$C_{\ref{exa:c2.5}}=w_0 $. $C_{\ref{exa:c2.5}}(U(-\lambda ))
- U(\lambda )=0$. Here we get:
\begin{eqnarray}\label{eq:c2.5.2}
p_{10}=q_{10},\quad p_{11}=q_{11}, \quad p_1=q_1,\quad p_{21}=q_{21}.
\end{eqnarray}
Then we obtain the following $4$--wave system, see Remark~\ref{rem:3}:
\begin{eqnarray}\label{eq:c2.5.3}
&&i(a_1-a_2)q_{10,t}-i(b_1-b_2)q_{10,x} - 2\kappa
(q_{21}q_{11}+q_{1}q_{11})=0, \nonumber\\
&&ia_2q_{1,t}-ib_2q_{1,x} - 2\kappa q_{10}q_{11} =0, \\
&&i(a_1+a_2)q_{11,t}-i(b_1+b_2)q_{11,x} + 2\kappa
(q_{21}q_{10}-q_{1}q_{10})=0, \nonumber\\
&&ia_1q_{21,t}-ib_1q_{21,x} + 2\kappa q_{10}q_{11} =0. \nonumber
\end{eqnarray}
with $\kappa = a_1b_2 - a_2b_1 $.
Note that this reduction doesn't restrict the Cartan elements.
\end{example}

\subsection{${\frak g} \simeq {\bf G}_2$}\label{ssec:3.4}

${\bf G}_2 $ has six positive roots $\Delta ^+=\{10, 01,
11, 21, 31, 32\} $ where again $km=k\alpha _1 + m\alpha _2 $,
$\alpha _1= (e_1- e_2+2e_3)/3 $, $\alpha _2=e_2-e_3 $
and the interaction Hamiltonian contains the set ot triples of indices
${\cal M} \equiv \{ [11,1,10]$, $[21,11,10]$, $[31,21,10]$, $[32,31,1]$,
$[32,21,11] \}$.

Note that here if the Cartan elements are real then the $N $--wave
equations after the reduction become trivial except one case
\ref{exa:g2.t}, see Remark~\ref{rem:5}.

\begin{example}\label{exa:g2.t}
$C_{\ref{exa:g2.t}}=\openone $.
$C_{\ref{exa:g2.t}}(U^T(-\lambda )) +U(\lambda )=0 $.
This does not restrict the Cartan elements and for the potential matrix
gives:  \begin{eqnarray}\label{eq:g2.t.1} p_{\alpha }=q_{\alpha }, \quad
\alpha \in \Delta _+ \end{eqnarray} and a $6 $--wave system, see Remark 3:
\begin{eqnarray}\label{eq:g2.t.3}
&&i(2a_1-a_2)q_{10,t}-i(2b_1-b_2)q_{10,x}+
\kappa (q_1q_{11}+2q_{21}q_{11}+q_{31}q_{21})=0 \nonumber\\
&&i(3a_1-a_2)q_{1,t}-i(3b_1-b_2)q_{1,x}-
3\kappa (q_{10}q_{11}+q_{32}q_{31})=0 \nonumber\\
&&i(a_1-a_2)q_{11,t}-i(b_1-b_2)q_{11,x}+
\kappa (q_{1}q_{10}-2q_{21}q_{10}+q_{32}q_{21})=0 \nonumber\\
&&ia_1q_{21,t}-ib_1q_{21,x}-
\kappa (2q_{11}q_{10}-q_{31}q_{10}+q_{32}q_{11})=0 \\
&&i(3a_1-a_2)q_{31,t}-i(3b_1-b_2)q_{31,x}+
3\kappa (q_{32}q_{1}-q_{21}q_{10})=0 \nonumber\\
&&ia_2q_{32,t}-ib_2q_{32,x}+
3\kappa (q_{21}q_{11}-q_{31}q_{1})=0 \nonumber
\end{eqnarray}
with
$\kappa = a_1b_2 - a_2b_1 $. Due to the reduction conditions for the
potential matrix (\ref{eq:g2.t.1}) the terms $H(\alpha ,\beta ,\gamma ) $
in (\ref{eq:H-om}) vanish.

\end{example}

\begin{example}\label{exa:g2.3}
$C_{\ref{exa:g2.3}}=w_0 $.
$C_{\ref{exa:g2.3}}(U^*(\eta \lambda ^* )) +U(\lambda )=0 $ and $\eta =
\pm 1 $.
This gives:
\begin{eqnarray}\label{eq:g2.3.1}
&&a_1^*=\eta a_1,\quad a_2^*=\eta a_2,\quad b_1^*=\eta b_1,\quad
b_2^*=\eta b_2; \quad p_{\alpha } = -\eta q_{\alpha }^*
\end{eqnarray}
and a $6 $--wave system described by the Hamiltonian:
\begin{eqnarray}\label{eq:g2.3.3}
\fl H^{(0)} = 3(H_{0*}(10) + H_{0*}(11) + H_{0*}(21)) + H_{0*}(1) +
H_{0*}(31) + H_{0*}(32) \\
\fl +3 \kappa  [ H_*(32,31,1) +H_*(32,21,11)- H_*(31,21,10)
- 2H_*(21,11,10) + H_*(11,10,1) ],\nonumber
\end{eqnarray}
with $\kappa = a_1b_2 - a_2b_1 $.

\end{example}

\begin{example}\label{exa:g2.hh}
$\Sigma^{-1} U^{\dagger}(\eta \lambda ^*)\Sigma -U(\lambda )=0 $, $\eta
=\pm1 $ where $\Sigma  $ belongs to the Cartan subgroup and equals $\Sigma
=\mathrm{diag}\,(s_1s_2,s_1,s_2,1,{1 /s_2},{1 /s_1}, {1/(s_1s_2)})$.
Then all Cartan elements become real (purely imaginary) for $\eta =1 $
($\eta =-1 $) and
\begin{eqnarray}\label{eq:g2.hh-1}
p_{10}=-\eta {1\over s_2}q_{10}^* \qquad
&p_{1}=-\eta {s_2\over s_1}q_{1}^* \qquad
&p_{11}=-\eta {1 \over s_1}q_{11}^* , \\
p_{21}=-\eta {1\over s_1s_2}q_{21}^* \qquad
&p_{31}=-\eta {1 \over s_1s_2^2}q_{31}^* \qquad
&p_{32}=-\eta {1 \over s_1^2s_2}q_{32}^* ,\nonumber
\end{eqnarray}
which leads to a $6 $-wave system with Hamiltonian
\begin{eqnarray}\label{eq:g2.3.31}
\fl H^{(0)} &=& 3\left({1\over s_2}H_{0*}(10) + {1\over s_1}
H_{0*}(11) + {1\over s_1s_2}H_{0*}(21)) + {s_2\over s_1}H_{0*}(1) +
{1\over s_1s_2^2}H_{0*}(31) \right.\\
\fl  &+& \left.{1\over s_1^2s_2}H_{0*}(32) \right)
+3 \kappa  \left[ {1\over s_1^2s_2}H_*(32,31,1) +
{1\over s_1^2s_2}H_*(32,21,11)     \right. \nonumber\\
\fl &-& \left.{1\over s_1s_2^2}H_*(31,21,10) -
{2\over s_1s_2}H_*(21,11,10) + {1\over s_1}H_*(11,10,1) \right],
\nonumber
\end{eqnarray}
with $\kappa = a_1b_2 - a_2b_1 $ and $H_*(\alpha ,\beta ,\gamma ) $ is
defined by (\ref{eq:H*}).  In the particular case $s_1=s_2=1 $ we obtain
the result of example \ref{exa:g2.3}, $\eta =1 $.  \end{example}

\begin{example}\label{exa:g2.1}
$C_{\ref{exa:g2.1}}=S_{\alpha _1} $. $C_{\ref{exa:g2.1}}(U^*(\eta
\lambda ^* )) +U(\lambda )=0 $ and $\eta = \pm 1 $.
Then:
\begin{eqnarray}\label{eq:g2.1}
&&a_2=a_1-\eta a_1^*, \quad b_2=b_1-\eta b_1^* \nonumber\\
&&q_{31} = \eta q_{1}^*,\quad p_{10}=\eta q_{10}^*, \quad q_{21}
= \eta q_{11}^*,\quad q_{32}^* = \eta q_{32}, \nonumber\\
&&p_{31} = \eta p_{1}^*,\quad
p_{21} = \eta p_{11}^*, \quad p_{32}^* =\eta p_{32}.
\end{eqnarray}
so we obtain the $7 $--wave ($2$ real and $5$ complex) system with the
Hamiltonian:
\begin{eqnarray}\label{eq:g2.3}
&&H^{(0)} = -H_{0*}(10) + 2\re H_0(31) + 2\re H_0(21) +
H_0(32) \\
&& + \left\{ \begin{array}{c} 32 \\ 21, 11'  \end{array} \right\}+
\left\{ \begin{array}{c} 32 \\ 31, 01'  \end{array} \right\}  +
\left\{ \begin{array}{c} 21 \\ 11', 10  \end{array} \right\}   +
2 \re \left\{ \begin{array}{c} 31 \\ 21, 10  \end{array} \right\}
\nonumber \end{eqnarray}

\end{example}

\begin{example}\label{exa:g2.2}
$C_{\ref{exa:g2.2}}=S_{\alpha _2} $. $C_{\ref{exa:g2.2}}(U^*(\eta \lambda
^* )) +U(\lambda )=0 $ and $\eta = \pm 1 $.
Then:
\begin{eqnarray}\label{eq:g2.2.1}
&&a_1 = {1 \over 3}(a_2-\eta a_2^*), \quad b_1 = {1 \over 3}(b_2-\eta
b_2^*), \nonumber\\ &&q_{11} = -\eta q_{10}^*,\quad p_{1}=\eta q_{1}^*,
\quad q_{21}^* = -\eta q_{21},\quad q_{32} = \eta q_{31}^*, \nonumber\\
&&p_{11} = -\eta p_{10}^*,\quad
p_{21}^* = -\eta p_{21}, \quad p_{32} = \eta p_{31}^*.
\end{eqnarray}
so we obtain the  $7 $--wave ($2$ real and $5$ complex) system which is
described by the Hamiltonian:
\begin{eqnarray}\label{eq:g2.2.3}
&&H^{(0)} = 2\re H_0(11)- H_{0*}(1) +H_0(21) + 2\re H_0(32)\\
&& + 2 \re \left\{ \begin{array}{c} 32 \\ 21, 11  \end{array} \right\}+
\left\{ \begin{array}{c} 32 \\ 31', 01  \end{array} \right\}  +
\left\{ \begin{array}{c} 21 \\ 11, 10'  \end{array} \right\}   +
\left\{ \begin{array}{c} 11 \\ 01, 10'  \end{array} \right\}
\nonumber
\end{eqnarray}
with $\kappa =a_2b_2^* - a_2^*b_2 $.
\end{example}

\subsection{${\frak g} \simeq {\bf A}_3 = {\it sl(4)}$}\label{ssec:a.3}

This algebra has 6 positive roots: $\Delta _+= \{100 $, $010 $, $001$,
$110 $, $011 $, $111 \} $ where again $ijk= i\alpha _1 + j\alpha _2
+ k\alpha _3 $ and $\alpha _1=e_1-e_2 $; $\alpha _2=e_2-e_3 $;
$\alpha _3=e_3-e_4 $ are the simple roots of the ${\bf A}_3 $-algebra.
The set ${\cal  M} $ consists of
\[
{\cal  M}=\{[111,011,100],[111,110,001],[011,001,010],[110,010,100]\}.
\]

\begin{example}\label{exa:a3.4}
$C_{\ref{exa:a3.4}}=S_{e_1-e_2} $.
$C_{\ref{exa:a3.4}}(U(\lambda )) - U(\lambda )=0 $.
This reduction gives:
\begin{eqnarray}\label{eq:a3.1}
a_2=a_1, \quad b_2=b_1, \quad p_{100}=q_{100},\quad q_{110}=-q_{10} \\
q_{111}=-q_{11}, \quad p_{110}=-p_{10}, \quad p_{111}=-p_{11}
\nonumber
\end{eqnarray}
and leaves $q_1 $ and $p_1 $ unrestricted. Thus we obtain the 6-wave
system with the Hamiltonian:
\begin{eqnarray}\label{eq:a3.2}
H^{(0)} = 2H_0(11)+2H_0(10)+H_0(1)+
2\left\{ \begin{array}{c} 011 \\ 010, 001  \end{array} \right\},
\end{eqnarray}
which is related to ${\bf A}_2 $-- subalgebra.

\end{example}

\begin{example}\label{exa:a3.cc}
$C_{\ref{exa:a3.cc}}=\openone $.
$U^*(\eta \lambda ^*) + U(\lambda )=0 $.
This reduction gives that all Cartan elements must be purely imaginary
(real) for $\eta=1 $ ($\eta=-1 $) and
\begin{eqnarray}\label{eq:ac.1}
p_{\alpha }^*=\eta p_{\alpha }, \qquad q_{\alpha }^*=\eta q_{\alpha }.
\end{eqnarray}
Thus we get $12$ 'real' fields and $12$-wave system with the Hamiltonian in
general position with the upper restrictions.
This reduction leads to the non-compact real form $sl(4, \bbbr) $ of the
${\bf A}_3$-algebra.

\end{example}

\begin{example}\label{exa:a3.7}
$C_{\ref{exa:a3.7}}=S_{e_1-e_2} $.
$C_{\ref{exa:a3.7}}(U^*(\eta \lambda ^*)) + U(\lambda )=0 $.
Then:
\begin{eqnarray}\label{eq:a7.1}
a_2=-\eta a_1^*, \quad a_{3,4}^*=-\eta a_{3,4}, \quad
b_{2}=-\eta b_{1}^*,\quad b_{3,4}^*=-\eta b_{3,4} \\
q_{110}=-\eta q_{10}^*, \quad q_{111}=-\eta q_{11}^*, \quad
q_1=\eta q_1^*, \quad p_{100}=\eta q_{100}^* \nonumber\\
p_{110}=-\eta p_{10}^*, \quad
p_{111}=-\eta p_{11}^*, \quad p_{1}^*=\eta p_{1}. \nonumber
\end{eqnarray}
This leads to the 7-wave (2 real and 5 complex) system with the
Hamiltonian:
\begin{eqnarray}\label{eq:a3.2a}
&&H^{(0)} = -H_{0*}(100) +H_0(1)+2\re H_0(110) +2\re H_0(111)\\
&&+
\left\{ \begin{array}{c} 111 \\ 011', 100  \end{array} \right\} +
2\re \left\{ \begin{array}{c} 111 \\ 110, 001  \end{array} \right\} +
\left\{ \begin{array}{c} 110 \\ 010', 100  \end{array} \right\}
\nonumber
\end{eqnarray}

\end{example}

\begin{example}\label{exa:a3.8}
$C_{\ref{exa:a3.8}}=S_{e_1-e_2}S_{e_3-e_4} $.
$C_{\ref{exa:a3.8}}(U^*(\eta \lambda ^*)) + U(\lambda )=0 $.
Therefore:
\begin{eqnarray}\label{eq:a8.1}
a_2=-\eta a_1^*, \quad a_{4}=-\eta a_{3}^*, \quad
b_{2}=-\eta b_{1}^*,\quad b_{4}=-\eta b_{3}^* \\
q_{110}=\eta q_{11}^*, \quad q_{111}=\eta q_{10}^*, \quad
p_1=\eta q_1^*, \quad p_{100}=\eta q_{100}^* \nonumber\\
p_{110}=\eta p_{11}^*, \quad p_{111}=\eta p_{10}^*, \nonumber
\end{eqnarray}
and we obtain the 6-wave (complex) system with the following
Hamiltonian:
\begin{eqnarray}\label{eq:a3.8.2}
&&H^{(0)} = -H_{0*}(100) -H_{0*}(1)+2\re H_0(111) +2\re H_0(11) \\
&&+
2\re \left(\left\{ \begin{array}{c} 111 \\ 011, 100  \end{array} \right\}
+ \left\{ \begin{array}{c} 111 \\ 110', 001  \end{array} \right\} \right)
\nonumber
\end{eqnarray}

\end{example}

\begin{example}\label{exa:a3.t}
$C_{\ref{exa:a3.t}}=\openone $.
$C_{\ref{exa:a3.t}}(U^T(-\lambda )) + U(\lambda )=0 $.
This reduction does not restrict the Cartan elements. For the elements of
the potential matrix we have the following restrictions:
\begin{eqnarray}\label{eq:a8.t.1}
p_{\alpha }=q_{\alpha }
\end{eqnarray}
and this leads to the 6-wave system :
\begin{eqnarray}\label{eq:a3.t.2}
&&i(a_1-a_2)q_{100,t}-i(b_1-b_2)q_{100,x} +
\kappa _4q_{11}q_{111} + \kappa _2 q_{10}q_{110} = 0; \nonumber\\
&&i(a_2-a_3)q_{10,t}-i(b_2-b_3)q_{10,x} +
\kappa _2q_{100}q_{110} - \kappa _3 q_{1}q_{11} = 0; \nonumber\\
&&i(a_3-a_4)q_{1,t}-i(b_3-b_4)q_{1,x} +
\kappa _1q_{110}q_{111} - \kappa _3 q_{11}q_{10} = 0; \\
&&i(a_1-a_3)q_{110,t}-i(b_1-b_3)q_{110,x} +
\kappa _1q_{1}q_{111} - \kappa _2 q_{10}q_{100} = 0; \nonumber\\
&&i(a_2-a_4)q_{11,t}-i(b_2-b_4)q_{11,x} -
\kappa _4q_{100}q_{111} + \kappa _3 q_{1}q_{10} = 0; \nonumber\\
&&i(a_1-a_4)q_{111,t}-i(b_1-b_4)q_{111,x} -
\kappa _1q_{1}q_{110} - \kappa _4 q_{11}q_{100} = 0; \nonumber
\end{eqnarray}
where
$\tilde{\kappa }_i $, $i=1,\dots ,4 $ are given in Appendix A. The
Hamiltonian vanishes, see Remark 3.

\end{example}

\begin{example}\label{exa:a3.3}
$C_{\ref{exa:a3.3}}=S_{e_3-e_4} $.
$C_{\ref{exa:a3.3}}(U^T(-\lambda )) + U(\lambda )=0 $.
This gives:
\begin{eqnarray}\label{eq:a8.1.1}
a_4=a_3, \quad b_{4}=b_{3}, \quad p_{100}=q_{100}, \quad p_{110}=-
q_{111}, \\
p_{11}=- q_{10} \quad p_{10}=-q_{11}, \quad p_{111}=-q_{110}, \nonumber
\end{eqnarray}
while the fields $q_1 $ and $p_1 $ are both unrestricted and  redundant.
This gives the 5-wave (complex) system :
\begin{eqnarray}\label{eq:a3.8.3}
&&i(a_1-a_2)q_{100,t}-i(b_1-b_2)q_{100,x} -
\kappa _3(q_{10}q_{111} + q_{11}q_{110}) = 0; \nonumber\\
&&i(a_2-a_3)q_{10,t}-i(b_2-b_3)q_{10,x} + \kappa _3q_{110}q_{100} = 0;\\
&&i(a_1-a_3)q_{110,t}-i(b_1-b_3)q_{110,x} + \kappa _3q_{10}q_{100} = 0;
\nonumber\\
&&i(a_2-a_3)q_{11,t}-i(b_2-b_3)q_{11,x} + \kappa _3q_{111}q_{100} = 0;
\nonumber\\
&&i(a_1-a_3)q_{111,t}-i(b_1-b_3)q_{111,x} - \kappa _3q_{11}q_{100} = 0;
\nonumber
\end{eqnarray}
The Hamiltonian vanishes, see Remark 3.
\end{example}

\begin{example}\label{exa:a3.10}
$C_{\ref{exa:a3.10}}=S_{e_2-e_3}S_{e_1-e_4} $.
$C_{\ref{exa:a3.10}}(U^T(-\lambda )) + U(\lambda )=0 $.
Therefore:
\begin{eqnarray}\label{eq:a8.10.1}
a_4=-a_1, \quad a_3=-a_2, \quad b_4=-b_1, \quad b_3=-b_2; \nonumber\\
q_{1}=-q_{100}, \quad q_{11}=q_{110}, \quad p_{1}=-p_{100},\quad
p_{11}=p_{110};
\end{eqnarray} and we obtain the 8-wave system with the Hamiltonian:
\begin{eqnarray}\label{eq:a3.10.2}
&& H^{(0)}= H_0(1)+H_0(10)+H_0(11)+H_0(111) \nonumber\\
&&+2 \left(\left\{ \begin{array}{c} 111 \\ 011, 100'  \end{array} \right\}
+ \left\{ \begin{array}{c} 011 \\ 010, 001  \end{array} \right\} \right)
\nonumber
\end{eqnarray}
Here $p_{10}, q_{10} $ and $p_{111}, q_{111} $ are unrestricted fields.

\end{example}

\begin{example}\label{exa:a3.0}
$C_{\ref{exa:a3.0}}=\openone $.
$C_{\ref{exa:a3.0}}(U^{\dag}(\eta \lambda ^*)) - U(\lambda )=0 $.
This reduction gives that the Cartan elements must be real (purely
imaginary) for $\eta=1 $ ($\eta =-1 $) and for the potential matrix:
\begin{eqnarray}\label{eq:a3.0.1}
p_{\alpha }=-\eta q_{\alpha }^*, \qquad \alpha \in \Delta _+.
\end{eqnarray}
Thus we get the 6-wave system with the general Hamiltonian for this algebra
and $\tilde{\kappa }_i $; $i=1,\dots, 4 $ are real. The case $\eta=1 $
leads to the compact real form $su(4) $ for the ${\bf A}_3 $- algebra.

\end{example}

\begin{example}\label{exa:a3.hh}
$C_{\ref{exa:a3.hh}}=\Sigma = \mbox{{\rm diag}}\,(s_1, s_2, s_3, s_4) $.
$\Sigma U^{\dag}(\eta \lambda ^*)\Sigma ^{-1} -
U(\lambda )=0 $ and $s_1s_2s_3s_4=1 $. This reduction gives that the
Cartan elements must be real (purely imaginary) for $\eta=1 $ ($\eta =-1
$) and for the potential matrix:
\begin{eqnarray}\label{eq:a3.h.1}
p_{100 }=-\eta {s_2 \over s_1 } q_{100 }^*,\quad p_{10 }=-\eta {s_3 \over
s_2}q_{10 }^*,\quad p_{1}=-\eta {s_4 \over s_3 }q_{1}^*, \nonumber\\
p_{110}=-\eta {s_3 \over s_1}q_{110}^*,\quad
p_{11}=-\eta {s_4 \over s_2}q_{11}^*, \quad p_{111}=-\eta
{s_4 \over s_1}q_{111}^*,
\end{eqnarray}
Thus we get the 6-wave system with the Hamiltonian
\begin{eqnarray}\label{eq:H-rfa}
\fl H^{(0)} = {s_2 \over s_1}H_{0*}(100) + {s_3 \over s_2}H_{0*}(10)
+ {s_4\over s_3}H_{0*}(1) + {s_3 \over s_1}H_{0*}(110) + {s_4 \over
s_2}H_{0*}(11) \nonumber\\
+ {s_4 \over s_1}H_{0*}(111)+  {s_4 \over s_2} \tilde{\kappa} _1
H_*(11,1,10) + {s_4 \over s_1} \left(\tilde{\kappa} _2 H_*(111,110,1)
\right. \\
\left. + \tilde{\kappa} _3 H_*(111,11,100) \right) +{s_3 \over s_1}
\tilde{\kappa}_4 H_*(110,10,100) , \nonumber
\end{eqnarray}
where $H_*(\alpha ,\beta ,\gamma )$ is given by (\ref{eq:H*})
and $\tilde{\kappa }_i $; $i=1,\dots, 4 $ are real. The case $\eta=1 $,
$s_1=s_2=s_3=s_4 $ leads to the compact real form $su(4) $ for the ${\bf
A}_3 $- algebra, see the result of Example \ref{exa:a3.0}.
{}For $\eta=1 $ the choice $s_1=-s_2=-s_3=-s_4 $
gives us the non-compact real form $su(3,1) $ and the choice
$s_1=s_2=-s_3=-s_4 $ leads to another non-compact real form $su(2,2) $
for the ${\bf A}_3$- algebra.

\end{example}

\begin{example}\label{exa:a3.1}
$C_{\ref{exa:a3.1}}=S_{e_3-e_4} $.
$C_{\ref{exa:a3.1}}(U^{\dag}(\eta \lambda ^* )) - U(\lambda )=0 $.
Therefore:
\begin{eqnarray}\label{eq:a8.3.1}
a_{1,2}^*=\eta a_{1,2}, \quad a_{4}=\eta a_{3}^*, \quad
b_{1,2}^*=\eta b_{1,2},\quad b_{4}=\eta b_{3}^* \\
p_{100}=-\eta q_{100}^*, \quad p_{111}=\eta q_{110}^*, \quad
p_{110}=\eta q_{111}^*, \quad p_{10}=\eta q_{11}^* \nonumber\\
p_{11}=\eta q_{10}^*, \quad
p_{1}^*=-\eta p_{1}, \quad q_{1}^*=-\eta q_{1},\nonumber
\end{eqnarray}
and we obtain the 7-wave (2 real and 5 complex) system with the
Hamiltonian:
\begin{eqnarray}\label{eq:a3.1.2}
&&H^{(0)} = H_0(1)+H_{0*}(100)+2\re H_{0*}(11) +
2\re H_{0*}(111) \nonumber\\
&& + 2 \re \left(\left\{ \begin{array}{c} 111 \\ 011, 100  \end{array}
\right\} + \left\{ \begin{array}{c} 011 \\ 010, 001  \end{array} \right\}
\right) \nonumber
\end{eqnarray}

\end{example}

\begin{example}\label{exa:a3.2}
$C_{\ref{exa:a3.2}}=S_{e_2-e_3}S_{e_1-e_4} $.
$C_{\ref{exa:a3.2}}(U^{\dag}(\eta \lambda ^*)) - U(\lambda )=0 $.
Therefore:
\begin{eqnarray}\label{eq:a8.2.1}
a_{4}=\eta a_{1}^*, \quad a_{3}=\eta a_{2}^*, \quad
b_{4}=\eta b_{1}^*,\quad b_{3}=\eta b_{2}^* \\
q_{1}=\eta q_{100}^*, \quad q_{10}^*=\eta q_{10}, \quad q_{11}=-\eta
q_{110}^*, \quad q_{111}^*=\eta q_{111}, \nonumber\\
p_{1}=\eta p_{100}^*, \quad p_{10}^*=\eta p_{10}, \quad
p_{11}=-\eta p_{110}^*, \quad p_{111}^*=\eta p_{111},\nonumber
\end{eqnarray}
and we obtain the 8-wave (4 real and 4 complex) system with the following
Hamiltonian:
\begin{eqnarray}\label{eq:a3.1.3}
&&H^{(0)} = H_0(10)+H_{0}(111)+2\re H_{0*}(100)+
2\re H_{0*}(110) \\
&& + 2 \re \left(\left\{ \begin{array}{c}
111 \\ 110, 001'  \end{array} \right\} + \left\{ \begin{array}{c} 110 \\
010, 100  \end{array} \right\} \right). \nonumber
\end{eqnarray}
This system is related to the noncompact real form $su^*(4) $ of ${\bf
A}_3 $.

\end{example}

\subsection{${\frak g} \simeq {\bf B}_3 = so(7)$}\label{ssec:b.3}

In this case there are nine positive roots $\Delta _+=\{ 100, 010$, $
001$, $ 110$, $ 011$, $111$, $012, 112, 122\} $ where again
$ijk=i\alpha _1 + j\alpha _2 +k\alpha _3 $ and $\alpha _1=e_1-e_2 $,
$\alpha _2=e_2-e_3 $, $\alpha _3=e_3 $. The interaction Hamiltonians below
are given by (\ref{eq:H-om}) where the set ot triples of indices ${\cal
M}$ is $\{ [122,112,10]$, $[122,111,11]$, $[122,12,110]$, $ [112,111,1]$,
$[112,12,100]$, $ [111,110,1]$, $[111,11,100]$, $[12,11,1]$, $[11,1,10]$,
$[110,10,100] \}$.

\begin{example}\label{exa:b3.9}
$C_{\ref{exa:b3.9}}=w_0 $.
$C_{\ref{exa:b3.9}}(U(-\lambda )) - U(\lambda )=0 $. This reduction doesn't
restrict the Cartan elements. Therefore:
\begin{eqnarray}\label{eq:b3.9.1}
p_{\alpha }=q_{\alpha }
\end{eqnarray}
thus we get $9$- wave system. The Hamiltonian vanishes, see Remark 3.
\end{example}

\begin{example}\label{exa:b3.1}
$C_{\ref{exa:b3.1}}=S_{e_1-e_2} $. $C_{\ref{exa:b3.1}}(U(\lambda )) -
U(\lambda )=0 $. Then
\begin{eqnarray}\label{eq:b3.1.1}
&&p_{100}=q_{100},\quad q_{110}=q_{10}, \quad q_{111}=q_{11}, \quad
q_{112}=q_{12}, \nonumber\\
&&q_{122}=0, \quad p_{110}=p_{10}, \quad p_{111}=p_{11}, \\
&&p_{112}=p_{12}, \quad p_{122}=0, \quad a_2=a_1 \quad
b_2=b_1. \nonumber
\end{eqnarray}
The interaction reduces to the $8 $-wave system with the Hamiltonian:
\begin{eqnarray}\label{eq:3.2.5}
H^{(0)} &=& 2H_0(10)+H_0(1)+2H_0(11)+2H_0(12) \nonumber\\
&+& 2 \left(\left\{ \begin{array}{c} 012 \\ 011, 001  \end{array} \right\}
+ \left\{ \begin{array}{c} 011 \\ 010, 001  \end{array} \right\} \right)
\end{eqnarray}
In addition $q_{100} $ becomes redundant, see Remark~\ref{rem:2} and $p_1,
q_1$ are unrestricted ones.  This system is related to the
${\bf B}_2 $--subalgebra.
\end{example}

\begin{example}\label{exa:b3.2}
$C_{\ref{exa:b3.2}}=S_{e_3} $. $C_{\ref{exa:b3.2}}(U(\lambda )) -
U(\lambda)=0 $. Here we have
\begin{eqnarray}\label{eq:b3.2.1}
&&q_{112}=q_{110}, \quad q_{12}=q_{10},\quad q_{111}=q_{11}=0, \quad
p_1=-q_1, \\
&&p_{112}=p_{110}, \quad p_{12}=p_{10},\quad p_{111}=p_{11}=0, \quad
a_3=b_3=0.\nonumber
\end{eqnarray}
The Hamiltonian reduces to
\begin{eqnarray}\label{eq:3.2.6}
H^{(0)} &=& H_0(100)+2H_0(12)+2H_0(112)+H_0(122) \nonumber\\
&+& 2 \left(\left\{ \begin{array}{c} 122 \\ 012, 110'  \end{array} \right\}
+ \left\{ \begin{array}{c} 112 \\ 012, 100  \end{array} \right\} \right)
\end{eqnarray}
Here $q_1, p_1 $ are redundant fields and $p_{100}, q_{100} $, $p_{122},
q_{122} $ are unrestricted. This system is related to the
${\bf D}_3 $--subalgebra.

\end{example}

\begin{example}\label{exa:b3.3}
$C_{\ref{exa:b3.3}}=S_{e_1-e_2}S_{e_3} $. $C_{\ref{exa:b3.3}}(U(-\lambda
)) - U(\lambda )=0 $. Then
\begin{eqnarray}\label{eq:b3.3.1}
\fl &&p_{100}=-q_{100}, \quad q_{12}=-q_{110}, \quad q_{111}=q_{11}, \quad
q_{112}=-q_{10}, \quad p_1=q_1, \nonumber\\
\fl &&p_{12}=-p_{110}, \quad p_{111}=p_{11}, \quad p_{112}=-p_{10}, \quad
a_2 =-a_1, \quad b_2=-b_1.
\end{eqnarray}
However this choice means that $C_{\ref{exa:b3.3}}(J)=-J $ and therefore
Remark~\ref{rem:3} applies. This automorphism reduces (\ref{eq:1.4})
to the following $8 $-wave equations:
\begin{eqnarray}\label{eq:b3.3.1a}
&&ia_1q_{100,t}-ib_1q_{100,x}+\kappa (q_{10}p_{110}-q_{110}p_{10}) = 0,
\nonumber\\
&&i(a_1 + a_3)q_{10,t} - i(b_1 + b_3)q_{10,x} + 2\kappa (q_1q_{11} -
q_{100}q_{110}) = 0, \nonumber\\
&&ia_3q_{1,t} - ib_3q_{1,x} + \kappa (q_{11}p_{110} -
q_{11}p_{10} + p_{11}q_{110} - p_{11}q_{10}) = 0, \nonumber\\
&&i(a_1 - a_3)q_{110,t} - i(b_1 - b_3)q_{110,x} + 2\kappa (q_1q_{11} +
q_{100}q_{10}) = 0,
\nonumber\\
&&ia_1q_{11,t} - ib_1q_{11,x} - \kappa (q_1q_{110} + q_1q_{10}) = 0, \\
&&i(a_1 + a_3)p_{10,t} - i(b_1 + b_3)p_{10,x} + 2\kappa (p_{11}q_1 +
q_{100}p_{110}) = 0, \nonumber\\
&&i(a_1 - a_3)p_{110,t} - i(b_1 - b_3)p_{110,x} + 2\kappa (p_{11}q_1 -
q_{100}p_{10}) = 0, \nonumber\\
&&ia_1p_{11,t} - ib_1p_{11,x} - \kappa (q_1p_{110} + q_1p_{10}) = 0,
\nonumber
\end{eqnarray}
where $\kappa = a_1b_3 - a_3b_1$ and $q_{122} $, $p_{122} $ are redundant,
see Remark~\ref{rem:2}.
\end{example}

\begin{example}\label{exa:b3.4}
$C_{\ref{exa:b3.4}}=S_{e_1}S_{e_2} $. $C_{\ref{exa:b3.4}}(U(-\lambda ))
- U(\lambda )=0 $.
The reduction conditions give $C_{\ref{exa:b3.4}}(J)=-J $ and:
\begin{eqnarray}\label{eq:b3.4.1}
&&p_{100}=q_{100},\quad p_{112}=q_{110},\quad p_{111}=-q_{111},\quad
p_{110}=q_{112},\nonumber\\
&&p_1=0,\quad p_{122}=q_{122},\quad p_{12}=q_{10},\quad p_{11}=-q_{11},
\nonumber\\
&&p_{10}=q_{12},\quad q_1=0 \quad a_3=0,\quad b_3=0.
\end{eqnarray}
Again Remark~\ref{rem:3} applies and we obtain the next $8 $--wave system:
\begin{eqnarray}\label{eq:b3.4.3}
\fl &&i(a_1-a_2)q_{100,t}-i(b_1-b_2)q_{100,x}+\kappa
(q_{10}q_{112}+q_{12}q_{110}-2q_{11}q_{111})=0, \nonumber\\
\fl &&ia_2q_{10,t}-ib_2q_{10,x}-\kappa (q_{100}+q_{122})q_{110} = 0,
\nonumber\\
\fl &&ia_1q_{110,t}-ib_1q_{110,x}-\kappa (q_{100}-q_{122})q_{10} = 0,
\nonumber\\
\fl &&ia_2q_{11,t}-ib_2q_{11,x}-\kappa (q_{100}+q_{122})q_{111} = 0, \\
\fl &&ia_1q_{111,t}-ib_1q_{111,x}-\kappa (q_{100}-q_{122})q_{11} = 0,
\nonumber\\
\fl &&ia_2q_{12,t}-ib_2q_{12,x}-\kappa (q_{100}+q_{122})q_{112} = 0,
\nonumber\\
\fl &&ia_1q_{112,t}-ib_1q_{112,x}-\kappa (q_{100}-q_{122})q_{12} = 0,
\nonumber\\
\fl &&i(a_1+a_2)q_{122,t}-i(b_1+b_2)q_{122,x}+\kappa (q_{10}q_{112}+
q_{12}q_{110}-2q_{11}q_{111})=0, \nonumber
\end{eqnarray}
where $\kappa =a_1b_2-a_2b_1 $.
\end{example}

\begin{example}\label{exa:b3.00}
$C_{\ref{exa:b3.00}}=\openone $. $U^{\dagger}(\eta \lambda ^*)-U(\lambda
)=0 $, $\eta =\pm 1 $. This reduction makes all Cartan elements real for
$\eta =1 $ and purely imaginary for $\eta =-1 $ and
\begin{eqnarray}\label{eq:b3.00-1}
p_{\alpha }=-\eta q_{\alpha }^*.
\end{eqnarray}
Thus we get the $9 $-wave system with Hamiltonian (\ref{eq:H-om}) with the
upper restrictions.

\end{example}

\begin{example}\label{exa:b3.hh}
$\Sigma =\mathrm{diag}\,(s_1,s_2,s_3,1,{1/s_3},{1 /s_2},{1 /s_1}) $,
$\Sigma^{-1} U^{\dagger}(\eta \lambda ^*)\Sigma -U(\lambda )=0 $, $\eta
=\pm 1 $. After this reduction all Cartan elements become real for $\eta =1
$ and purely imaginary for $\eta =-1 $ and
\begin{eqnarray}\label{eq:b3.hh-1}
p_{100}=-\eta {s_1\over s_2}q_{100}^*, \quad p_{10}=-\eta {s_2\over
s_3}q_{10}^* ,\quad p_{1}=-\eta s_3q_{1}^* , \nonumber\\
p_{110}=-\eta {s_1\over s_3}q_{110}^*, \quad p_{11}=-\eta s_2 q_{11}^* ,
\quad p_{111}=-\eta s_1 q_{111}^* ,\\
p_{12}=-\eta s_2s_3q_{12}^*, \quad p_{112}=-\eta s_1s_3q_{112}^*, \quad
p_{122}=-\eta s_1s_2q_{122}^* ,\nonumber
\end{eqnarray}
Thus we get the $9 $-wave system with real Hamiltonian:
\begin{eqnarray}\label{eq:b3.hh.3}
\fl &&H^{(0)}={s_1 \over s_2}H_{0*}(100) + {s_2 \over s_3}H_{0*}(10) +
s_3H_{0*}(1) + {s_1 \over s_3}H_{0*}(110) + s_2H_{0*}(11) \nonumber\\
\fl &&+ s_1H_{0*}(111)+s_2s_3H_{0*}(12)+s_1s_3H_{0*}(112)
+s_1s_2H_{0*}(122) \\
\fl &&+ (\kappa _1-\kappa _2-\kappa _3)s_1s_2H_*(122,112,10)-\kappa
_3s_1s_2H_*(122,111,11) \nonumber\\
\fl &&+(\kappa _1-\kappa _2+\kappa _3)s_1s_2H_*(122,12,110) +
\kappa _2s_1s_3H_*(112,111,1) \nonumber\\
\fl &&+(-\kappa_1-\kappa _2+\kappa _3)s_1s_3H_*(112,12,100) +\kappa
_2s_1H_*(111,110,1) \nonumber\\
\fl && +\kappa _3s_1H_*(111,11,100) -\kappa _1s_2s_3H_*(12,11,1)+\kappa
_1s_2H_*(11,1,10)\nonumber\\
\fl && +(\kappa_1+\kappa _2+\kappa _3){s_1 \over s_3}H_*(110,10,100)
\nonumber
\end{eqnarray}
where the terms $H_*(\alpha ,\beta ,\gamma ) $ are given in
(\ref{eq:H*}). In the particular case $s_1=s_2=s_3=1 $ the result
coincides with example \ref{exa:b3.00}.
\end{example}

\begin{example}\label{exa:b3.10}
$C_{\ref{exa:b3.10}}=w_0 $.
$C_{\ref{exa:b3.10}}(U^{\dagger}(\eta \lambda ^*)) - U(\lambda )=0 $,
$\eta =\pm1 $. This reduction requires that all fields $p_{\alpha },
q_{\alpha } $ are real while all Cartan elements must be purely
imaginary for $\eta =1 $ and vice versa for $\eta =-1 $. Thus we get $18$-
wave system with the Hamiltonian (\ref{eq:H-om}).
The case with $\eta =1 $ leads to the real form $so(7, \bbbr)\simeq
so(7,0) $ for the ${\bf B}_3 $-algebra.

\end{example}

\begin{example}\label{exa:b3.7}
$C_{\ref{exa:b3.7}}=S_{e_1-e_2} $.
$C_{\ref{exa:b3.7}}(U^{\dagger}(\eta \lambda ^*)) - U(\lambda )=0 $,
$\eta =\pm1 $. Therefore:
\begin{eqnarray}\label{eq:b3.7.1}
\fl &&q_{100}^*=-\eta q_{100},\quad p_{10}=-\eta q_{110}^*,\quad
p_{11}=-\eta q_{111}^*,\quad p_{12}=-\eta q_{112}^*,\nonumber\\
\fl &&p_{100}^*=-\eta p_{100},\quad p_{110}=-\eta q_{10}^*,\quad
p_{111}=-\eta q_{11}^*,\quad p_{112}=-\eta q_{12}^*,\\
\fl &&p_{122}=\eta q_{122}^*,\quad p_1=-\eta q_1^*,\quad a_2 = \eta
a_1^*,\quad b_2 = \eta b_1^*,\quad a_3^*=\eta a_3,\quad b_3^*=\eta
b_3,\nonumber
\end{eqnarray}
which gives the $10 $--wave (2 real and 8 complex) system with the following
Hamiltonian:
\begin{eqnarray}\label{eq:b3.7.5}
\fl H^{(0)} = H_0(100) +H_{0*}(1) +H_{0*}(122) +2\re H_{0*}(110)
\nonumber\\
\fl + \left\{ \begin{array}{c} 122 \\ 111, 011  \end{array} \right\}
+ \left\{ \begin{array}{c} 112 \\ 012, 100  \end{array} \right\}
+ \left\{ \begin{array}{c} 111 \\ 011, 100  \end{array} \right\}
+ \left\{ \begin{array}{c} 110 \\ 010, 100  \end{array} \right\}
\nonumber\\
\fl +2\re \left(\left\{ \begin{array}{c} 122 \\ 012, 110  \end{array}
\right\} + \left\{ \begin{array}{c} 112 \\ 111, 001  \end{array} \right\}
+ \left\{ \begin{array}{c} 111 \\ 110, 001  \end{array} \right\}\right)
\end{eqnarray}
\end{example}

\begin{example}\label{exa:b3.8}
$C_{\ref{exa:b3.8}}=S_{e_3} $.
$C_{\ref{exa:b3.8}}(U^{\dagger}(\eta \lambda ^*)) - U(\lambda ) $, $\eta =
\pm1$. Then:
\begin{eqnarray}\label{eq:b3.7.1a}
&&p_{100}=-\eta q_{100}^*,\quad p_{112}=-\eta q_{110}^*,\quad
p_{111}=\eta q_{111}^*,\quad p_{110}=-\eta q_{112}^*,\nonumber\\
&&p_{122}=-\eta q_{122}^*,\quad p_{12}=-\eta q_{10}^*,\quad
p_{11}=\eta q_{11}^*,\quad p_{10}=-\eta q_{12}^*,\nonumber\\
&&q_1^*=\eta q_1,\quad p_1^*=\eta p_1,\quad a_3^* =-\eta a_3,
\quad b_3^* = -\eta b_3,\nonumber\\
&&a_1^*=\eta a_1, \quad a_2^*=\eta a_2, \quad b_1^* = \eta b_1,\quad
b_2^*=\eta b_2,
\end{eqnarray}
so we obtain the $10 $--wave (2 real and 8 complex) system with the
Hamiltonian:
\begin{eqnarray}\label{eq:b3.8.5}
\fl &&H^{(0)} = H_{0*}(100) +H_{0*}(11)+H_{0*}(111) +H_{0*}(122)+ H_0(1)
+2\re H_{0*}(112) \nonumber\\
\fl && + \left\{ \begin{array}{c} 122 \\ 111, 011  \end{array} \right\}
+ \left\{ \begin{array}{c} 111 \\ 011, 100  \end{array} \right\}
+2\re \left(\left\{ \begin{array}{c} 122 \\ 012, 110  \end{array} \right\}
+ \left\{ \begin{array}{c} 110 \\ 010, 100  \end{array} \right\}
\right. \nonumber\\
\fl && + \left. \left\{ \begin{array}{c} 111 \\ 110, 001  \end{array}
\right\} + \left\{ \begin{array}{c} 012 \\ 011, 001  \end{array} \right\}
\right)
\end{eqnarray}
\end{example}

\begin{example}\label{exa:b3.5}
$C_{\ref{exa:b3.5}}=S_{e_1-e_2}S_{e_3} $.
$C_{\ref{exa:b3.5}}(U^{\dagger}(\eta \lambda ^* )) - U(\lambda )=0 $,
$\eta =\pm1 $. Then:
\begin{eqnarray}\label{eq:b3.5.1}
\fl &&q_{100}=-\eta q_{100}^*, \quad p_{12}=-\eta q_{110}^*,\quad
p_{11}=\eta q_{111}^*,\quad p_{10}=-\eta q_{112}^*,\nonumber\\
\fl &&p_{122}=\eta q_{122}^*,\quad p_{112}=-\eta q_{10}^*,\quad
p_{111}=\eta q_{11}^*,\quad p_{110}=-\eta q_{12}^*, \nonumber\\
\fl &&q_1^*=\eta q_1,\quad p_1^*=\eta p_1,\quad p_{100}^*=-\eta p_{100}, \\
\fl &&a_3^*=-\eta a_3, \quad b_3^*=-\eta b_3, \quad
a_2 = \eta a_1^*,\quad b_2 = \eta b_1^*, \quad \nonumber
\end{eqnarray}
and we obtain the $11 $-wave (4 real and 7 complex) system with Hamiltonian:
\begin{eqnarray}\label{eq:b3.5.5}
\fl &&H^{(0)} =
H_{0}(100) +H_{0}(1)+H_{0*}(122) +2\re H_{0*}(112) +
2\re H_{0*}(12) \nonumber\\
\fl &&+2\re H_{0*}(111) + \left\{ \begin{array}{c} 122 \\ 012, 110
\end{array} \right\} + \left\{ \begin{array}{c} 122 \\ 111, 011
\end{array} \right\} + \left\{ \begin{array}{c} 122 \\ 112, 010
\end{array} \right\} + \left\{ \begin{array}{c} 111 \\ 011, 100
\end{array} \right\} \nonumber\\
\fl &&+2\re \left(\left\{ \begin{array}{c} 110 \\ 010, 100  \end{array}
\right\} + \left\{ \begin{array}{c} 112 \\ 111, 001  \end{array} \right\}
+ \left\{ \begin{array}{c} 012 \\ 011, 001  \end{array} \right\}\right)
\end{eqnarray}
\end{example}

\begin{example}\label{exa:b3.6}
$C_{\ref{exa:b3.6}}=S_{e_1}S_{e_2} $.
$C_{\ref{exa:b3.6}}(U^{\dagger}(\eta \lambda ^*)) - U(\lambda )=0 $.
As a result:
\begin{eqnarray}\label{eq:b3.6.1}
\fl &&a_3^* = \eta a_3,\quad b_3^* = \eta b_3,\quad
a_{1,2}^*=-\eta a_{1,2}, \quad b_{1,2}^*=- \eta b_{1,2}, \nonumber\\
\fl &&q_{112}=\eta q_{110}^*,\quad q_{12}=\eta q_{10}^*,\quad
p_{112}=\eta p_{110}^*,\quad p_{12}=\eta p_{10}^*,\nonumber\\
\fl &&q_{100}^*=\eta q_{100},\quad p_{100}^*=\eta p_{100},\quad
q_{111}^*=-\eta q_{111},\quad p_{111}^*=-\eta p_{111}, \\
\fl &&q_{122}^*=\eta q_{122},\quad p_{122}^*=\eta p_{122},\quad
p_1 = -\eta q_1^*, \quad q_{11}^*=-\eta q_{11},\quad
p_{11}^*=-\eta p_{11}, \nonumber
\end{eqnarray}
which leads to the $13 $-wave (8 real and 5 complex) system with
Hamiltonian:
\begin{eqnarray}\label{eq:b3.6.5}
\fl &&H^{(0)} = H_{0}(100) +H_{0}(122)+H_{0*}(1)+H_{0*}(11) + H_{0*}(111)
+2\re H_{0*}(12) \nonumber\\
\fl && +2\re H_{0*}(112)+  \left\{ \begin{array}{c} 122 \\ 111, 011
\end{array} \right\} + \left\{ \begin{array}{c} 111 \\ 011, 100
\end{array} \right\} \nonumber\\
\fl &&+2\re \left(\left\{ \begin{array}{c} 122 \\ 012, 110'  \end{array}
\right\} + \left\{ \begin{array}{c} 112 \\ 012, 100  \end{array} \right\}
+ \left\{ \begin{array}{c} 112 \\ 111, 001  \end{array} \right\}
+ \left\{ \begin{array}{c} 012 \\ 011, 001  \end{array} \right\}\right)
\end{eqnarray}
\end{example}

\subsection{${\frak g} \simeq {\bf C}_3 = {\it sp(6)}$}\label{ssec:c3.3}

In this case there are nine positive roots $\Delta _+=\{ 100, 010$,
$001$, $110$, $ 011$, $ 111$, $021$, $121, 221\} $ where again
$ijk=i\alpha _1 + j\alpha _2 +k\alpha _3 $ and $\alpha _1=e_1-e_2 $,
$\alpha _2=e_2-e_3 $, $\alpha _3=2e_3 $,
the set ot triples of indices is ${\cal M} \equiv \{ [110,10,100]$,
$[111,11,100]$, $[121,21,100]$, $ [121,11,110]$, $[21,11,10]$, $
[111,110,1]$, $[11,1,10]$, $[121,111,10]$, $[221,121,100]$,\\
$[221,111,110] \}$.

\begin{example}\label{exa:c3.16}
$C_{\ref{exa:c3.16}}=w_0 $.
$C_{\ref{exa:c3.16}}(U(-\lambda )) - U(\lambda )=0 $. This reduction
doesn't restrict the Cartan elements. Therefore:
\begin{eqnarray}\label{eq:c3.16.1}
p_{\alpha } = q_{\alpha }, \quad \alpha \in \Delta
\end{eqnarray}
and leads to the $9 $--wave system with vanishing Hamiltonian, see Remark 3.
This system is similar to the general one without reductions but with the
fields $p_\alpha  $ replaced by $q_\alpha  $.

\end{example}

\begin{example}\label{exa:c3.1}
$C_{\ref{exa:c3.1}}=S_{e_1-e_2} $. $C_{\ref{exa:c3.1}}(U(\lambda ))
- U(\lambda )=0 $. Therefore:
\begin{eqnarray}\label{eq:c3.1.1}
&&q_{110}=q_{10},\quad q_{111}=q_{11},\quad q_{221}=-q_{21},\quad
p_{100}=q_{100},\nonumber\\
&&p_{110}=p_{10} ,\quad p_{111}=p_{11},\quad p_{221}=-p_{21},\quad a_2 =
a_1,\quad b_2 = b_1,
\end{eqnarray}
and we obtain the $10 $--wave system which is described by the Hamiltonian:
\begin{eqnarray}\label{eq:c3.1.3}
 H^{(0)} &=& 2H_0(110)+H_0(1)+2H_0(111)+2H_0(221) +H_0(121)\nonumber\\
&+& 2 \left( \left\{ \begin{array}{c} 221 \\ 111, 110  \end{array} \right\}
+ \left\{ \begin{array}{c} 121 \\ 011', 110  \end{array} \right\} +
\left\{ \begin{array}{c} 011' \\ 010', 001  \end{array} \right\} \right)
\end{eqnarray}
Note that the functions
$q_{121} $, $p_{121} $ and $ q_1 $, $p_1 $ remain unrestricted and
$q_{100}$ is redundant, see Remark \ref{rem:2}.

\end{example}

\begin{example}\label{exa:c3.2}
$C_{\ref{exa:c3.2}}=S_{2e_3} $. $C_{\ref{exa:c3.2}}(U(\lambda ))
- U(\lambda )=0 $. Then:
\begin{eqnarray}\label{eq:c3.2.1}
&&q_{111}=-q_{110},\quad q_{11}=-q_{10},\quad p_{111}=-p_{110},\quad
p_{11}=-p_{10},\\
&&p_{221}=p_{121}=p_{21}=0 ,\quad q_{221}=q_{121}=q_{21}=0,\quad
p_1=q_1, \nonumber\\
&& a_3 = 0,\quad b_3 = 0;\nonumber
\end{eqnarray}
giving the $6 $--wave (complex) system with the Hamiltonian:
\begin{eqnarray}\label{eq:c3.2.3}
H^{(0)} &=& H_0(100)+2H_0(10)+2H_0(110) + 2 \left\{ \begin{array}{c} 110
\\ 010, 100  \end{array} \right\},
\end{eqnarray}
related to ${\bf A}_2 $--subalgebra. Here $\kappa = a_1b_2 - a_2b_1 $,
$q_{100} $ and $p_{100} $ are unrestricted fields and $q_1 $ is redundant,
see Remark 2.
\end{example}

\begin{example}\label{exa:c3.3}
$C_{\ref{exa:c3.3}}=S_{e_1-e_2}S_{2e_3} $. $C_{\ref{exa:c3.3}}(U(-\lambda
)) - U(\lambda )=0 $. This gives:
\begin{eqnarray}\label{eq:c3.3.1}
\fl &&a_2=-a_1,\quad b_2=-b_1, \quad p_{100}=-q_{100},\quad
q_{110}=q_{11},\quad q_{111}=q_{10},\\
\fl &&q_{221}=-q_{21},\quad p_{1}=-q_{1},\quad p_{110}=p_{11},\quad
p_{111}=p_{10},\quad p_{221}=-p_{21}, \nonumber
\end{eqnarray}
and the next $8 $--wave system, see Remark~\ref{rem:3}:
\begin{eqnarray}\label{eq:c3.3.3}
\fl &&ia_1q_{100,t}-ib_1q_{100,x}+\kappa (p_{10}q_{11}-p_{11}q_{10})=0,
\nonumber\\
\fl &&i(a_1+a_3)q_{10,t}-i(b_1+b_3)q_{10,x}-2\kappa (q_{21}
p_{11}+q_{100}q_{11}+q_1q_{11}) = 0, \nonumber\\
\fl &&ia_3q_{1,t}-ib_3q_{1,x}-\kappa (p_{10}q_{11}-p_{11}q_{10}) = 0.
\nonumber\\
\fl &&i(a_1-a_3)q_{11,t}-i(b_1-b_3)q_{11,x}-2\kappa
(q_{21}p_{10}-q_{100}q_{10}+q_1q_{10}) = 0, \nonumber\\
\fl &&ia_1q_{21,t}-ib_1q_{21,x}-2\kappa q_{10}q_{11} = 0, \\
\fl &&i(a_1+a_3)p_{10,t}-i(b_1+b_3)p_{10,x}+2\kappa (q_{1}p_{11}
+q_{100}p_{11}-p_{21}q_{11}) = 0, \nonumber\\
\fl &&i(a_1-a_3)p_{11,t}-i(b_1-b_3)p_{11,x}+2\kappa (q_{1}p_{10}
-q_{100}p_{10}-p_{21}q_{10}) = 0, \nonumber\\
\fl &&ia_1p_{21,t}-ib_1p_{21,x}+2\kappa p_{10}p_{11}=0, \nonumber
\end{eqnarray}
where $\kappa =a_1b_3-a_3b_1 $ and $q_{121} $ and $p_{121} $ are
redundant fields, see Remark \ref{rem:2}.
\end{example}

\begin{example}\label{exa:c3.4}
$C_{\ref{exa:c3.4}}=S_{2e_1}S_{2e_3} $.
$C_{\ref{exa:c3.4}}(U(-\lambda )) - U(\lambda )=0 $. Then:
\begin{eqnarray}\label{eq:c3.4.1}
\fl &&a_2=0,\quad b_2=0, \quad p_{121}=-q_{100},\quad
p_{100}=-q_{121},\quad p_{111}=iq_{111},\\
\fl &&p_{110}=iq_{110},\quad p_{221}=-q_{221},\quad p_{1}=q_{1},\quad
q_{11}=iq_{10},\quad p_{11}=-ip_{10}. \nonumber
\end{eqnarray}
so we get the next $8 $--wave system, see Remark~\ref{rem:3}:
\begin{eqnarray}\label{eq:c3.4.3}
ia_1q_{100,t}-ib_1q_{100,x}-\kappa (q_{111}p_{10}+ip_{10}q_{110})=0,
\nonumber\\
ia_3q_{10,t}-ib_3q_{10,x}+\kappa (q_{121}q_{111}+iq_{121}q_{110}) = 0,
\nonumber\\
ia_3q_{1,t}-ib_3q_{1,x}+2i\kappa q_{110}q_{111} = 0, \nonumber\\
\fl i(a_1-a_3)q_{110,t}-i(b_1-b_3)q_{110,x} + \kappa (2iq_{221}
q_{111}+iq_{121}p_{10}+2q_{1}q_{111}+q_{100}q_{10}) = 0, \nonumber\\
\fl i(a_1+a_3)q_{111,t}-i(b_1+b_3)q_{111,x} +\kappa (2iq_{221}
q_{110}-2q_{1}q_{110}-iq_{100}q_{10}-q_{121}p_{10}) = 0, \nonumber\\
ia_1q_{121,t}-ib_1q_{121,x}+\kappa q_{111}q_{10} = 0, \\
ia_1q_{221,t}-ib_1q_{221,x}-2\kappa (q_{111} q_{110}+iq_{110}q_{10})= 0,
\nonumber\\
ia_3p_{10,t}-ib_3p_{10,x}+\kappa (iq_{100}q_{110}-q_{100}q_{111})=0,
\nonumber
\end{eqnarray}
where $\kappa =a_1b_3-a_3b_1 $ and $q_{21} $ and $p_{21} $ are redundant
fiels, see Remark~\ref{rem:2}.
\end{example}

\begin{example}\label{exa:c3.00}
$C_{\ref{exa:c3.00}}=\openone $. $U^{\dagger}(\eta \lambda ^*)-U(\lambda
)=0 $, $\eta =\pm 1 $. This reduction means that all Cartan elements are
real for $\eta =1 $ and purely imaginary for $\eta =-1 $ and
\begin{eqnarray}\label{eq:c3.00-1}
p_{\alpha }=-\eta q_{\alpha }^*.
\end{eqnarray}
Thus we get the  $9 $-wave system with Hamiltonian (\ref{eq:H-om}) with the
restrictions given above.
\end{example}

\begin{example}\label{exa:c3.hh}
$\Sigma =\mathrm{diag}\,(s_1,s_2,s_3,{1 /s_3},{1 /s_2},{1 /s_1}) $,
$\Sigma^{-1} U^{\dagger}(\eta \lambda ^*)\Sigma -U(\lambda )=0 $.
After this reduction all Cartan elements are real for $\eta =1 $ and
purely imaginary for $\eta =-1 $ and
\begin{eqnarray}\label{eq:c3.hh-1}
p_{100}=-\eta {s_1\over s_2}q_{100}^*, \quad p_{10}=-\eta {s_2\over
s_3}q_{10}^* ,\quad p_{1}=-\eta s_3^2q_{1}^* , \nonumber\\
p_{110}=-\eta {s_1\over s_3}q_{110}^* ,\quad p_{11}=-\eta s_2s_3q_{11}^*,
\quad p_{111}=-\eta s_1s_3q_{111}^* ,\\
p_{21}=-\eta s_2^2q_{12}^*, \quad p_{121}=-\eta s_1s_2q_{121}^* ,\quad
p_{221}=-\eta s_1^2q_{221}^* ,\nonumber
\end{eqnarray}
which leads to a $9 $-wave system with the Hamiltonian:
\begin{eqnarray}\label{eq:c3.hh.3}
\fl &&H^{(0)}={s_1 \over s_2}H_{0*}(100) + {s_2 \over s_3}H_{0*}(10) +
s_3^2H_{0*}(1) + {s_1 \over s_3}H_{0*}(110) + s_2s_3H_{0*}(11) \nonumber\\
\fl &&+ s_1s_3H_{0*}(111)+s_2^2H_{0*}(21)+s_1s_2H_{0*}(121)
+s_1^2H_{0*}(221) \\
\fl &&+ (\kappa _1+\kappa _2+\kappa _3){s_1\over s_3}H_*(110,10,100)
+(-\kappa _1-\kappa _2+\kappa _3)s_1s_3H_*(111,11,100) \nonumber\\
\fl &&+2\kappa _3 s_1s_2H_*(121,21,100) +(\kappa _1-\kappa _2+\kappa
_3)s_1s_2H_*(121,11,110) \nonumber\\
\fl &&+2\kappa _1 s_2^2H_*(21,11,10) +2\kappa _2s_1s_3H_*(111,110,1)
+2\kappa _1s_2s_3H_*(11,1,10)\nonumber\\
\fl &&+(\kappa _1-\kappa _2-\kappa _3) s_1s_2H_*(121,111,10)+2\kappa
_3s_1^2H_*(221,121,100)\nonumber\\
\fl &&-2\kappa _2s_1^2H_*(221,111,110) , \nonumber
\end{eqnarray}
where the terms $H_*(\alpha ,\beta ,\gamma ) $ are given in (\ref{eq:H*}).
In the particular case $s_1=s_2=s_3=1 $ we obtain the result of example
\ref{exa:c3.00}.
\end{example}

\begin{example}\label{exa:c3.17}
$C_{\ref{exa:c3.17}}=w_0 $.
$C_{\ref{exa:c3.17}}(U^{\dagger}(\eta \lambda^* )) - U(\lambda )=0 $
, $\eta =\pm 1 $. This reduction means that all fields $q_{\alpha },
p_{\alpha } $ are real and the Cartan elements are purely imaginary for
$\eta =1 $ and vice versa for $\eta =-1 $. Thus we get the $18 $--wave
system with the Hamiltonian (\ref{eq:H-om}). The case $\eta =1 $ leads to
the non-compact real form $sp(6, \bbbr) $ for the ${\bf C}_3 $-algebra.

\end{example}

\begin{example}\label{exa:c3.5}
$C_{\ref{exa:c3.5}}=S_{e_1-e_2}$.
$C_{\ref{exa:c3.5}}(U^{\dagger}(\eta \lambda^* )) - U(\lambda )=0 $, $\eta
=\pm 1 $. Therefore:
\begin{eqnarray}\label{eq:c3.5.1}
\fl &&q_{100}^*=-\eta q_{100}, \quad p_{100}^*=-\eta p_{100},\quad
p_{1}=-\eta q_{1}^*,\quad p_{121}=-\eta q_{121}^*,\nonumber\\
\fl &&p_{10}=-\eta q_{110}^*,\quad p_{110}=-\eta q_{10}^*,\quad
p_{11}=-\eta q_{111}^*,\quad p_{111}=-\eta q_{11}^*,\\
\fl &&p_{21}=\eta q_{221}^*,\quad p_{221}=\eta q_{21}^*,
\quad a_2=\eta a_1^*,\quad b_2=\eta b_1^*, \quad a_3^*=\eta a_3,
\quad b_3^*=\eta b_3. \nonumber
\end{eqnarray}
so this leads to the $10 $-wave (2 real and 8 complex) system with the
Hamiltonian:
\begin{eqnarray}\label{eq:c3.5.5}
\fl &&H^{(0)} =H_{0}(100) +H_{0*}(121)+H_{0*}(1) +2\re H_{0*}(221)
+2\re H_{08}(111) +2\re H_{0*}(110) \nonumber\\
\fl  && + \left\{ \begin{array}{c} 111 \\ 011, 100  \end{array} \right\}
+\left\{ \begin{array}{c} 110 \\ 010, 100  \end{array} \right\}+ 2\re
\left(\left\{ \begin{array}{c} 221 \\ 121, 100  \end{array} \right\}
+\left\{ \begin{array}{c} 221 \\ 111, 110  \end{array} \right\} \right.
\nonumber\\
\fl && +\left. \left\{ \begin{array}{c} 121 \\ 011, 110  \end{array}
\right\} +\left\{ \begin{array}{c} 011 \\ 010, 001  \end{array} \right\}
\right)
\end{eqnarray}
\end{example}

\begin{example}\label{exa:c3.6}
$C_{\ref{exa:c3.6}}=S_{2e_3}$. $C_{\ref{exa:c3.6}}(U^{\dagger}( \eta
\lambda^*)) - U(\lambda )=0 $, $\eta =\pm 1 $. Then:
\begin{eqnarray}\label{eq:c3.6.1}
&&q_{1}^*=-\eta q_1,\quad p_{1}^*=-\eta p_1,\quad p_{100}=- \eta
q_{100}^*,\quad p_{111}=\eta q_{110}^*,\nonumber\\
&&p_{110}=\eta q_{111}^*,\quad p_{11}=\eta q_{10}^*,
\quad p_{10}=\eta q_{11}^*,\quad p_{121}=\eta q_{121}^*, \\
&&p_{221}=\eta q_{221}^*,\quad p_{21}=\eta q_{21}^*,
\quad a_{1,2}^*=\eta a_{1,2},\quad b_{1,2}^*=\eta b_{1,2}, \nonumber\\
&&a_3^*=-\eta a_3, \quad b_3^*=-\eta b_3.\nonumber
\end{eqnarray}
Thus we obtain the $10 $-wave (2 real and 8 complex) system with the
Hamiltonian:
\begin{eqnarray}\label{eq:c3.6.h}
\fl &&H^{(0)} = H_{0*}(100)+H_0(1) +H_{0*}(21)+
H_{0*}(121) +H_{0*}(221)+2\re H_{0*}(111) \nonumber\\
\fl && +2\re H_{0*}(11)
+ \left\{ \begin{array}{c} 221 \\ 121, 100  \end{array} \right\}
+\left\{ \begin{array}{c} 221 \\ 111, 110  \end{array} \right\}
+\left\{ \begin{array}{c} 121 \\ 021, 100  \end{array} \right\}
+\left\{ \begin{array}{c} 111 \\ 110, 001  \end{array} \right\} \nonumber\\
\fl && +\left\{ \begin{array}{c} 021 \\ 011, 010  \end{array} \right\}
+\left\{ \begin{array}{c} 011 \\ 010, 001  \end{array} \right\}
+ 2\re \left(\left\{ \begin{array}{c} 121 \\ 011, 110  \end{array} \right\}
+\left\{ \begin{array}{c} 111 \\ 011, 100  \end{array} \right\} \right).
\end{eqnarray}
\end{example}

\begin{example}\label{exa:c3.14}
$C_{\ref{exa:c3.14}}=S_{e_1-e_2}S_{2e_3} $.
$C_{\ref{exa:c3.14}}(U^{\dagger}(\eta \lambda ^*)) - U(\lambda )=0 $,
$\eta = \pm 1 $. Then:
\begin{eqnarray}\label{eq:c3.14.1}
\fl &&p_{100}^*=-\eta p_{100},\quad q_{100}^*=-\eta q_{100},\quad
p_{11}=\eta q_{110}^*, \quad p_{111}=\eta q_{10}^*, \quad p_{10}=\eta
q_{111}^* \nonumber\\
\fl &&p_{110}=\eta q_{11}^* ,\quad p_{121}=\eta q_{121}^*,\quad
p_{21}=-\eta q_{221}^*,\quad p_{221}=-\eta q_{21}^* , \\
\fl && q_1^*=-\eta q_1,\quad p_1^*=-\eta p_1, \quad a_2=\eta a_1^*,\quad
b_2=\eta b_1^*, \quad a_3^*=-\eta a_3, \quad b_3^*=-\eta b_3 ;\nonumber
\end{eqnarray}
giving the $11 $--wave (4 real and 7 complex) system with Hamiltonian:
\begin{eqnarray}\label{eq:c3.14.h}
\fl &&H^{(0)} =H_{0}(100) +H_{0}(1) +H_{0*}(121)+2\re H_{0*}(11)
+2\re H_{0*}(111)\nonumber\\
\fl && +2\re H_{0*}(221)
+ \left\{ \begin{array}{c} 121 \\ 011, 110  \end{array} \right\}
+\left\{ \begin{array}{c} 121 \\ 111, 010  \end{array} \right\}
+ 2\re \left(\left\{ \begin{array}{c} 221 \\ 121, 100  \end{array} \right\}
+\left\{ \begin{array}{c} 221 \\ 111, 110  \end{array} \right\} \right.
\nonumber\\
\fl && +\left. \left\{ \begin{array}{c} 111 \\ 110, 001  \end{array}
\right\} +\left\{ \begin{array}{c} 111 \\ 011, 100  \end{array} \right\}
\right) .
\end{eqnarray}
\end{example}

\begin{example}\label{exa:c3.15}
$C_{\ref{exa:c3.15}}=S_{e_1-e_2} S_{e_1+e_2} $.
$C_{\ref{exa:c3.15}}(U^{\dagger}(\eta \lambda ^*)) -U(\lambda )=0 $,
$\eta =\pm 1 $. Therefore:
\begin{eqnarray}\label{eq:c3.15.1}
\fl &&q_{100}^*=-\eta q_{100},\quad q_{111}=\eta iq_{110}^*,\quad
q_{11}=-\eta iq_{10}^*,\quad q_{21}^*=\eta q_{21},\nonumber\\
\fl &&q_{121}^*=-\eta q_{121} ,\quad q_{221}^*=\eta q_{221},\quad
p_{100}^*=-\eta p_{100},\quad p_{111}=-\eta ip_{110}^* \\
\fl &&p_{11}=\eta ip_{10}^*, \quad p_{21}^*=\eta p_{21}, \quad
p_{121}^*=-\eta p_{121}, \quad p_{221}^*=\eta p_{221} ,\quad p_1=\eta
q_1^* \nonumber\\
\fl &&a_{1,2}^*=-\eta a_{1,2}, \quad b_{1,2}^*=-\eta b_{1,2},
\quad a_3^*=\eta a_3, \quad b_3^*=\eta b_3. \nonumber
\end{eqnarray}
Thus we get the $13 $-wave (8 real and 5 complex) system with Hamiltonian:
\begin{eqnarray}\label{eq:c3.15.h}
\fl &&H^{(0)} = H_{0}(100) +H_{0}(121) +H_0(21)+ H_{0*}(1)
+2\re H_{0*}(11) +2\re H_{0*}(111)     \nonumber\\
\fl &&  +H_0(221)+ \left\{
\begin{array}{c} 221 \\ 121, 100  \end{array} \right\} +\left\{
\begin{array}{c} 121 \\ 021, 100  \end{array} \right\} +\left\{
\begin{array}{c} 021 \\ 011, 010'  \end{array} \right\} +\left\{
\begin{array}{c} 011 \\ 010', 001  \end{array} \right\} \nonumber\\
\fl && +\left\{ \begin{array}{c} 221 \\ 111, 110'  \end{array} \right\}
+\left\{ \begin{array}{c} 111 \\ 110', 001  \end{array} \right\}
+ 2\re \left(\left\{ \begin{array}{c} 121 \\ 111, 010'  \end{array}
\right\} +\left\{ \begin{array}{c} 111 \\ 011, 100  \end{array} \right\}
\right).
\end{eqnarray}
\end{example}

\section{Real forms of ${\frak g} $ as $\bbbz_2 -$ reductions }
\label{rf}

As we already mentioned, in several of the examples above the
$\bbbz_2 $-reductions act as Cartan involutions, i.e. $iU(x,\lambda ) $
belongs to a real form $\fr{g}^{\bbbr} $ of the corresponding complex
simple Lie algebra $\fr{g} $. As a result the scattering matrix $T(\lambda
)$ (see eq. (\ref{eq:5.2}) below) belongs to the corresponding compact or
non-compact Lie group.

It is well known that $X\in {\frak g}^{\bbbr} $ if $X\in \fr{g} $ and
(see e.g. \cite{Helg}):
\[
\sigma (\theta(X)) \equiv \theta (\sigma(X)) =X ,\qquad \theta
(X)=-X^{\dagger} , \qquad X\in {\frak g} ,  \]
where $\sigma $ is an involutive Cartan automorphism: $\sigma
^2=\openone$. The related $\bbbz_2 $-reduction acts in addition on
the complex spectral parameter $\lambda  $ via complex
conjugation:  $\kappa (\lambda )=  \lambda ^* $.  The compact real form
$\tilde{{\frak g}}^{\bbbr} $ of ${\frak g} $ is obtained with $\sigma =
\openone $.  For the non-compact cases the Cartan involution splits the
roots of $\fr{g} $ into compact and non-compact ones as follows:

1) If $\sigma (E_{\alpha })=E_{\alpha } $, where $E_{\alpha } $ is the
Weyl generator for the root $\alpha $, we say that $\alpha $ is a
compact root.

The non-compact roots are of two types depending on the orbit-size of
$\sigma  $:

2) If $\sigma (E_{\alpha })=\varepsilon E_{\alpha } $, $\varepsilon =\pm 1
$ the orbit of $\sigma  $ consist of only one element;

3) If $\sigma (E_{\alpha })= \varepsilon E_{-\beta } $, $\alpha \neq \beta
>0$ and $\varepsilon =\pm 1$ then $\{\alpha ,\beta \} $ is a two-element
orbit of $\sigma  $.

Let $\pi$ be the system of simple roots of the algebra and $\pi _0 $ be
the set of the compact simple roots. The (inner) Cartan involution which
extracts the non-compact real form ${\frak g}^{\bbbr} $ from the compact
one is given by:
\begin{eqnarray}\label{eq:rf-1}
\sigma = \exp \left(\sum_{\alpha_k \in \pi \backslash \pi _0} {2\pi i
\over (\alpha _k,\alpha _k) } H_{\omega _k}\right)
\end{eqnarray}
Here $H_{\omega_k}=\sum_{i=1}^{r}(\omega _k,e_i)h_i $, where $\{h_i\} $
is the basis in the Cartan subalgebra ${\frak h} \subset {\frak g} $ dual
to the orthogonal basis $ \{e_i\} $ in the root space and $\{\omega _k\} $
are the fundamental weights of the algebra.

Note that in the examples above for the ${\bf A}_2 $ and ${\bf A}_3 $
algebras the corresponding normal real forms $sl(3, \bbbr) $, $sl(4,
\bbbr) $ and $su^*(4) $ are extracted with external automorphisms. We
leave more details about non-compact real forms generated by external
involutive automorphisms to the second paper of this sequence.

The list of the Cartan involutions for the considered real forms of the
simple Lie algebras and the relevant examples is given in
table~\ref{tab:1}.

\begin{table}[t,h]
\begin{tabular}{|c|c|c|l|}
\hline \hline
$\fr{g} $ & Compact real & non-compact real & Cartan involution \\
\qquad & form & form &\qquad \\ \hline \hline
${\bf A}_2 $ & $su(3)$,  Ex. \ref{exa:a2.00} and  & $sl(3,\bbbr)$, Ex. 2
& \mbox{external}   \\ \cline{3-4}
\qquad & Ex. \ref{exa:a2.hh}, $s_1=s_2=s_3 $.
& $su(2,1)$, Ex. \ref{exa:a2.hh}  & $s_1=-s_2=-s_3$ \\
\hline
${\bf C}_2$ & $sp(4)$, Ex. 8 and & $sp(4,\bbbr)$, Ex. 7  & \qquad
 \\ \cline{3-4} \qquad & Ex. \ref{exa:c2.hh}, $s_1=s_2=1 $. &
$sp(2,2)$, Ex. \ref{exa:c2.hh}  &  $s_1=-1, s_2=1 $ \\
\hline
${\bf G}_2 $ & ${\frak g}_2$, Ex. \ref{exa:g2.3} and  &
${\frak g}_2^{\prime} $, Ex. \ref{exa:g2.hh} & $s_1=s_2=-1 $ \\
\qquad
& Ex. 15, $s_1=s_2 =1 $.  &\qquad   &\qquad   \\
\hline
${\bf A}_3 $ & $su(4)$, Ex. \ref{exa:a3.0} and  & $sl(4,\bbbr)$, Ex. 19  &
$\mbox{external} $  \\ \cline{3-4}
& Ex. 26, $s_1=s_2=s_3=s_4$ & $su(1,3) $, Ex. \ref{exa:a3.hh} &
$s_1=-s_2=-s_3=-s_4$ \\ \cline{3-4} \qquad &
& $su(2,2) $, Ex. \ref{exa:a3.hh} & $s_1=s_2=-s_3=-s_4$ \\
\cline{3-4} \qquad & \qquad & $su^*(4) $, Ex. 28 &
$\sigma =S_{e_2-e_3}\circ S_{e_1-e_4} $ \\
\hline
${\bf B}_3 $ & $so(7)\simeq so(7,\bbbr) $
& $so(2,5) $, Ex. \ref{exa:b3.hh} & $s_1=-1, s_2= s_3=1$ \\
\cline{3-4}
\qquad & Ex. \ref{exa:b3.00},  Ex. 36 and & $so(3,4)$, Ex. \ref{exa:b3.hh}
& $s_1=s_2=-1, s_3=1$ \\ \cline{3-4} \qquad & Ex. 35, $s_1=s_2=s_3=1 $ &
$so(1,6)$, Ex. \ref{exa:b3.hh}  & $s_1=s_2=s_3=-1$ \\ \hline ${\bf C}_3 $
& $sp(6) $, Ex.  \ref{exa:c3.00} and & $sp(6,\bbbr)$, Ex. 48  & \qquad  \\
\cline{3-4}
\qquad & Ex. \ref{exa:c3.hh}, $s_1=s_2=s_3=1 $.  & $sp(2,4) $, Ex.
\ref{exa:c3.hh} & $s_1=-s_2=-s_3=1$ \\
\hline
\end{tabular}
\caption{List of reductions related to real forms of
$\fr{g}$. In all examples we assume $\eta=1 $.}\label{tab:1} \end{table}

\section{Scattering data and the $\bbbz_2 $-reductions.}
\label{5}

In order to determine the scattering data of the Lax operator
(\ref{eq:1.1}) we start from the Jost solutions
\begin{equation}\label{eq:5.1}
\lim_{x\to\infty } \psi (x,\lambda )e^{i\lambda Jx} = \openone ,\qquad
\lim_{x\to-\infty } \phi (x,\lambda )e^{i\lambda Jx} = \openone ,
\end{equation}
and the scattering matrix
\begin{equation}\label{eq:5.2}
T(\lambda )=(\psi (x,\lambda ))^{-1} \phi (x,\lambda ).
\end{equation}

Let us start with the simplest case when $J $ has purely real and
pair-wise distinct eigenvalues. Since the classical papers of Zakharov
and Shabat \cite{Sh,Za*Sh} the most efficient way to construct the minimal
set of scattering data of (\ref{eq:1.1}) and to study its properties is to
make use of the equivalent Riemann-Hilbert problem (RHP).

We treat below only the simplest non-trivial case when $J $ has real
pair-wise distinct eigenvalues, i.e. when $(a,\alpha _j)>0 $ for
$j=1,\dots,r $, see \cite{G*86}. Then one is able to construct the
fundamental analytic solutions (FAS) of (\ref{eq:1.1}) $\chi ^\pm
(x,\lambda ) $ by using the Gauss decomposition of $T(\lambda ) $:
\begin{equation}\label{eq:5.3}
T(\lambda )= T^-(\lambda ) D^+(\lambda ) \hat{S}^+(\lambda )
= T^+(\lambda ) D^-(\lambda ) \hat{S}^-(\lambda ),
\end{equation}
where by "hat" above we denote the inverse matrix $\hat{S}\equiv S^{-1} $
and
\begin{eqnarray}\label{eq:5.4}
\fl && S^\pm (\lambda ) = \exp \left( \sum_{\alpha \in\Delta _+}^{}
s_{\pm}^{\pm\alpha }(\lambda ) E_{\pm\alpha }\right) , \quad
T^\pm (\lambda ) = \exp \left( \sum_{\alpha \in\Delta _+}^{}
t_{\pm}^{\pm\alpha }(\lambda ) E_{\pm\alpha }\right) , \\
\label{eq:5.5}
\fl && D^+ (\lambda ) = I\exp \left( \sum_{j=1}^{r}
{2d_j^{+}(\lambda ) \over (\alpha _j,\alpha _j)} H_{j}\right) ,
\quad D^- (\lambda ) = I\exp \left( \sum_{j=1}^{r}
{2d_j^{-}(\lambda ) \over (\alpha _j,\alpha _j)}H_{j}^-\right) , \\
\fl && H_j \equiv H_{\alpha _j}, \qquad H_j^- = w_0(H_j).\nonumber
\end{eqnarray}

Here $I $ is an element from the universal center of ${\frak G} $ and the
superscript $+ $ (or $- $) in $D^\pm (\lambda ) $ shows that
$D_j^+(\lambda) $ (or $D_j^-(\lambda) $) are analytic functions of $\lambda
$ for $\im \lambda >0 $ (or $\im \lambda <0 $ respectively). Then we can
prove that \cite{G*86}
\begin{equation}\label{eq:5.6}
\chi ^\pm(x,\lambda ) =\phi (x,\lambda ) S^\pm (\lambda ) =
\psi (x,\lambda ) T^\mp (\lambda ) D^\pm(\lambda )
\end{equation}
are FAS of (\ref{eq:1.1}) for $\im
\lambda \gtrless 0 $. On the real axis $\chi ^+(x,\lambda ) $ and
$\chi ^-(x,\lambda ) $ are linearly related by
\begin{equation}\label{eq:5.7}
\chi ^+(x,\lambda )= \chi ^-(x,\lambda ) G_0(\lambda ), \qquad
G_0(\lambda ) =S^+(\lambda ) \hat{S}^-(\lambda ),
\end{equation}
and the sewing function $G_0(\lambda ) $ may be considered as a minimal
set of scattering data provided the Lax operator (\ref{eq:1.1}) has no
discrete eigenvalues. The presence of discrete eigenvalues $\lambda _k^\pm
$ means that some of the functions
\[
D_j^\pm (\lambda ) = \langle \omega _j^\pm |D^\pm (\lambda )|\omega^\pm
_j\rangle = \exp \left( d_j^\pm (\lambda )\right),
\]
will have zeroes and poles at $\lambda _k^\pm $, for more details see
\cite{G*86,G*87}. Equation (\ref{eq:5.7}) can be easily rewritten in the
form:  \begin{equation}\label{eq:5.8} \xi^+(x,\lambda )= \xi^-(x,\lambda )
G(x,\lambda ), \qquad G(x,\lambda ) = e^{-i\lambda Jx} G_0(\lambda )
e^{i\lambda Jx}.  \end{equation} Then (\ref{eq:5.8}) together with
\begin{equation}\label{eq:5.9}
\lim_{\lambda \to\infty } \xi^\pm(x,\lambda ) = \openone
\end{equation}
can be considered as a RHP with canonical normalization condition.

The solution $\xi^+(x,\lambda )$, $ \xi^-(x,\lambda )$ to (\ref{eq:5.8}),
(\ref{eq:5.9}) is called regular if $\xi^+(x,\lambda )$ and
$\xi^-(x,\lambda )$ are nondegenerate and non-singular functions of
$\lambda $ for all $\im \lambda >0 $ and $\im \lambda <0 $ respectively.
To the class of regular solutions of RHP there correspond Lax operators
(\ref{eq:1.1}) without discrete eigenvalues. The presence of discrete
eigenvalues $\lambda _k^\pm $ leads to singular solutions of the RHP;
their explicit construction can be done by the Zakharov-Shabat dressing
method \cite{Za*Mi,Za*Sh}.

If the potential $q(x,t) $ of the Lax operator (\ref{eq:1.1}) satisfies
the $N $-wave equation (\ref{eq:1.4}) then $S^\pm(t,\lambda ) $ and
$T^\pm(t,\lambda ) $ satisfy the linear evolution equations
\begin{equation}\label{eq:5.10}
i {dS^\pm \over dt } - \lambda [I,S^\pm(t,\lambda ) ] =0, \qquad
i {dT^\pm \over dt } - \lambda [I,T^\pm(t,\lambda ) ] =0,
\end{equation}
while the functions $D^\pm (\lambda ) $ are time-independent. In other
words $D_j^\pm(\lambda ) $ can be considered as the generating functions
of the integrals of motion of (\ref{eq:1.4}).

If we now impose a reduction on $L $ it will reflect also on the
scattering data. It is not difficult to check that if $L $ satisfies
(\ref{eq:2.1}) then the scattering matrix will satisfy
\begin{equation}\label{eq:5.11}
C_k\left( T(\Gamma _k(\lambda )\right) = T(\lambda ), \qquad \lambda \in
\bbbr.
\end{equation}
Note that strictly speaking (\ref{eq:5.11}) is valid only for real values
of $\lambda $ (more generally, for $\lambda $ on the continuous spectrum
of $L $). If we choose reductions with automorphisms of the form
(\ref{eq:C-1}), (\ref{eq:A-1}) and (\ref{eq:C_2}) for the FAS and for the
Gauss factors $S^\pm(\lambda ) $, $T^\pm(\lambda ) $ and $D^\pm(\lambda )
$ we will get:
\begin{eqnarray}\label{eq:ra}
\fl & S^+(\lambda ) = A_1 \left(\hat{S}^-(\lambda ^*)\right)^\dag
A_1^{-1}, \qquad &T^+(\lambda ) = A_1 \left(\hat{T}^-(\lambda
^*)\right)^\dag A_1^{-1}, \nonumber\\
\fl & D^+(\lambda ) =  \left(\hat{D}^-(\lambda^*)\right)^* , \qquad
&F(\lambda ) = \left(F(\lambda ^*)\right)^*,\qquad \eta=1,\\
\label{eq:rb}
\fl & S^+(\lambda ) = A_2 \left(\hat{S}^-(-\lambda)\right)^T A_2^{-1},
\qquad &T^+(\lambda ) = A_2 \left(\hat{T}^-(-\lambda)\right)^T A_2^{-1},
\nonumber\\
\fl & D^+(\lambda ) =  \hat{D}^-(-\lambda),\qquad \eta=-1,\\
\label{eq:rc}
\fl & S^+(\lambda ) = A_3 \left(S^-(-\lambda ^*)\right)^*A_3^{-1},
\qquad &T^+(\lambda ) = A_3 \left(T^-(-\lambda^*)\right)^* A_3^{-1},
\nonumber\\
\fl & D^+(\lambda ) =  \left(D^-(-\lambda ^*)\right)^*,
\qquad &F(\lambda ) =  \left(F(-\lambda ^*)\right)^* , \qquad \eta=-1,
\end{eqnarray}
where we also used the fact that $A_i $ belong to the Cartan subgroup of
${\frak  g} $. Next we  make use of the integral representations for
$d_j^\pm(\lambda )$ allowing one to reconstruct them as analytic functions
in their regions of analyticity $\bbbc_\pm $. In the case of absence of
discrete eigenvalues we have \cite{G*86,VYa}:
\begin{equation}\label{eq:5.12}
{\cal D}_j(\lambda ) = {i \over 2\pi }\int_{-\infty }^{\infty } {d\mu
\over \mu -\lambda } \ln \langle \omega _j | \hat{T}^+(\mu ) T^-(\mu ) |
\omega _j\rangle ,
\end{equation}
where $\omega _j $ and $|\omega _j\rangle $ are the $j $-th fundamental
weight of $\fr{g} $ and the highest weight vector in the corresponding
fundamental representation $R(\omega _j) $ of $\fr{g} $. The function
${\cal D}_j(\lambda ) $ is a piece-wise analytic function of $\lambda $
equal to:
\begin{equation}\label{eq:5.13}
{\cal D}_j(\lambda ) = \left\{ \begin{array}{ll}
d_j^+(\lambda ), & \mbox{for} \quad \lambda \in \bbbc_+ \\
(d_j^+(\lambda )-d_{j'}^-(\lambda ))/2, & \mbox{for} \quad \lambda \in
\bbbr, \\
-d_{j'}^-(\lambda ), & \mbox{for} \quad \lambda \in \bbbc_- ,
\end{array} \right.
\end{equation}
where $d_j^\pm (\lambda ) $ were introduced in (\ref{eq:5.5}) and the
index $j' $ is related to $j $ by $w_0(\alpha _j)=-\alpha _{j'} $.
Here $w_0 $ is the Weyl reflection that maps the highest weight in
$R(\omega _j) $ into the lowest weight of $R(\omega _j) $, see \cite{LA}.

The functions ${\cal D}_j(\lambda ) $ can be viewed also as generating
functions of the integrals of motion. Indeed, if we expand
\begin{equation}\label{eq:5.14}
{\cal D}_j(\lambda ) = \sum_{k=1}^{\infty } {\cal D}_{j,k} \lambda^{-k},
\end{equation}
and take into account that $D^\pm(\lambda ) $ are time independent we find
that $d{\cal D}_{j,k}/dt =0 $ for all $k =1,\dots, \infty $ and
$j=1,\dots r $. Moreover it can be checked that ${\cal D}_{j,k} $
expressed as functional of $q(x,t) $ has a kernel that is local in $q $,
i.e. depends only on $q $ and its derivatives with respect to $x $.

{}From (\ref{eq:5.12}) and (\ref{eq:ra})-(\ref{eq:rc}) we easily obtain the
effect of the reductions on the set of integrals of motion; namely, for
the reduction (\ref{eq:ra}):
\begin{equation}\label{eq:5.15a}
{\cal D}_j(\lambda )=- {\cal D}_j^*( \lambda ^*), \qquad
\mbox{i.e.}, \qquad {\cal D}_{j,k} = (-1)^{k+1} {\cal D}_{j,k}^*,
\end{equation}
with $\eta =1 $; for (\ref{eq:rb}) we have
\begin{equation}\label{eq:5.15b}
{\cal D}_j(\lambda )=-{\cal D}_j(-\lambda ), \qquad
\mbox{i.e.}, \qquad {\cal D}_{j,k} = (-1)^{k+1} {\cal D}_{j,k},
\end{equation}
and for (\ref{eq:rc})
\begin{equation}\label{eq:5.15c}
{\cal D}_j(\lambda )={\cal D}_j^*(-\lambda ^*), \qquad
\mbox{i.e.}, \qquad {\cal D}_{j,k} = (-1) ^{k} {\cal D}_{j,k}^*.
\end{equation}

In particular from (\ref{eq:5.15b}) it follows that al integrals of motion
with even $k $ become degenerate, i.e. ${\cal D}_{j,2k}=0 $. The
reductions (\ref{eq:5.15a}) and (\ref{eq:5.15c}) mean that "half" of the
integrals ${\cal D}_{j,2k} $ become real and the other "half" ${\cal
D}_{j,2k} $ - purely imaginary.

We finish this section with a few comments on the simplest local integrals
of motion. To this end we write down the first two types of integrals of
motion ${\cal D}_{j,1} $ and ${\cal D}_{j,2} $ as functionals of the
potential $Q $ of (\ref{eq:1.1}). Skipping the details (see \cite{G*86})
we get:
\begin{eqnarray}\label{eq:Dj1}
&& {\cal D}_{j,1} = -{ i \over 4 } \int_{-\infty }^{\infty } dx\, \langle
[J, Q], [H_j^{\vee} ,Q] \rangle ,
\end{eqnarray}
and
\begin{eqnarray}\label{eq:Dj2}
\fl && {\cal D}_{j,2} = -{ 1\over 2 } \int_{-\infty }^{\infty } dx\,
\langle Q, [H_j^{\vee} ,Q_x] \rangle -{ i \over 3 } \int_{-\infty
}^{\infty } dx\, \langle [J, Q], [Q, [H_j^{\vee} ,Q]] \rangle ,
\end{eqnarray}
where $H_j^{\vee} =2H_{\omega _j}/(\alpha _j,\alpha _j) $.

The fact that ${\cal D}_{j,1} $ are integrals of motion for $j=1,\dots, r
$, can be considered as natural analog of the Manley--Rowe relations
\cite{ZM,K}. In the case when the reduction is of the type (\ref{eq:C-1}),
i.e. $p_\alpha =s_\alpha q_\alpha ^* $ then (\ref{eq:Dj1}) is equivalent
to
\begin{eqnarray}\label{eq:M-R}
\sum_{\alpha >0} {2 (\vec{a},\alpha )(\omega _j,\alpha ) \over
(\alpha ,\alpha )}\int_{-\infty }^{\infty }dx\,
s_{\alpha } |q_{\alpha }(x)|^2 = \mbox{const} ,
\end{eqnarray}
and can be interpreted as relations between the densities $|q_\alpha |^2 $
of the `particles' of type $\alpha $. For the other types of reductions
such interpretation is not so obvious.

The integrals of motion ${\cal D}_{j,2} $ are directly related to the
Hamiltonian of the $N $-wave equations (\ref{eq:1.4}), namely:
\begin{eqnarray}\label{eq:H-D}
H_{N\rm -wave} = -\sum_{j=1}^{r} {2(\alpha _j, \vec{b})\over (\alpha _j,
\alpha _j)} {\cal D}_{j,2} ={1 \over 2i} \left\langle \left\langle
\dot{{\cal D}}(\lambda ), {}F(\lambda ) \right\rangle \right\rangle _{0},
\end{eqnarray}
where $\dot{{\cal D}}(\lambda )=d{\cal D}/d\lambda $ and
$F(\lambda )=\lambda I $ is the dispersion law of the $N $-wave equation
(\ref{eq:1.4}). In (\ref{eq:H-D}) we used just one of the hierarchy of
scalar products in the Kac-Moody algebra $\widehat{\fr{g}} \equiv \fr{g}
\otimes \bbbc[\lambda ,\lambda ^{-1}] $:
\begin{equation}\label{eq:ScPr}
\fl \left\langle \left\langle X(\lambda ), Y(\lambda ) \right\rangle
\right\rangle _{k} = \mbox{Res} \, \lambda^{k+1} \left\langle
\hat{D}^+(\lambda ) X(\lambda ),Y(\lambda )\right\rangle , \qquad
X(\lambda ), Y(\lambda ) \in \widehat{\fr{g}},
\end{equation}
see \cite{RST}.

\section{Hamiltonian structures of the reduced $N $-wave
equations}\label{sec:6}

The generic $N $-wave interactions (i.e., prior to any reductions) possess
a hierarchy of Hamiltonian structures. As mentioned in the Introduction
the simplest one is $\{ H^{(0)}, \Omega ^{(0)}\} $; the symplectic form
$\Omega^{(0)} $ after simple rescaling
\[
q_\alpha \to {q'_\alpha \over \sqrt{(a,\alpha )} }, \qquad
p_\alpha \to {p'_\alpha \over \sqrt{(a,\alpha )} }, \qquad \alpha \in
\Delta _+, \]
becomes canonical with $q'_\alpha $ being canonically
conjugated to $p'_\alpha $.

The hierarchy of symplectic forms is generated by the so-called
generating (or recursion) operator $\Lambda =(\Lambda_+ + \Lambda _-)/2 $:
\begin{eqnarray}\label{eq:g-op}
\fl && \Lambda _\pm Z(x) =\ad_J^{-1} \left( i {dZ \over dx} + P_0 \cdot
\left( [q(x), Z(x)] \right) + i \left[ q(x), I_\pm \left(\openone
-P_0\right) [q(y),Z(y)] \right] \right), \nonumber\\
\fl && P_0 S \equiv \ad_J^{-1} \cdot \ad_J \cdot S, \qquad (I_\pm S)(x)
\equiv \int_{\pm\infty }^{x} dy\, S(y),
\end{eqnarray}
as follows:
\begin{eqnarray}\label{eq:hs-k}
\Omega^{(k)} = {i c_k\over 2} \int_{-\infty }^{\infty } dx\, \left\langle
[J, \delta Q(x,t)] \wedgecomma \Lambda ^k \delta Q(x,t) \right\rangle ,
\end{eqnarray}
where $q(x,t)=[J,Q(x,t)] $.
Using the completeness relation for the "squared" solutions which is
directly related to the spectral decomposition of $\Lambda $ we can
recalculate $\Omega ^{(k)} $ in terms of the scattering data of $L $ with
the result:
\begin{eqnarray}\label{eq:5.17}
&& \Omega ^{(k)} = {c_k \over 2\pi } \int_{-\infty }^{\infty } d\lambda
\lambda ^k \left( \Omega _0^+(\lambda ) - \Omega _0^-(\lambda )\right),
\nonumber\\
&& \Omega _0^\pm(\lambda ) = \left\langle \hat{D}^\pm(\lambda )
\hat{T}^\mp(\lambda ) \delta T^\mp(\lambda ) D^\pm(\lambda ) \wedgecomma
\hat{S}^\pm(\lambda ) \delta S^\pm(\lambda ) \right\rangle .
\end{eqnarray}
The first consequence from (\ref{eq:5.17}) is that the kernels of
$\Omega^{(k)} $ differs only by the factor $\lambda ^k $; i.e., all of
them can be cast into canonical form simultaneously. This is quite
compatible with the results of \cite{ZM,ZM1,BS} for the action-angle
variables.

Again it is not difficult to find how the reductions influence $\Omega
^{(k)} $. Using the invariance of the Killing form, from (\ref{eq:5.17})
and (\ref{eq:ra})--(\ref{eq:rc}) we get respectively:
\begin{eqnarray}\label{eq:5.18a}
&& \Omega ^+_0(\lambda ) = \left(\Omega _0^-(\lambda ^*)\right)^*, \\
\label{eq:5.18b}
&& \Omega ^+_0(\lambda ) = \Omega _0^-(-\lambda ),\\
\label{eq:5.18c}
&& \Omega ^+_0(\lambda ) = \left(\Omega _0^-(-\lambda ^*)\right)^*.
\end{eqnarray}
Then for $\Omega ^{(k)} $ from (\ref{eq:rb}) we find:
\begin{equation}\label{eq:5.19b}
\Omega ^{(k)} = (-1)^{k+1}\Omega ^{(k)}.
\end{equation}
Like for the integrals ${\cal D}_{j,k} $ we find that the reductions
(\ref{eq:ra}) and (\ref{eq:rc}) mean that $\Omega ^{(k)} $ become
real with a convenient choice for $c_k $.

Let us now briefly analyze the reduction (\ref{eq:rb}) which may lead to
degeneracies. We already mentioned that ${\cal D}_{j,2k}=0 $, see
(\ref{eq:5.15b}); in addition from (\ref{eq:5.19b}) it follows that
$\Omega ^{(2k)}\equiv 0 $. In particular this means that the canonical
2-form $\Omega ^{(0)} $ is also degenerate, so the $N $-wave equations
with the reduction (\ref{eq:rb}) do not allow Hamiltonian formulation with
canonical Poisson brackets. However they still possess a hierarchy of
Hamiltonian structures:
\begin{equation}\label{eq:5.20}
\Omega ^{(k)} \left( {dq \over dt }, \cdot \right) = \nabla H^{(k)},
\end{equation}
where $\nabla H^{(k)}= \Lambda \nabla_q H^{(k-1 )} $; by definition
$\nabla_q H=(\delta H)/(\delta q^T(x,t)) $. Thus we find that
while the choices $\left\{ \Omega ^{(2k)}, H^{(2k)} \right\}$ for the $N
$-wave equations are degenerate, the choices $\left\{ \Omega ^{(2k+1)},
H^{(2k+1)} \right\}$ provide us with correct nondegenerate (though
non-canonical) Hamiltonian structures, see \cite{2,G*86,VYa}.

This well known procedure for constructing the fundamental analytic
solutions of the Lax operators applies to the generic case when $J $ has
real and pair-wise distinct eigenvalues; such is the situation, e.g. in the
examples \ref{exa:g2.3} ($\eta =1$), and \ref{exa:b3.00}, \ref{exa:c3.00}.

However in several other examples $J $ has complex pair-wise distinct
eigenvalues. In such cases one should follow the procedure described in
\cite{VYa}. We do not have the space to do so here, but will mention the
basic differences. The most important one is that now the continuous
spectrum $\Gamma _L $ of $L $ is not restricted to the real axis, but
fills up a set of rays $\Gamma _L\equiv \cup_{\alpha }^{}l_\alpha $
which are determined by $l_\alpha \equiv \{\lambda : \im \lambda (\alpha,
\vec{a})=0\} $. Then it is possible to generalize the procedure described
above and to construct a fundamental analytic solution $\chi _\nu
(x,\lambda ) $ in each of the sectors closed between two neighboring
rays $l_\nu$. Then we can formulate again a RHP only now we will have
sewing function determined upon each of the rays $l_\nu $; possible
discrete eigenvalues will lie inside the sectors.

If we now impose the reduction the first consequence will be the symmetry
of $\Gamma _L $ with respect to it; more precisely, if $\lambda \in \Gamma
_L$ then also $\kappa (\lambda )\in \Gamma _L $.
{}Finally we just note that the consequences of imposing the reductions
(\ref{eq:ra})--(\ref{eq:rc}) will be similar to the ones already
described. In particular the reduction (\ref{eq:rb}) leads to the
degeneracy of "half" of the Hamiltonian structures, while the reductions
(\ref{eq:ra}) and (\ref{eq:rc}) make these structures real with
appropriate choices for $c_k $.

The last most difficult situations that takes place in many examples
above arises when two or more of the eigenvalues of $J $ become equal.
Then the construction of the FAS requires the use of the generalized
Gauss decompositions in which the factors $D^\pm(\lambda ) $ are
block-diagonal matrices while $T^\pm(\lambda ) $ and $S^\pm(\lambda ) $
are block-triangular matrices, see \cite{VSG*94}. These problems will be
addressed in subsequent papers.

\section{Conclusions}\label{sec:5}

We described the systems of $N $-wave type related to the low rank
simple Lie algebras. In section \ref{sec:descr} for any equivalence
class of the Weyl group for the corresponding Lie algebra we choose one
representative and we write down the corresponding reduced $N $-wave
system. The complete list of the reduced systems is given in the tables
below; two of the examples which we denote by $\ddag $ can formally be
listed in different locations of these tables, see Remark 1.

\vspace{0.1in}

\noindent
\begin{tabular}{|l|l|l||l|}
\hline \hline
${\frak g}\simeq {\bf A}_2 $ & $\openone $ & $\bbbz_2 $ &
$\mbox{Ad}_{{\frak  h}} $\\
\hline \hline
$C(U^T(\pm \lambda ))=-U(\lambda )
$ & Ex.  \ref{exa:a2.1} & \quad & \qquad \\ \hline $C(U^*(\pm \lambda
^*))=-U(\lambda ) $ & Ex. \ref{exa:a2.4} & Ex. \ref{exa:a2.5} &\qquad \\
\hline $C(U^{\dag}(\pm \lambda ^*))=U(\lambda ) $ & Ex. \ref{exa:a2.00} &
Ex. \ref{exa:a2.3} & Ex. 6\\ \hline \end{tabular}

\vspace{0.1in}

\noindent
\vspace{0.1in}
\begin{tabular}{|l|l|l|l|l||l|}
\hline \hline
${\frak g}\simeq {\bf C}_2 $ & $\openone $ & $-\openone$ & $\bbbz_2^{(1)}$ &
$\bbbz_2^{(2)}$ &$\mbox{Ad}_{{\frak  h}} $\\ \hline \hline
$C(U^*(\pm \lambda ^*))=-U(\lambda ) $ & Ex. 7 & Ex.
\ref{exa:c2.4}$^{\ddag} $ &
Ex. \ref{exa:c2.2} & Ex. \ref{exa:c2.3} &\qquad \\ \hline
$C(U(\pm \lambda ))=U(\lambda ) $ &\quad & Ex. \ref{exa:c2.5} &
\quad & \quad & \qquad \\ \hline
$C(U^\dag(\pm \lambda ^*))=U(\lambda ) $ & Ex. 8$^{\ddag} $  & & &
\quad & Ex.  9 \\ \hline

\end{tabular}

\vspace{0.1in}

\noindent
\begin{tabular}{|l|l|l|l|l||l|}
\hline \hline
${\frak g}\simeq {\bf G}_2 $ & $\openone $ & $-\openone$ &
$\bbbz_2^{(1)}$ & $\bbbz_2^{(2)}$ &$\mbox{Ad}_{{\frak  h}} $\\
\hline \hline $C(U^T(\pm \lambda )=-U(\lambda ) $ &  Ex.\ref{exa:g2.t}
& & \qquad & \qquad &\qquad \\
\hline $C(U^*(\pm \lambda ^*))=-U(\lambda ) $ & \qquad
& Ex. \ref{exa:g2.3}$^{\ddag} $ & Ex. \ref{exa:g2.1} & Ex.
\ref{exa:g2.2} &\qquad \\ \hline
\hline $C(U^\dag(\pm \lambda ^*))=U(\lambda ) $ & Ex. 14$^{\ddag} $
& \qquad  & \qquad  & \qquad
&Ex. 15 \\ \hline

\end{tabular}

\vspace{0.1in}

\noindent
\begin{tabular}{|l|l|l|l||l|}
\hline \hline
${\frak g}\simeq {\bf A}_3 $ & $\openone $ & $\bbbz_2^{(1)}$ &
$\bbbz_2^{(2)}$ &$\mbox{Ad}_{{\frak  h}} $\\ \hline \hline
$C(U(\pm \lambda ))=U(\lambda ) $ & \qquad
& Ex. \ref{exa:a3.4} & \qquad &\qquad \\ \hline
$C(U^*(\pm \lambda ^* ))=-U(\lambda ) $ & Ex. \ref{exa:a3.cc}
& Ex. \ref{exa:a3.7} & Ex. \ref{exa:a3.8} &\qquad \\ \hline
$C(U^T(\pm \lambda ))=-U(\lambda ) $ & Ex. \ref{exa:a3.t}
& Ex. \ref{exa:a3.3} & Ex. \ref{exa:a3.10} &\qquad \\ \hline
$C(U^{\dag}(\pm \lambda ^*))=U(\lambda ) $ & Ex. \ref{exa:a3.0}
& Ex. \ref{exa:a3.1} & Ex. \ref{exa:a3.2} & Ex.26\\ \hline
\end{tabular}

\vspace{0.1in}

\noindent
\begin{tabular}{|l|l|l|l|l|l|l||l|}
\hline \hline
${\frak g}\simeq {\bf B}_3 $ & $\openone $& $-\openone$ & $\bbbz_2^{(1)}$ &
$\bbbz_2^{(2)}$ & $\bbbz_2^{(3)}$ & $\bbbz_2^{(4)}$ &
$\mbox{Ad}_{{\frak  h}} $\\
\hline \hline
$C(U(\pm \lambda ))=U(\lambda ) $ &\quad & Ex.
\ref{exa:b3.9} & Ex. \ref{exa:b3.1} & Ex. \ref{exa:b3.2} & Ex.
\ref{exa:b3.3} & Ex. \ref{exa:b3.4} &\qquad \\ \hline
$C(U^{\dagger}(\pm
\lambda^* ))=U(\lambda ) $ & Ex. \ref{exa:b3.00} &
Ex. \ref{exa:b3.10} & Ex. \ref{exa:b3.7} &
Ex. \ref{exa:b3.8} & Ex. \ref{exa:b3.5} & Ex. \ref{exa:b3.6}
& Ex. 35\\ \hline
\end{tabular}

\vspace{0.1in}

\noindent
\begin{tabular}{|l|l|l|l|l|l|l||l|}
\hline \hline
${\frak g}\simeq {\bf C}_3 $& $\openone $ & $-\openone$ & $\bbbz_2^{(1)}$ &
$\bbbz_2^{(2)}$
& $\bbbz_2^{(3)}$ & $\bbbz_2^{(4)}$ & $\mbox{Ad}_{{\frak  h}} $ \\
\hline \hline
$C(U(\pm \lambda ))=U(\lambda ) $ & \quad& Ex.
\ref{exa:c3.16} & Ex. \ref{exa:c3.1} & Ex. \ref{exa:c3.2} & Ex.
\ref{exa:c3.3} & Ex. \ref{exa:c3.4} & \qquad \\ \hline
$C(U^{\dagger}(\pm
\lambda^* ))=U(\lambda ) $ & Ex. \ref{exa:c3.00}&
Ex. \ref{exa:c3.17} & Ex. \ref{exa:c3.5} &
Ex. \ref{exa:c3.6} & Ex. \ref{exa:c3.14} & Ex. \ref{exa:c3.15}
& Ex. 47\\ \hline
\end{tabular}

\vspace{0.1in}

The $N $-wave systems related to reductions from  the same equivalence
class will be equivalent. The empty boxes in the tables above
mean that the $N $-wave system after the reduction becomes trivial.

We end this paper with several remarks.

{\bf 1.} The $\bbbz_2 $-reductions which act on $\lambda $ by $\Gamma
_1(\lambda )=\lambda ^* $, combined with Cartan involutions on $\fr{g} $
lead in fact to restricting of the system to a specific real form of the
algebra $\fr{g} $.

{\bf 2.} To all reduced systems given above we can apply the analysis in
\cite{G*86,VYa} and derive the completeness relations for the
corresponding systems of ``squared'' solutions. Such analysis will allow
one to prove the pair-wise compatibility of the Hamiltonian structures and
eventually to derive their action-angle variables, see \cite{ZM,BS} for
the ${\bf A}_n $-series.

{\bf 3.} These results can be extended naturally in several directions:

\begin{itemize}

\item for NLEE with other dispersion laws. This would allow us to study
the reductions of the multicomponent NLS-type equations (see
\cite{Cho*99}), Toda type systems etc.

\item for Lax operators with more complicated $\lambda $-dependence, e.g.
\[
\fl L(\lambda )\psi = \left(i{d \over dx} + U_0(x,t)
+ \lambda U_1(x,t) + {1 \over \lambda }U_{-1}(x,t)
\right)\psi(x,t,\lambda ) = 0.
\]
This would allow us to investigate more complicated reduction groups as
e.g. $\bbbt $, $\bbbo $ (see \cite{Mi}) and the possibilities to imbed
them as subgroups of the Weyl group of $\fr{g} $.
\end{itemize}

{\bf 4.} In a series of papers Calogero \cite{Ca*89} demonstrated that a
number of integrable NLEE including ones of $N $-wave type are universal
and widely applicable in physics. Although the examples we have here do
not seem to coincide with the ones in \cite{Ca*89} there is a hope that
some of them might find applications in physics.

{\bf 5.} Some preliminary results concerning the $\bbbz_2\times\bbbz_2 $
reduction group are reported in \cite{Varna00}.

\section*{Acknowledgment}

We have the pleasure to thank Dr. L. Georgiev for valuable discussions and
one of the refferrees for numerous useful remarks.

\appendix
\section{}
\label{A}

{}For each of the algebras used above we list the sets ${\cal  M}_\alpha
$ which consist of all pairs of roots $\beta  $, $\gamma  $ such that
$\beta +\gamma =\alpha  $. The roots are denoted as usual by $j,k $ or
$i,j,k $; the negative roots are overlined. Obviously ${\cal  M}_\alpha  $
describes all possible decays of the $\alpha  $-type wave.

Next we write in more compact form the quantities $\omega_{jk} $
(\ref{eq:H-om}) using the following notations:
\begin{eqnarray}\label{eq:kap}
\kappa _1=a_2b_3-a_3b_2; \quad
\kappa _2=a_3b_1-a_1b_3; \quad
\kappa _3=a_1b_2-a_2b_1.
\end{eqnarray}

\medskip

{\bf 1. ${\frak  g}\simeq {\bf A}_2 $-- algebra.}
{}For this algebra there is only one "decay" of roots: $(11)=(10)+(01) $
and the corresponding coefficient is $\omega _{10,01}=6\kappa  $, where
$\kappa =a_1b_2-a_2b_1 $.

\medskip

{\bf 2. ${\frak  g}\simeq {\bf C}_2 $-- algebra.}
Here there are two "decays":
\[
(21)=(11)+(10), \qquad (11)=(10)+(01)
\]
and the corresponding coefficients $\omega _{jk} $ are:
\begin{eqnarray}\label{eq:c2.ap1}
\omega _{11,10}=-\omega _{10,01}=2\kappa , \qquad
\kappa =a_1b_2-a_2b_1.
\end{eqnarray}

\medskip

{\bf 3. ${\frak  g}\simeq {\bf G}_2 $-- algebra.} The sets ${\cal
M}_\alpha  $ are as follows:
\begin{eqnarray}\label{eq:g2.ma} \fl {\cal
M}_{10}=\{(11, \overline{1}),(21, \overline{11}), (31, \overline{21})\}
\qquad &{\cal M}_{1}=\{(11, \overline{10}),(32, \overline{31})\}
\nonumber\\ \fl {\cal M}_{11}= \{(10,1), (21, \overline{10}),(32,
\overline{21})\} \qquad &{\cal M}_{21}=\{(10, 11),(31, \overline{10}), (32,
 \overline{11})\} \\ \fl {\cal  M}_{31}=\{(21, 10),(32, \overline{1})\}
\qquad & {\cal M}_{32}=\{(31, 1),(11, 21)\}
\nonumber
\end{eqnarray}
and the coefficients $\omega _{jk} $ without reductions are
\begin{eqnarray}\label{eq:g2.ap1}
\omega _{10,01} = -2\omega _{10,11} = 2\omega _{21,10} =
2\omega _{31,01} = -2\omega _{21,11} =6\kappa
\end{eqnarray}

\medskip

{\bf 4. ${\frak  g}\simeq {\bf A}_3 $-- algebra.}
The sets ${\cal M}_{\alpha } $ for this algebra are as follows:
\begin{eqnarray}\label{eq:a3.ma}
{\cal  M}_{100}=\{(110, \overline{10}),(111, \overline{11})\} \qquad
&{\cal  M}_{10}=\{(110, \overline{100}),(11, \overline{1})\} \nonumber\\
{\cal  M}_{1}=\{(11, \overline{10}),(111, \overline{110})\} \qquad
&{\cal  M}_{110}=\{(100, 10),(111, \overline{1})\} \\
{\cal  M}_{11}=\{(10, 1),(111, \overline{100})\} \qquad
&{\cal  M}_{111}=\{(11, 100),(110, 1)\} \nonumber
\end{eqnarray}
and the general form of the quantities $\omega _{jk} $ are:
\begin{eqnarray}\label{eq:a3.om}
& \omega _{100,010} = 2(\kappa _1+\kappa _2+\kappa _3), \qquad
&\omega _{010,001} = 2(3\kappa _1-\kappa _2-\kappa _3), \nonumber\\
& \omega _{110,001} = 2(\kappa _1-3\kappa _2+\kappa _3),
&\omega _{100,011} = 2(-\kappa _1-\kappa _2+3\kappa _3),
\end{eqnarray}

{\bf 5. ${\frak  g}\simeq {\bf B}_3 $-- algebra.}
The sets ${\cal  M}_\alpha  $ here are given by:
\begin{eqnarray}
\label{eq:b3.ma}
{\cal  M}_{100} &=\{(110, \overline{10}),(111, \overline{11}), (112,
\overline{12})\} \nonumber\\
{\cal  M}_{10}&=\{(110, \overline{100}),(11, \overline{1}), (122,
\overline{112})\} \nonumber\\
{\cal  M}_{1}&=\{(11,\overline{10}), (12, \overline{11}), (111,
\overline{110}),(112, \overline{111})\} \nonumber\\
{\cal  M}_{110}&=\{(100, 10),(111, \overline{1}), (122,\overline{12})\} \\
{\cal  M}_{11}&=\{(1, 10),(12, \overline{11}),
(111, \overline{100}),(122, \overline{111})\} \nonumber\\
{\cal  M}_{111}&=\{(100, 11),(110, 1), (122, \overline{11}),
(112, \overline{1})\} \nonumber\\
{\cal  M}_{12}&=\{(1, 11), (112, \overline{100}), (122, \overline{110})\}
\nonumber\\
{\cal  M}_{112}&=\{(100, 12), (111, 1), (122, \overline{10})\} \nonumber\\
{\cal  M}_{122}&=\{(11, 111), (112, 10), (110, 12)\} \nonumber
\end{eqnarray}
and the coefficients $\omega _{jk} $ (\ref{eq:H-om}) are:
\begin{eqnarray}\label{eq:b3.o}
\fl \omega _{100,10}=2(\kappa _1+\kappa _2+\kappa _3), \quad
&\omega _{100,11}=4\kappa _3, \quad
&\omega _{100,12}=-2(\kappa _1+\kappa _2+\kappa _3), \nonumber\\
\fl \omega _{110,1}=2(\kappa _1-\kappa _2+\kappa _3), \quad
&\omega _{10,1}=4\kappa _1, \quad
&\omega _{10,112}=-2(\kappa _1-\kappa _2-\kappa _3), \\
\fl \omega _{1,110}=-4\kappa _2, \quad &\omega _{1,11}=-4\kappa _1, \quad
&\omega _{1,111}=-4\kappa _2, \quad \omega _{11,111}=-4\kappa _3 \nonumber
\end{eqnarray}
with $\kappa _1$, $\kappa _2 $ and $\kappa _3 $ given by (\ref{eq:kap}).

{\bf 6. ${\frak  g}\simeq {\bf C}_3 $-- algebra.} Here we have:
\begin{eqnarray}\label{eq:c3.ma}
{\cal  M}_{100}=\{(110, \overline{10}),(111, \overline{11}),
 (121, \overline{21}), (221, \overline{121})\} \nonumber\\
{\cal  M}_{10}=\{(110, \overline{100}),(11, \overline{1}),
(21, \overline{11}), (121, \overline{111}\} \nonumber\\
{\cal  M}_{1}=\{(11,\overline{10}), (111, \overline{110})\}
\nonumber\\
{\cal  M}_{110}=\{(100, 10),(111, \overline{1}), (121,\overline{11}),
(221, \overline{111}\} \\
{\cal  M}_{11}=\{(1, 10),(111, \overline{100}),
(21, \overline{10}),(121, \overline{110})\} \nonumber\\
{\cal  M}_{111}=\{(100, 11),(110, 1), (221, \overline{110}),
(121, \overline{10})\} \nonumber\\
{\cal  M}_{21}=\{(10, 11), (121, \overline{100})\} \nonumber\\
{\cal  M}_{121}=\{(100, 21), (110, 11), (111, 10), (221, \overline{100}\}
\nonumber\\
{\cal  M}_{122}=\{(121, 100), (110, 111)\} \nonumber
\end{eqnarray}
and the coefficients $\omega _{jk} $ (\ref{eq:H-om}) are:
\begin{eqnarray}\label{eq:c3.o}
\fl \omega _{100,10}=2(\kappa _1+\kappa _2+\kappa _3), \quad
\omega _{100,11}=-2(\kappa _1+\kappa _2-\kappa _3), \quad
\omega _{100,21}=4\kappa _3, \nonumber\\
\fl \omega _{110,11}=-2(\kappa _1-\kappa _2+\kappa _3), \quad
\omega _{10,11}=-4\kappa _1, \quad \omega _{1,110}=-4\kappa _2, \\
\fl \omega _{10,1}=4\kappa _1, \quad
\omega _{10,111}=-2(\kappa _1 -\kappa _2-\kappa _3),
\qquad \omega _{100,121}=-4\kappa _3, \quad \omega _{110,111}=-4\kappa _2
\nonumber
\end{eqnarray}
where again $\kappa _1$, $\kappa _2 $ and $\kappa _3 $ are given by
(\ref{eq:kap}).

\section{}
\label{B}

Here we show some typical interaction terms for each of the four types of
reductions. There we have also replaced the constant $c_0 $ by its
apropriate value.  The form of these terms crucially depends on
whether the roots $\alpha  $, $\beta  $ and $\gamma  $ belong to
$\Delta_+^0 $ or $\Delta_+^1 $.

{\bf $a)\; \alpha ,\beta ,\gamma \in \Delta _+^0  $.}
\begin{eqnarray}\label{eq:11.1}
\fl \mbox{1)} \qquad
\left\{ \begin{array}{cc} \alpha  \\ \beta , \gamma \end{array} \right\}
_R = {1  \over  \sqrt{\eta }}
n_{\alpha ,\alpha '} \omega _{\beta ,\gamma } \int_{-\infty
}^{\infty } dx\, \left( q_\alpha q_{\beta '}^{*} q_{\gamma '}^{*} + \eta
q_{\alpha '}^{*} q_\beta q_\gamma \right).  \\
\fl \mbox{2)} \qquad \left\{ \begin{array}{cc} \alpha  \\ \beta , \gamma
\end{array} \right\} _R = n_{\alpha ,\alpha '} \omega _{\beta ,\gamma }
\int_{-\infty }^{\infty } dx\, \left( q_\alpha q_{\beta '} q_{\gamma '} +
\eta q_{\alpha '} q_\beta q_\gamma \right).   \\
\fl \mbox{3)} \qquad
\left\{ \begin{array}{cc} \alpha  \\ \beta , \gamma \end{array} \right\}
_R = {1  \over  \sqrt{-\eta }} n_{\alpha ,\alpha '} \omega _{\beta ,\gamma
} \int_{-\infty }^{\infty } dx\, \left( q_\alpha p_{-\beta '}^* p_{-\gamma
'}^* - \eta p_{-\alpha '}^* q_\beta q_\gamma \right). \\
\fl \mbox{4)} \qquad
\left\{ \begin{array}{cc} \alpha  \\ \beta , \gamma \end{array} \right\}
_R =  n_{\alpha ,\alpha '} \omega _{\beta ,\gamma } \int_{-\infty
}^{\infty } dx\, \left( q_\alpha p_{-\beta '} p_{-\gamma '} - \eta
p_{-\alpha '} q_\beta q_\gamma \right).
\end{eqnarray}

{\bf $b)\; \alpha ,\gamma \in \Delta _+^0, \beta \in \Delta _+^1$.}
\begin{eqnarray}\label{eq:11.4}
\fl \mbox{1)} \qquad
\left\{ \begin{array}{cc} \alpha  \\ \beta , \gamma \end{array} \right\}
_R = {1  \over  \sqrt{\eta }} n_{\alpha ,\alpha '} \omega _{\beta ,\gamma
} \int_{-\infty }^{\infty } dx\, \left( q_\alpha p_{-\beta '}^{*}
q_{\gamma '}^{*} + \eta q_{\alpha '}^{*} q_\beta q_\gamma \right). \\
\fl \mbox{2)} \qquad \left\{ \begin{array}{cc} \alpha  \\ \beta , \gamma
\end{array} \right\} _R = n_{\alpha ,\alpha '} \omega _{\beta ,\gamma }
\int_{-\infty }^{\infty } dx\, \left( q_\alpha p_{-\beta '} q_{\gamma '} +
\eta q_{\alpha '} q_\beta q_\gamma \right).   \\
\fl \mbox{3)} \qquad
\left\{ \begin{array}{cc} \alpha  \\ \beta , \gamma \end{array} \right\}
_R = {1  \over  \sqrt{-\eta }} n_{\alpha ,\alpha '} \omega _{\beta
,\gamma } \int_{-\infty }^{\infty } dx\, \left( q_\alpha q_{-\beta '}^{*}
p_{-\gamma '}^{*} - \eta p_{-\alpha '}^{*} q_\beta q_\gamma \right).   \\
\fl \mbox{4)} \qquad \left\{ \begin{array}{cc} \alpha  \\ \beta , \gamma
\end{array} \right\} _R = n_{\alpha ,\alpha '} \omega _{\beta ,\gamma }
\int_{-\infty }^{\infty } dx\, \left( q_\alpha q_{-\beta '} p_{-\gamma '}
- \eta p_{\alpha '} q_\beta q_\gamma \right).
\end{eqnarray}
The right hand side of (\ref{eq:11.4}) coincides with the standard
interaction terms in (\ref{eq:hI-al}) only if $\beta '=\beta  $ and
$\gamma '=-\gamma  $, i.e. only if $\beta ,\gamma \in \Delta _+^\perp $.
There are  a special subcases of (\ref{eq:11.1}) when $\gamma '=\alpha  $;
then
\begin{eqnarray}\label{eq:11.2}
\fl \mbox{1)} \qquad
\left\{ \begin{array}{cc} \alpha  \\ \alpha ' , \gamma \end{array} \right\}
_R = {1  \over  \sqrt{\eta }} n_{\alpha ,\alpha '} \omega _{\beta ,\gamma
} \int_{-\infty }^{\infty } dx\, \left( |q_\alpha|^2 p_{-\beta '}^{*} +
\eta |q_{\alpha '}|^2  q_\beta \right). \\
\fl \mbox{2)}  \qquad \left\{ \begin{array}{cc} \alpha  \\ \alpha ' ,
\gamma \end{array} \right\} _R = n_{\alpha ,\alpha '} \omega _{\beta
,\gamma } \int_{-\infty }^{\infty } dx\, \left( q_\alpha ^2 p_{-\beta } +
\eta q_{\alpha '}^2  q_\beta \right).  \\
\fl \mbox{3)} \qquad
\left\{ \begin{array}{cc} \alpha  \\ \alpha ' , \gamma \end{array} \right\}
_R = {1  \over  \sqrt{-\eta }} n_{\alpha ,\alpha '} \omega _{\beta ,\gamma
} \int_{-\infty }^{\infty } dx\, \left( q_\alpha p_{-\alpha }^* q_{-\beta
'}^* -  \eta p_{-\alpha '}q_{\alpha '}^*  q_\beta \right).   \\
\fl \mbox{4)} \qquad
\left\{ \begin{array}{cc} \alpha  \\ \alpha ' , \gamma \end{array}
\right\} _R =  n_{\alpha ,\alpha '} \omega _{\beta ,\gamma } \int_{-\infty
}^{\infty } dx\, \left( q_\alpha p_{-\alpha } q_{\beta } -  \eta
p_{-\alpha '} q_{\alpha '} q_\beta \right).
\end{eqnarray}

{\bf $c)\; \alpha ,\beta ,\gamma \in \Delta _+^1, $.}
\begin{eqnarray}\label{eq:11.6}
\fl \mbox{1)} \qquad
\left\{ \begin{array}{cc} \alpha
\\ \beta , \gamma \end{array} \right\} _R ={1  \over  \sqrt{\eta }}
 n_{\alpha ,\alpha '} \omega_{\beta ,\gamma } \int_{-\infty }^{\infty }
dx\, \left( q_\alpha p_{-\beta '}^{*} p_{-\gamma '}^{*} + \eta p_{-\alpha
'}^{*} q_\beta q_\gamma \right)\\
\fl \mbox{2)} \qquad \left\{ \begin{array}{cc} \alpha \\ \beta , \gamma
\end{array} \right\} _R = n_{\alpha ,\alpha '} \omega _{\beta ,\gamma }
\int_{-\infty }^{\infty } dx\, \left( q_\alpha p_{-\beta '} p_{-\gamma '}
+ \eta p_{-\alpha '} q_\beta q_\gamma \right) \\
\fl \mbox{3)} \qquad \left\{ \begin{array}{cc} \alpha
\\ \beta , \gamma \end{array} \right\} _R ={1  \over  \sqrt{-\eta }}
n_{\alpha ,\alpha '} \omega _{\beta ,\gamma } \int_{-\infty
}^{\infty } dx\, \left( q_\alpha q_{-\beta '}^{*} q_{-\gamma '}^{*} - \eta
q_{-\alpha '}^{*} q_\beta q_\gamma \right).  \\
\fl \mbox{4)} \qquad \left\{ \begin{array}{cc} \alpha
\\ \beta , \gamma \end{array} \right\} _R =
 n_{\alpha ,\alpha '} \omega _{\beta ,\gamma } \int_{-\infty
}^{\infty } dx\, \left( q_\alpha q_{-\beta '} q_{-\gamma '} - \eta
q_{-\alpha '} q_\beta q_\gamma \right).
\end{eqnarray}

{\bf $d)\; \alpha ,\gamma \in \Delta _+^1, \beta \in \Delta _+^0$.}
\begin{eqnarray}\label{eq:11.8}
\fl \mbox{1)} \qquad
\left\{ \begin{array}{cc} \alpha  \\ \beta , \gamma \end{array} \right\}
_R ={1  \over  \sqrt{\eta }} n_{\alpha ,\alpha '} \omega _{\beta ,\gamma }
\int_{-\infty }^{\infty } dx\, \left( q_\alpha q_{\beta '}^{*} p_{-\gamma
'}^{*} + \eta p_{-\alpha '}^{*} q_\beta q_\gamma \right).   \\
\fl \mbox{2)} \qquad \left\{ \begin{array}{cc} \alpha  \\ \beta , \gamma
\end{array} \right\} _R = n_{\alpha ,\alpha '} \omega _{\beta ,\gamma }
\int_{-\infty }^{\infty } dx\, \left( q_\alpha q_{\beta '} p_{-\gamma '} +
\eta p_{-\alpha '} q_\beta q_\gamma \right).   \\
\fl \mbox{3)} \qquad
\left\{ \begin{array}{cc} \alpha  \\ \beta , \gamma \end{array} \right\}
_R = {1  \over  \sqrt{-\eta }} n_{\alpha ,\alpha '} \omega _{\beta ,\gamma
} \int_{-\infty }^{\infty } dx\, \left( q_\alpha p_{-\beta '}^*q_{-\gamma
'}^* - \eta q_{-\alpha '}^* q_{\beta }q_\gamma \right).  \\
\fl \mbox{4)} \qquad \left\{ \begin{array}{cc} \alpha  \\ \beta , \gamma
\end{array} \right\} _R = n_{\alpha ,\alpha '} \omega _{\beta ,\gamma }
\int_{-\infty }^{\infty } dx\, \left( q_\alpha p_{-\beta '} q_{-\gamma '}
- \eta q_{-\alpha '} q_\beta q_\gamma \right).
\end{eqnarray}
with a similar particular case when $\alpha =-\gamma ' $.

\end{document}